\documentclass{article}
\usepackage{float}
\usepackage{amsmath}

\usepackage[figuresright]{rotating}

\usepackage{times}
\usepackage{bm}
\usepackage[numbers]{natbib}

\usepackage{graphics,color,pict2e}
\usepackage{epsfig}
\usepackage{xcolor}

\date{July 30, 2026}

\begin{document}

\title{Comparing Two Survival Functions at a Fixed Timepoint}

\author{Michael P. Fay$^1$,  Allyson Mateja, and Megan C. Grieco  \\[4pt]
\textit{Biostatistics Research Branch, National Institute of Allergy and Infectious Diseases, Bethesda, MD (MPF)}, \\
\textit{Clinical Monitoring Research Program Directorate, Frederick National Laboratory for Cancer Research,} \\
\textit{Fredrick, MD  (AM), and Axle Informatics, Rockville, MD (MCG)}  \\
{mfay@niaid.nih.gov}}

\markboth%
{Fay, Mateja, and Grieco}
{Comparing Two Survival Functions}

\maketitle

\addtocounter{footnote}{1}
\footnotetext{Correspondence to M.P. Fay.}
\footnotetext{This is a work of the U.S. Government and is not subject to copyright protection in the United States}

\begin{abstract}
For comparing two survival curves at a fixed timepoint with right censoring, a standard method uses asymptotic methods on two Kaplan-Meier estimators with Greenwood variance estimators and transformations using the delta method. Confidence intervals on associated estimands can either breakdown or undercover due to Kaplan-Meier estimates of zero or one, small sample sizes, or heavy censoring. Although others have proposed adjustments to the Greenwood variance for a single sample, these adjustments have not been incorporated into the two-sample application of the delta method and we do that here as well as provide modifications when Kaplan-Meier estimates are zero or one. An alternative non-asymptotic existing method called melding on the beta product confidence procedure (BPCP) appears to have at least nominal coverage regardless of the sample size. Without censoring the problem reduces to comparing two independent binomials and melding on the BPCP gives p-values equivalent to Fisher's exact test p-values and further gives compatible confidence intervals for treatment comparison estimands; however, the melding can be very conservative. In this paper, we expand the melding on the BPCP intervals to include mid-p versions, which are designed to achieve coverage closer to nominal without guaranteeing coverage for all cases. We study these existing methods and our minor modifications of them primarily by numerical methods or simulation. Results suggest only the melding on the BPCP can guarantee coverage in all studied scenarios, and the mid-p version most often has closest to nominal coverage. We provide R functions in the {\sf bpcp} R package.
\end{abstract}


\maketitle

\section{Introduction}
\label{sec-intro}

This paper focuses on comparing survival functions for two groups at a specified time when there may be right censoring.
Let $S_1(t)$ and $S_2(t)$ be survival functions for the two groups at a milestone time $t$, and we consider estimands for
comparing them such as the difference: $S_2(t)-S_1(t)$, the ratio: $S_2(t)/S_1(t)$,
and the treatment effect using ratios of log survival: $1- \left[ \log \left\{ S_2(t) \right\}/\log
\left\{ S_1(t) \right\} \right]$.
The latter estimand is equivalent to one minus the hazard ratio under the proportional hazards assumption, but
is straightforwardly defined without that assumption.
These kinds of estimands may be applicable in situations where, for example, we want to compare the 6 month survival between two randomized treatment
regimens, and one regimen may have worse survival at the beginning but by the end of 6 months may have a better survival.  In that case, the proportional
hazards assumption fails and the hazard ratio estimated from a Cox model does not have a straightforward causal interpretation (see e.g.,
\cite{FayL:2024}).
Further, even the restricted mean survival or scaled versions of it
\citep{Karr:2025} may not capture the advantage of a treatment that has better survival at 6 months.

There are several contributions of this paper. First, we point out issues with the standard asymptotic methods when there is extreme data (small sample sizes, survival estimates at the boundary, and/or heavy censoring),
and we propose adjustments to handle that extreme data. Next, we  review the less well known melding on the beta product confidence procedure (BPCP) which more straightforwardly handles extreme data, and we introduce a 
mid-p version of it. Finally,  we compare the methods by either numeric calculation or simulation.  

The canonical example is comparing survival at milestone time $t$ 
in a randomized trial of a new  treatment compared  to 
standard care  for a serious disease. We are interested in if the new treatment is better, and if so, how much better it is than the standard care.   
Thus, we emphasize not just showing a significant difference with good power and bounding the type I error rate, but also supplying  a treatment effect parameter 
and its confidence interval with good coverage. 

For hypothesis tests comparing the two survival curves at a specified $t$, a traditional approach is to use the asymptotic normality of the Kaplan-Meier curves at $t$,
the Greenwood formula for the associated variances, and perhaps use a transformation with the delta method.
These traditional methods can create confidence intervals on comparison parameters (e.g., difference in [or ratio of] two survival functions at a fixed timepoint) that asymptotically have coverage that 
equals the nominal level (see e.g., \cite{Klei:2007}).
Klein, {\it et al}\cite{Klei:2007} performed simulations testing the null hypothesis of equal survival distributions at $t$. They tried different transformations  (identity, log, complementary log-log, logit,  and the arcsin function), and found that in general the best transformation in terms of bounding the type I error and maximizing power was the  complementary log-log transformation.
One issue with these traditional applications of the delta method is that the Greenwood formula can give zero variances if the Kaplan-Meier curve at $t$ is either $0$ or $1$.
Klein {\it et al.} noted that problem but did not address that issue, and based their simulated power estimates only on samples where the Kaplan-Meier estimates at $t$ are not equal to $0$ or $1$. Another issue is the small sample bias of the Greenwood variance formula.
Borkowf\cite{Bork:2005} introduced less biased estimates.  Unlike Greenwood's formula which
only changes when the Kaplan-Meier curve changes, Borkowf's variance estimators may increase as more individuals are censored between failure times, since fewer at risk correctly  implies
less precision on the survival estimate.
Borkowf\cite{Bork:2005} proposed several adjustments for the Greenwood variance, although not all produce non-zero variance estimates.
We propose and study the delta method using Borkowf's adjustments applied to the two-sample problem, and provide {\it ad hoc } modifications to those versions and the standard version of the delta method so that all may be applied
when Kaplan-Meier estimators are $0$ or $1$.

Fay {\it et al}\cite{FayP:2015} proposed  {\it melding}, a confidence interval procedure for two-sample estimands
 that melds together lower and upper confidence distributions associated with the one-sample estimands.
 They applied it to the problem of testing the null hypothesis of $S_2(t)=S_1(t)$ by using confidence intervals that
  melded  two beta product confidence procedures (BPCPs)\citep{FayB:2013}. The BPCP gives confidence intervals on a survival function at a specified $t$
  that: (1) seamlessly provide intervals when Kaplan-Meier estimates are $0$ or $1$, (2) reduce to exact binomial intervals \citep{Clop:1934} when there is no censoring,  (3) guarantee coverage under progressive type II censoring, and
  (4) have coverage that asymptotically approaches the nominal level with independent censoring
  (see \cite{FayB:2013}, Theorem~2),
  and (5) appear by simulation to have at least nominal level with independent censoring\citep{FayB:2016}. When there is no censoring, melding on the BPCP gives intervals that are
  compatible with the central Fisher's exact test (compatibility means a $100(1-\alpha)\%$ interval excludes parameters in the null hypothesis space if and only if the test rejects at level $\alpha$)\cite{FayP:2015}.
Simulations by Fay {\it et al}\cite{FayP:2015} estimated the proportion where the true parameter was either higher or lower than the confidence interval given data generated with $S_2(t)=S_1(t)$ with samples sizes of $n=50$ and $n=100$ with moderate and heavy censoring. The simulated error for the 95\% intervals showed that testing based on melding on the BPCP were  conservative (errors less than $1\%$ for a one-sided error with nominal 2.5\% target on each side for a central interval), while the delta method using a complementary log-log transformation could have an inflated one-sided error rate (almost 10\%).   Those simulations were only a `preliminary assessment' of melding on the BPCP, and this paper provides a more in-depth assessment, studying their coverage on different estimands. Further, we take  the  mid-p version of the BPCP \citep{FayB:2016}, and propose melding it to obtain confidence intervals on two-sample estimands that are closer to nominal on average. Although that proposal is 
a straightforward application of melding, it has not previously been proposed or studied, and we  study it by simulation in this paper.  

Other methods have been used for confidence intervals  for estimands comparing $S_1(t)$ and $S_2(t)$
which will not be explored here.
Andersen, {\it et al.}\cite{Ande:2003} proposed a pseudo-observation approach, but for the two-sample case,
Klein {\it et al}\cite{Klei:2007}  has shown by simulation that it has relatively poor power compared to the delta method with the complementary log-log transformation.
Tang\cite{Tang:2021} explored using the MOVER method, which uses specific one-sample confidence limits to create two sample tests of differences or ratios of parameters.
For the difference it is equivalent to standard asymptotic methods, and
 we expect the MOVER methods for the ratio to be similar to the delta methods.
We do not explore  melding using other one-sample intervals, since many of the methods developed  for one-sample Kaplan-Meier confidence interval  \citep[e.g.,][]{Thom:1975,Barb:1999,Stra:1997,Akri:1986} have previously been shown to be anti-conservative in some cases\cite{FayB:2013}.
Two-sample tests may be done by empirical likelihood\cite{Thom:1975,Owen:2001}, but they require specialized constrained maximization for each estimand.
Zhao {\it et al}\cite{Zhao:2012}  explore the difference and ratio of two survival functions using pertubation-resampling, but since that is a resampling-based method that may be slower especially in simulations.

Here is an outline of the paper. In Section~\ref{sec-inferential} we introduce notation and the class of estimands of interest, together with our overall inferential goals, which focus on central two-sided confidence intervals and one-sided directional tests of effects.
Section~\ref{sec-motivating} describes some motivating applications.
In Section~\ref{sec-delta} we describe the delta method and modifications to it including Borkowf's adjustments and our {\it ad hoc } fixes when Kaplan-Meier estimates are $0$ or $1$.  In Section~\ref{sec-melding} we review melding and the beta product
confidence procedure and propose melding on the mid-p BPCP. 
Section~\ref{sec-no-censoring} numerically examines
type I error rates and coverage when there is no censoring,
Section~\ref{sec-simulations} describes  simulations with censoring,
Section~\ref{sec-applications} returns to the motivating applications, and
Section~\ref{sec-discussion} gives final recommendations.

\section{Inferential Goals}
\label{sec-inferential}

We are interested directional inferences, how much larger (or smaller)   $S_1(t)$ is  than $S_2(t)$. To measure the size of the effect, we choose an estimand from a class that has the following 
properties:
\begin{enumerate}
\item The estimand is $\beta=b \left\{ S_1(t), S_2(t) \right\}$,  defined by a function $b$ that inputs $S_1(t)$ and $S_2(t)$
and outputs a scalar.
\item There is a unique value $\beta_e$, where if $S_1(t)=S_2(t)$ then $b \left\{ S_1(t),S_2(t) \right\} = \beta_e$.
\item The estimand $\beta=b \left\{ S_1(t), S_2(t) \right\}$ is monotonically decreasing in 
$S_1(t)$ and monontonically increasing in $S_2(t)$.
\item The estimand $\beta=b \left\{ S_1(t), S_2(t) \right\}$ may be rewritten in the form
$b \left\{ S_1(t), S_2(t) \right\} = g \left[ h \left\{ S_2(t) \right\} - h \left\{ S_1(t) \right\}  \right]$, where $h(\cdot)$ is a differentiable monotonic function and $g(\cdot)$ is a monotonic function. 
 \end{enumerate}
The first three properties imply that 
if $S_2(t) > S_1(t)$ then $b \left\{ S_1(t), S_2(t) \right\} > \beta_e$, and
if  $S_2(t) < S_1(t)$ then $b \left\{ S_1(t), S_2(t) \right\} < \beta_e$.
This ensures that all estimands in the class have the same direction.
The third property is needed for using the melded intervals.
The last property is additionally needed for intervals that use the delta method.

\begin{figure}[tb]
\includegraphics[width=6in]{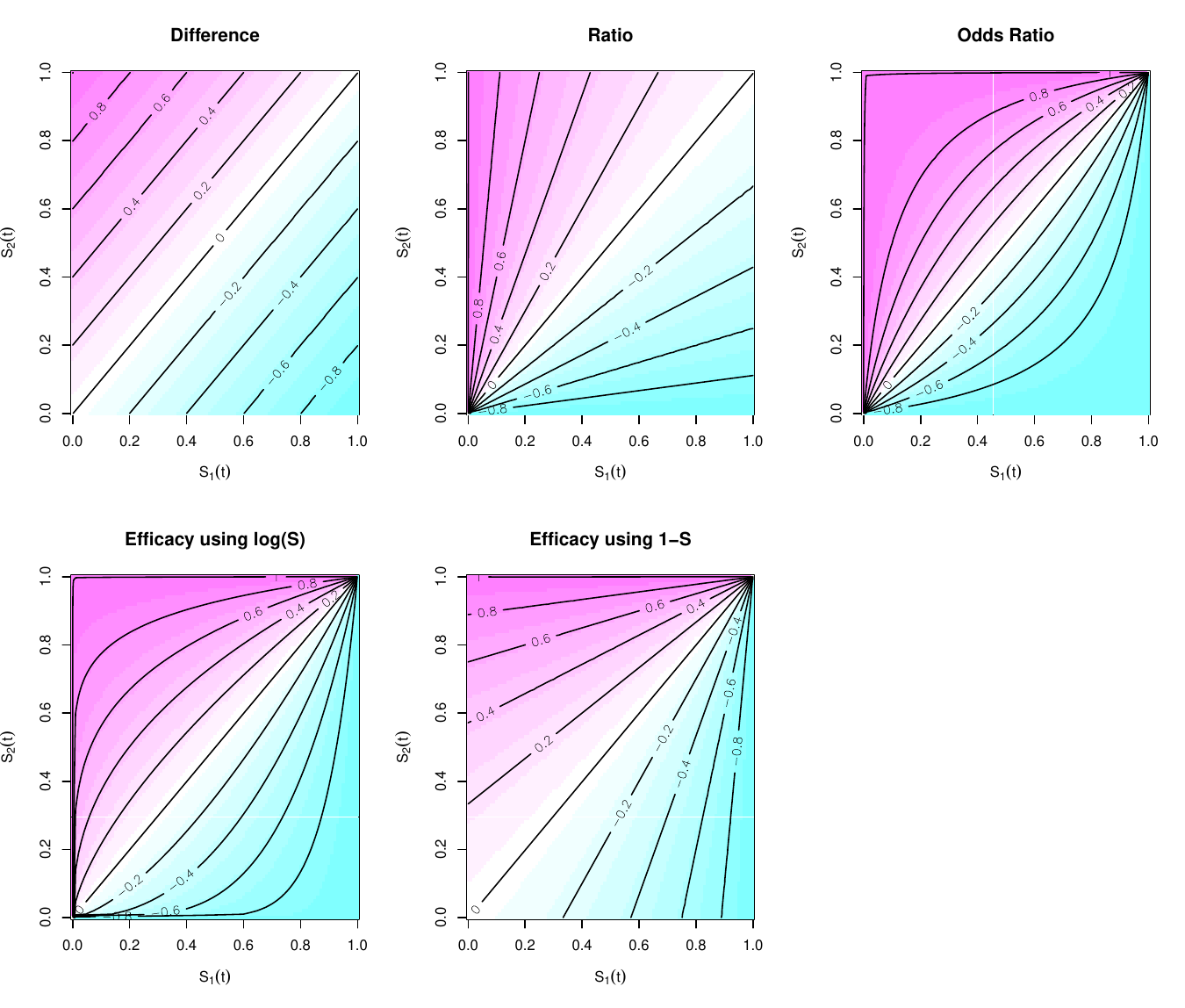}
\caption{ Five estimands: difference, $\beta_d=S_2(t)-S_1(t)$;
ratio, $\beta_r=S_2(t)/S_1(t)$; odds ratio, $\beta_{or}=\left\{ S_2(t) \left[ 1- S_1(t) \right] \right\}/\left\{ S_1(t) \left[ 1- S_2(t) \right] \right\}$; efficacy using $\log(S)$, $\beta_{els}=1- \log \left\{ S_2(t) \right\}/\log \left\{ S_1(t) \right\}$; and efficacy using $1-S$, $\beta_{ecd}=1- \left\{ 1- S_2(t) \right\}/\left\{ 1-S_1(t) \right\}$. Colors and contours use possibily different plotting transformations, $m(\beta)$, so that all estimands have $m(\beta_e)=0$, are symmetric about $0$, and range from $-1$ to $1$. 
The transformations are: for difference, $m(\beta)=\beta$; for
ratio and odds ratio, $m(\beta) = \frac{\beta-1}{\beta+1}$;
and for efficacy using $\log(S)$ or $1-S$, $m(\beta) = \frac{\beta}{2-\beta}$.
 \label{fig:B5panels}
 }
\end{figure}

We focus on 5 estimands in this paper (see Figure~\ref{fig:B5panels}).  We use different plotting transformations so that the effects are
all on the same $-1$ to $1$ scale and are symmetric so that switching group labels changes only the sign of the effect.  
The plotting scales have the similar motivation as plotting ratios on the log scale, which ensure that $S_2(t)/S_1(t) = 2$ is the same distance from $1$ as $1/2$, which one would get by switching the group labels. 
Unlike the log transformation on the ratio,  the transformation we use for the ratio,  $m(\beta_{r}) = \left\{ \beta_r-1 \right\}/ \left\{ \beta_r+1 \right\}$,  has the additional property that the extreme values are finite (specifically, 
 $m(b(1,0) )=-1$ and $m(b(0,1))=1$).  We use different transformations for the different $\beta$ parameters so that all are on the same scale (see Figure~\ref{fig:B5panels} caption).  This allows us to visually compare the different 
comparison parameters over their full ranges (including at their extrema) in a fair way. 

The choice of the estimand depends on the application. For example, for individually randomized natural exposure vaccine studies, typically the two efficacy estimands (i.e., the estimand plots in the lower panels of Figure~\ref{fig:B5panels} )
are used partly because the number of events is usually a small proportion of the sample size and it is untenable to control the natural exposure rate. Under those situations, those two estimands can be approximately invariant to the natural exposure rate, and hence the causal effects are more transportable to different environments \citep{FayF:2026}.

In terms of causal interpretation, $S_2(t) - S_1(t)$ has both an individual-level interpretation and a population-level interpretation, but the other estimands have only a population-level interpretation \citep{FayL:2024}. Efficacy using log(S) (i.e., $\beta_{els}$, see Figure~1 caption) is defined nonparametrically, but under the proportional hazards assumption may be interpreted as
1 minus the population-level hazard ratio. Since the proportional hazards assumption may not be true (and is falsifiable),
it is easier interpretationally to pre-specify the efficacy using log(S) as the estimand rather than the hazard ratio from a Cox model, which is complicated to interpret causally without that assumption  \citep[see][]{FayL:2024,WeiS:2008}.

Once we choose the estimand, we focus on testing two one-sided hypotheses at the $\alpha/2$ level, because we typically care more about whether the treatment is better or worse than the control, rather than only showing if they are different. We use central $100(1-\alpha)\%$  confidence intervals, which can be inverted to create tests that are compatible with those two hypotheses.
Central intervals may not be the smallest confidence intervals \citep{Agre:2001}; however, they are better than other two-sided intervals for getting powerful directional inferences
 \citep[see e.g.,][Section 4.3]{FayH:2021}.

We use plug-in estimators, $\hat{\beta} = b \left\{ \hat{S}_1(t), \hat{S}_2(t) \right\}$,
 where $\hat{S}_1(t)$ and  $\hat{S}_2(t)$ are Kaplan-Meier estimators.
 Notationally, we write $b \left\{ S_1(t),S_2(t) \right\}$ as
 $g \left[  h \left\{ S_2(t) \right\} - h \left\{ S_1(t) \right\} \right]$,
 where $h(\cdot)$ is the function used in the delta method, which is useful when
 $h(\hat{\beta})$ is closer to asymptotic normality than $\hat{\beta}$.  The difference
 $D =  h \left\{ \hat{S}_2(t) \right\} - h \left\{ \hat{S}_1(t) \right\}$ is asymptotically normal
 and $g(\cdot)$ transforms $D$ into our $\beta$ estimator.
 Table~\ref{tab:betas} in the Supplement gives those functions for the 5 estimands.
If there is ambiguity in the definition when $\hat{S}_i(t)=0$ or $1$,  then we use limits.

\section{Motivating Applications}
\label{sec-motivating}

We consider two studies that highlight the problems with the standard delta method approach.

Swennes\cite{Swen:2007} studied cross-immunity in vaccinations of channel catfish with two different
serotypes of a parasitic ciliate. Two vaccinations based on either isolate G5 or G12, where challenged with either of the two isolates. The factorial experiment had 6 arms of 20 fish each: one of three vaccines (G5, G12 or control) where challenged with either the G5 or G12 isolate.
Swennes\cite{Swen:2007} Figure~1  gives the Kaplan-Meier curves for the arms, and the only censoring
occurs at 26 days  after challenge due to planned end of the follow-up.
The proportion survived to Day $t=26$ are
20/20 (vaccination=G12, challenge=G12),
20/20 (V=G12, C=G5),
13/20 (V=G5, C=G12),
16/20 (V=G5, C=G5),
0/20 (V=control, C=G12), and
0/20 (V=control, C=G5). 
The original study used logrank tests to compare the arms but gave no estimands for the size of the effects. Since the logrank test is a score test on the Cox model, the natural estimand
for the study is the hazard ratio. As stated previously, the efficacy using $\log(S)$ is equivalent to 1 minus the hazard ratio under the proportional hazards assumption.  
Other estimands may have a more straightforward interpretation. We apply the standard delta method to three cases when $S_1(t)=0$ and/or $S_2(t)=1$. 
Results  are in Table~\ref{tab:delta.method.no.mods} of the Supplement. In all three cases, the standard delta method cannot give inferences for the odds ratio or the two efficacy estimands due to undefined variances (from $0/0$ expressions).
In the case comparing $S_1(t)=0/20$ for (V=control, C=G12) to
$S_2(t) =20/20$ for (V=G12, C=G12) there is a clear protective effect of vaccination with G12 compared to control when both are challenged with G12 (central Fisher's exact p-value is p=$1.5 \times 10^{-11}$), but
only the difference estimand gives a numeric confidence interval and p-value. However, the inferences on the difference estimand are invalid, giving a confidence interval of $(1,1)$ and a two-sided (central) p-value of $0$.
Following Fay and Brittain\cite{FayB:2022}, we define a {\it valid} confidence interval as one that gives at least nominal coverage and a {\it valid} p-value as one that bounds the type I error rate at the $\alpha$-level.
Comparing $S_1(t)=0/20$ for (V=control, C=G5) to
$S_2(t) =16/20$ for (V=G5, C=G5), gives inferences on the effect of vaccination with G5 against challenge with G5 for the difference and the ratio, but they are likely invalid for the ratio because the variance estimate for $h(S_1(t))$ is $0$.  Similarly,
comparing $S_1(t)=16/20$ for (V=G5, C=G5) to
$S_2(t) =20/20$ for (V=G12, C=G5) is likely invalid for the ratio because the variance estimate for
 $h(S_2(t))$ is $0$.

Moertel {\it et al}\cite{Moer:1995} describe a randomized study of patients with stage III colon cancer recovered from surgery comparing three arms: levamisole alone (Lev) [$n=310$], flourouracil plus levamisole (Lev+5FU) [$n=304$], and control (Obs) [$n=315$]. The data set is available in the colon data.frame of the survival R package \citep{Ther:2024}. All patients were followed for at least 5 years, but some patients were followed for more than 8 years. In Supplementary Figure~\ref{lev3} we plot the Kaplan-Meier curves for time to death
together with the 95\% confidence intervals by two delta methods (the default method from the survival package that uses the Greenwood variance and a log transformation, and the adjusted hybrid method of \cite{Bork:2005}),
and two BPCP methods (the original BPCP method \citep{FayB:2013}, and its mid-p version \citep{FayB:2016}). When the number at risk is large, there is little difference between the confidence interval methods, but as fewer individuals remain
the BPCP lower limits diverge from the other methods substantially.  Fay and Brittain\cite{FayB:2016} show that
in the one-sample case, the two delta methods can substantially undercover the survival
curve when there are few remaining at risk. This suggests that there may be issues with coverage for the two sample case using those methods at later time points, even though there are substantial sample sizes in each arm early on. Alternatively, some subsets of the population may be rare and the treatment may have
a differential effect on them. In practice, care must be made when making inferences on subsets of a population, but to highlight the differences in the methods only, we consider the population
that had a perforation in the colon prior to randomization and compare the two treatment arms.
In  Supplementary Figure~\ref{Perf2panel}
we compare the two active treatment arms.

\section{Asymptotic Normality with the Two-Sample Delta Method}
\label{sec-delta}

We review the standard two-sample delta method applied to our problem in 
Section~\ref{sec-delta.description},  and review Borkowf's modifications to the one-sample variance 
estimation in Section~\ref{sec-borkowf}. Our minor contribution is Section~\ref{sec-zero-one} where we propose 
applying Borkowf's adjustments to the two-sample problem and propose our own {\it ad hoc} adjustments when 
$\hat{S}_a(t)=0$ or $1$. 

\subsection{Standard Two-Sample Delta Method}
\label{sec-delta.description}

Let the Kaplan-Meier estimate and its Greenwood variance estimate for group $i=1,2$  at time $t$ be
$\hat{S}_i(t)$ and $\hat{V}_i(t)$, respectively.
Under regularity conditions
(e.g., independent censoring, $0<S_i(t)<1$, and the expected proportion of the study at risk at $t$ approaches a constant greater than $0$), $\hat{S}_i(t)$ is asymptotically normal
with mean $S_i(t)$ and variance, say $V_i(t)$ (for details see e.g.,
\cite{Ande:1993}),
which we write as
\begin{eqnarray*}
\hat{S}_i(t) \sim AN \left\{ S_i(t), V_i(t) \right\}.
\end{eqnarray*}
Although the variance $V_i(t)$ depends on the sample size and censoring process, we suppress that dependence in the notation.
Suppose our estimand is
$\beta =
g \left\{   h(S_2(t)) - h(S_1(t)) \right\}$
 (see Table~\ref{tab:betas}).
The estimator  $h(\hat{S}_i(t))$
 is asymptotically normal by the delta method (see e.g.,
\cite{Lehm:1999}),
\begin{eqnarray*}
h \left\{ \hat{S}_i(t) \right\} & \sim & AN \left\{ h \left\{ {S}_i(t) \right\}, \Sigma_i(t) \right\},
\end{eqnarray*}
where $\Sigma_i(t) = \left\{ h'(S_i(t)) \right\}^2 V_i(t)$ and
 $h'(S) = \left[ \frac{ \partial h(u) }{ \partial u } \right]_{u=S}$.
Then, by the independence of $\hat{S}_1(t)$ and $\hat{S}_2(t)$, we have
a two-sample implementation of the delta method,
\begin{eqnarray*}
h \left\{ \hat{S}_2(t) \right\} - h \left\{ \hat{S}_1(t) \right\} & \sim &
AN \left\{ h \left( {S}_2(t) \right)- h \left( {S}_1(t) \right),
 \Sigma_1(t) + \Sigma_2(t) \right\},
\end{eqnarray*}
so that a $100(1-\alpha)\%$ confidence interval for $\beta$
is
\begin{eqnarray}
g \left\{ h \left( \hat{S}_2(t) \right) - h \left( \hat{S}_1(t) \right) \pm z_{\alpha/2} \sqrt{ \hat{\Sigma}_1(t) + \hat{\Sigma}_2(t) } \right\}
\label{eq:deltaMethodCIs}
\end{eqnarray}
where $z_{\alpha/2}$ is the $(1-\alpha/2)$th quantile of the standard normal distribution,
and $\hat{\Sigma}_i(t) = \left\{ h'(\hat{S}_i(t)) \right\}^2 \hat{V}_i(t)$.
Equation~(\ref{eq:deltaMethodCIs}) can breakdown when $\hat{S}_j(t)=0$ or $1$,
and we address that in Section~\ref{sec-zero-one}.
We first consider other estimators of $V_i(t)$.

\subsection{Borkowf One-Sample Variance Estimator Modifications}
\label{sec-borkowf}

The Greenwood formula tends to underestimate $V_i(t)$.
Besides the Greenwood formula, another estimator of $V_i(t)$ has been proposed that handles ties by grouping and rounding,
but it has slightly more bias than the Greenwood formula \citep[see][pp.259-260]{Ande:1993}.
Both those estimators  only change over time when the Kaplan-Meier estimate changes. This can be an issue when there is censoring after failures.
In that case, as more censoring occurs between failures, there are fewer individuals at risk and logically less precision in the Kaplan-Meier estimator and yet the above traditional variance estimators of $\hat{S}_i(t)$ do not show increasing variance.
Peto {\it et al.}\cite{Peto:1977} and
Borkowf\cite{Bork:2005}  have noted this and proposed alternative variance estimators.

Borkowf\cite{Bork:2005} proposed two small sample variance estimators for $V_j(t)$
which he called the regular hybrid variance estimator, say $\hat{V}_{j}^{RH}(t)$,
and the adjusted hybrid variance estimator, $\hat{V}_{j}^{AH}(t)$.
Borkowf\cite{Bork:2005} showed these estimators  each have less
simulated bias than the Greenwood estimator or that of Peto {\it et al}\cite{Peto:1977}.

Borkowf\cite{Bork:2005} noted that when there is no censoring, we have essentially a binomial problem, and
\begin{eqnarray}
\hat{V}_i(t) & = &  \frac{\hat{S}_i(t) \left\{ 1 - \hat{S}_i(t) \right\} }{n_i},
\label{eq:binomVariance.noCens}
\end{eqnarray}
where $n_i$ is the number of individuals in group $i$. Borkowf's regular hybrid variance estimator has a similar form,
\begin{eqnarray}
\hat{V}_i^{RH}(t) & = &  \frac{w_i(t) \left\{ 1 - w_i(t) \right\} }{n_{i,eff}(t)},
\label{eq:VRH}
\end{eqnarray}
where $w_i(t)=\hat{S}_i(t)$ when $\hat{S}_i(t) \geq 0.5$ and is modified otherwise
\citep[see][eqn. 9]{Bork:2005}, and  $n_{i,eff}(t)$ is the
 `effective sample size',
$n_{i,eff}(t) = \max \left\{ 0.5, n_i - c_i(t) \right\},$  where $c_i(t)$ is the number censored in group $i$ before $t$.
The $\hat{V}_i^{AH}(t)$ adjusts $w_i(t)$ slightly in equation~(\ref{eq:VRH}) and
ensures that the adjusted $w_i(t)$, say $w^*_i(t)$, has $0< w^*_i(t) < 1$.

For confidence intervals about $S_i(t)$, Borkowf\cite{Bork:2005} proposed three choices: (1)  regular hybrid,
$\hat{S}_i(t) \pm z_{\alpha/2} \hat{V}_i^{RH}(t)$; (2) adjusted hybrid,
$\hat{S}_i(t) \pm z_{\alpha/2} \hat{V}_i^{AH}(t)$; or (3) shrunken adjusted hybrid,
$\hat{S}_i^*(t) \pm z_{\alpha/2} \hat{V}_i^{AH}(t)$, where
$\hat{S}_i^*(t)  =  (1-n_i^{-1}) \hat{S}_i(t) + \frac{1}{2 n_i },$
is a shrunken estimator of $S_i(t)$.
Borkowf\cite{Bork:2005} recommends either the adjusted hybrid estimator or the shrunken adjusted hybrid estimator. We focus on the adjusted hybrid estimator in this paper, since we are interested in
estimators of $\beta$ that use the Kaplan-Meier estimators.

Borkowf\cite{Bork:2005} considered using the delta method using only the $\log$ transformation,
and we consider the other transformations of Table~\ref{tab:betas} as well.

\subsection{Two-Sample Delta Method with Borkowf and Zero-One Modifications}
\label{sec-zero-one}

In this section, we propose minor modifications to the standard two-sample delta method.
The standard delta method confidence interval was defined in equation~(\ref{eq:deltaMethodCIs}).
Let the regular hybrid (adjusted hybrid) interval be equation~(\ref{eq:deltaMethodCIs}) except with
$\hat{V}_i(t)$ replaced by $\hat{V}_i^{RH}(t)$ (by  $\hat{V}_i^{AH}(t)$).
Let the shrunken adjusted hybrid interval be equation~(\ref{eq:deltaMethodCIs}) with
$\hat{S}_i(t)$ replaced by $\hat{S}^*_i(t)$ (following \cite{Bork:2005}, we do not replace $\hat{S}_i(t)$ when inside $h'$),
and $\hat{V}_i(t)$ replaced by $\hat{V}_i^{AH}(t)$.

For either the standard delta method or a modification using one of Borkowf's methods,
there are cases when $\hat{S}_i(t)=0$ or $1$ where breakdowns occur.
By breakdowns, we mean either $h(\hat{S}_i(t))$ being $-\infty$ or $+\infty$ (see Table~\ref{tab:delta.method.no.mods}), or $\hat{\Sigma}_i(t)$ equals $0$ or is undefined (from $0/0$ or $0 \times \infty$).

When pre-specifying an analysis method, it should be defined for any outcome. This is also helpful for  simulations. Thus, for the simulations of this paper we propose `zero-one' adjustments that may be needed when $\hat{S}_i(t)=0$ or $1$.  The need and type of adjustment needed depend on the estimand and which (if any) of the modifications of Borkowf\cite{Bork:2005} are used.

Here are the details.
We describe adjustments for the delta methods using either the standard method or one of Borkowf's three methods.
For any of the four methods, let $\tilde{S}_i^{h}(t)$, $\tilde{S}_i^{h'}(t)$ and $\tilde{V}_i(t)$ be estimators of respectively, $S_i(t)$ used in $h(\cdot)$,
$S_i(t)$ used in $h'(\cdot)$, and $V_i(t)$.
Then, as long as the estimators of $S_i(t) \neq 0,1$ and $\tilde{V}_i(t) \neq 0$, the
 $100(1-\alpha)\%$ confidence interval for $\beta$
is
\begin{eqnarray}
g \left\{ h \left( \tilde{S}_2^h(t) \right) - h \left( \tilde{S}_1^h(t) \right) \pm \phi z_{\alpha/2} \sqrt{ \dot{\Sigma} \left( \tilde{S}_1^{h'}(t), \tilde{V}_1(t) \right) +
\dot{\Sigma} \left( \tilde{S}_2^{h'}(t), \tilde{V}_2(t) \right) } \right\}
\label{eq:genericDeltaMethodCIs}
\end{eqnarray}
where when $\phi=1$ then $\pm$ represents $'+'$ for the upper limit and $'-'$ for the lower limit, and  $\phi=-1$ switches the  order of the limits.
Specifically, $\phi=1$ when $h$ is an increasing function (transformation for s is none, $\log(s)$, or $\log(s/(1-s))$), and
$\phi=-1$ when $h$ is a decreasing function (transformation for s is $\log(1-s)$ or $\log(-\log(s))$).

We classify  the cases with $\hat{S}_i(t)=0$ or $1$ into three types,
where the  $(\hat{S}_1,\hat{S}_2)$ values are either:
\begin{description}
\item[Type~1:] $(0,0)$ or $(1,1)$
\item[Type 2:] $(0,1)$ or $(1,0)$
\item[Type 3:]  $(0,s)$ or $(s,0)$ or $(1,s)$ or $(s,1)$, where $0<s<1$.
\end{description}

\subsubsection{Type 1: (0,0) or (1,1) }

Consider the cases when $\hat{S}_1(t) = \hat{S}_2(t)=s$ and $s=0$ or $1$.
We address 10 cases within Type 1: for $s=0$ or $1$ in each of the five estimands of Figure~1.
It makes sense to describe the zero-one adjustment in two subtypes.

For the first subtype, consider the four cases where $h(0)$ or $h(1)$ give finite values (see Table~\ref{tab-fdelta}).
In that case, the plug-in estimators straightforwardly give $b(s,s)=\beta_e$ (defined in Section~\ref{sec-inferential}):
for $\beta_d$, $b_d(0,0)=0$ and $b_d(1,1)=0$,
for $\beta_r$, $b_r(1,1)=1$, and
for $\beta_{ecd}$,
 $b_{ecd}(0,0) =0$.
Since for these cases $h'(0)$ or $h'(1)$ also give finite values,
the only adjustment to equation~\ref{eq:genericDeltaMethodCIs} needed is when $\tilde{V}_i(t)=0$. When $\hat{S}_i=1$ then $\hat{V}_i(t)=0$ and $\hat{V}^{RH}(t)=0$.
When  $\hat{S}_i(t)=0$ then $\hat{V}_i(t)=0$ but depending on the censoring $\hat{V}_i^{RH}(t)$ may or may not equal $0$.
Since $0<\hat{V}_i^{AH}(t)<1$, we need no adjustment for $\hat{V}_i^{AH}(t)$.
Whenever $\tilde{V}_i(t)=0$, replace $\tilde{V}_i(t)$ with $\dot{V}_i(t)$, where
\begin{eqnarray}
\dot{V}_i(t) & = & \left\{
\begin{array}{ll}
\hat{V}_i^{AH}(t) & \mbox{ for Borkowf's methods}  \\
\mbox{ } &  \\
\frac{ \hat{S}^*_i(t) \left\{ 1 - \hat{S}^*_i(t)  \right\} }{n_i} & \mbox{for the standard method,}
\end{array}
\right.
\label{eq:Vdot.replacement.defn}
\end{eqnarray}
 where $\hat{S}^*_i(t)$ is the shrunken estimator of $S_i(t)$ and $n_i$ is the sample size for group~$i$.

\begin{table}[hbt!]
\begin{center}
\caption{Functions for delta method. The five rows are functions associated with the five different estimands.
For this table assume $\dot{V}>0$.
\label{tab-fdelta}}

\begin{tabular}{lcccccccc} 
$\beta$ & $h(S)$  & $h'(S)$ &  $h(0)$ & $h(1)$  & $h'(0)$ & $h'(1)$  & $\dot{\Sigma}(0,\dot{V})$ & $\dot{\Sigma}(1,\dot{V})$ \\ \hline
$\beta_d$ &  $S$ & $1$   & $0$ & $1$ & $1$ & $1$  & $\dot{V}$ & $\dot{V}$ \\
$\beta_r$ &  $\log(S)$ & $\frac{1}{S}$   & $-\infty$ & $0$ & $\infty$ & $1$ & $\infty$ & $\dot{V}$  \\
$\beta_{or}$ &  $\log \left\{ \frac{S}{1-S} \right\}$ & $\frac{1}{S(1-S)}$   & $-\infty$ & $\infty$ & $\infty$ & $\infty$ & $\infty$ & $\infty$ \\
$\beta_{ecd}$ & $\log(1-S)$ &   $\frac{-1}{1-S}$   & $0$ &  $-\infty$ & $-1$ & $-\infty$ &  $\dot{V}$ & $\infty$  \\
$\beta_{els}$ & $\log \left\{ - \log( S ) \right\}$ & $\frac{1}{S \log(S)}$   & $\infty$ & $-\infty$ &  $-\infty$ & $-\infty$ & $\infty$ & $\infty$ \\
 \hline
\end{tabular}
\end{center}
\end{table}

Now consider the rest of the cases in Type 1.
One approach is to just use a limit assuming equal survival values at $t$ for the estimates,
$\lim_{s \rightarrow 0} b(s,s) = \beta_e$ or $\lim_{s \rightarrow 1} b(s,s) = \beta_e$; however, it is possible
to approach $(0,0)$ or $(1,1)$ from different directions in the two-dimensional space.
Consider the ratio when $\hat{S}_1(t)=\hat{S}_2(t)=0$. Let $s_2 = r s_1$, where
    $0 < r< \infty$ so that $\lim_{s_1 \rightarrow 0} b(s_1, r s_1)= r$. In this case,  the limit $r$ could be any value of $\beta$, so $\hat{S}_1(t)=\hat{S}_2(t)=0$ essentially has no information about $\beta$.  Therefore, we define the estimate as $\beta_e$ and in place of equation~\ref{eq:genericDeltaMethodCIs}, we define    the $100(1-\alpha)\%$ central confidence interval as $(\beta_{\wedge}, \beta_{\vee})$,
where $\beta_{\wedge}$ is the  smallest possible value  for $\beta$ (and $\beta_{\wedge}=-\infty$ is allowed)
and   $\beta_{\vee}$ is the  largest possible value  for $\beta$ (and $\beta_{\vee}=\infty$ is allowed).
      We do the same for any value where $S_1(t)=S_2(t)=s$ for $s=0$ or $1$  gives no information about $\beta$. Specifically, the scenarios when $S_1(t)=S_2(t)=0$ for
     $\beta_r$, $\beta_{or}$, and $\beta_{els}$,
      and when $S_1(t)=S_2(t)=1$ for
      $\beta_{or}$, $\beta_{ecd}$, and $\beta_{els}$.

\subsubsection{Type 2: (0,1) and (1,0)}

First, consider the cases when $(\hat{S}_1(t), \hat{S}_2(t))=(0,1)$.
In these cases, our estimates of $\beta$ and upper confidence limit for $\beta$ are set to $\beta_{\vee}$.
For the lower confidence limits, we use equation~\ref{eq:genericDeltaMethodCIs} except do the following replacements:
\begin{eqnarray}
& \begin{array}{ll}
\mbox{replace both   $\tilde{S}_i^{h}(t)$ and $\tilde{S}_i^{h'}(t)$  with   $\hat{S}^*_i(t)$  } & \mbox{ when $h(\hat{S}_i(t)) = \pm \infty$, and } \\
\mbox{replace $\tilde{V}_i(t)$ with $\dot{V}_i(t)$ (equation~\ref{eq:Vdot.replacement.defn})} & \mbox{ whenever $\hat{S}_i(t)=0$ or $1$. }
\end{array}
\label{defn:replacements}
\end{eqnarray}

The cases when $(\hat{S}_1(t), \hat{S}_2(t))=(1,0)$ are analogous.
In these cases, our estimates of $\beta$ and lower confidence limit for $\beta$ are set to $\beta_{\wedge}$.
For the upper confidence limits, we use equation~\ref{eq:genericDeltaMethodCIs} except using the replacements~\ref{defn:replacements}.

\subsubsection{Type 3: (0,s) or (s,0) or (1,s) or (s,1)}

For Type 3, we consider four scenarios, $(\hat{S}_1(t),\hat{S}_2(t))$ equals either
 $(0,s),$ $(s,0),$ $(1,s),$ or $(s,1)$, where $0<s<1$.

Our strategy is as follows. In each case, for the value that is either $0$ or $1$, we define the estimand as the limit (e.g., for $(0,s)$, the estimand is
$\lim_{a \rightarrow 0} b(a,s)=\hat{\beta}$ ($\hat{\beta}=\infty$ or $-\infty$ allowed).

For the confidence intervals, we do the following:
\begin{description}
\item[ If $\hat{\beta}=\beta_{\wedge}$,] then
\begin{itemize}
\item the lower confidence limit is $\beta_{\wedge}$, and
\item for the upper confidence limit
use equation~\ref{eq:genericDeltaMethodCIs} except
using the replacements~\ref{defn:replacements}.
\end{itemize}
\item[ If $\hat{\beta}=\beta_{\vee}$,] then
\begin{itemize}
\item for the lower confidence limit
use equation~\ref{eq:genericDeltaMethodCIs} except
using the replacements~\ref{defn:replacements}.
\item the upper confidence limit is $\beta_{\vee}$.
\end{itemize}
\item[ If $\beta_{\wedge} < \hat{\beta} < \beta_{\vee}$ ] (these are 4 cases where the limiting value of the $h$ function is finite, see Table~\ref{tab-fdelta})  then
\begin{itemize}
\item for both confidence limits, use
equation~\ref{eq:genericDeltaMethodCIs} except
using the replacements~\ref{defn:replacements}.
\end{itemize}
\end{description}

P-values are calculated assuming asymptotic normality on the difference in the $h$ functions, except in the cases when confidence limits are set to the extremes, in which case the associated one-sided p-values are set to $1$.

\section{Melding using the Beta Product Confidence Procedure }
\label{sec-melding}

In Section~\ref{sec-melding.overview} we first review the general melding procedure, focusing on its application to our problem, which melds two one-sample BPCP intervals. Section~\ref{sec-bpcp} reviews the one-sample BPCP and its mid-p version. Section~\ref{sec-alg} reviews algorithms necessary to implement the melding on the BPCP, and shows how they work with the mid-p BPCP. Although the proposal to meld on the mid-p BPCP is new, it is a straightforward 
use of melding and so is presented within the reviews of this section.

\subsection{General Melding Method}
\label{sec-melding.overview}

Fay, Proschan, and Brittain\cite{FayP:2015} developed {\it melding}, a general method for combining one-sample confidence interval procedures in the two-sample scenario. If the one-sample parameters are $S_1(t)$ and $S_2(t)$, then melding can be applied to estimands such as those in Table~\ref{tab:betas}, because they are all  decreasing functions of $S_1(t)$ and increasing functions of $S_2(t)$.
Suppose that for group $i$, there is a central $100(1-\alpha)\%$ confidence interval for $S_i(t)$ created from two one-sided $100(1-\alpha/2)\%$  confidence intervals,
\begin{eqnarray}
\left(  L_{S_i(t)}(1-\alpha/2), U_{S_i(t)}(1-\alpha/2) \right).  \label{eq:thetajCIs}
\end{eqnarray}
Then the melded  $100(1-\alpha)\%$ central confidence interval for $\beta= b(S_1(t),S_2(t))$ is
\begin{eqnarray}
\left( q \left\{ \alpha/2, b(W_{1U},W_{2L} ) \right\}, q \left\{ 1-\alpha/2, b(W_{1L},W_{2U}) \right\} \right),
\label{eq:meldedCI}
\end{eqnarray}
where $q(a,X)$ is the $a$th quantile of a random variable $X$, and
$W_{iL} \equiv W_{iL}(t)$ and  $W_{iU} \equiv W_{iU}(t)$ are  lower and upper confidence distribution random variables (CDRVs)
associated with $S_i(t)$.
CDRVs are draws from a confidence distribution, which is a frequentist analog to a Bayesian posterior
distribution \citep{XieS:2013,Vero:2015}.
Specifically, when the confidence intervals are based on discrete random variables (e.g., $\hat{S}_i(t)$), the lower and upper CDRVs may be derived from the one-sided confidence
interval procedures using
$W_{iL} = L_{S_i(t)}(A_{iL})$ and
$W_{iU}  = U_{S_i(t)}(A_{iU})$, where $A_{iL}$ and $A_{iU}$ are independent uniform random variables.

For testing the null hypothesis $H_0: \beta \leq \beta_0$ against the alternative hypothesis $H_1: \beta > \beta_0$,
 the melding p-value is 
\begin{eqnarray}
Pr \left[ b(W_{1U}, W_{2L}) \leq \beta_0 \right].  \label{eq:p.alt.is.greater}
\end{eqnarray}
If the p-value equals $\alpha/2$, then the lower limit of equation~\ref{eq:meldedCI} would equal $\beta_0$.  
Analogously, for testing $H_0: \beta \geq \beta_0$ against $H_1: \beta <  \beta_0$,
 the melding p-value is 
\begin{eqnarray}
Pr \left[ b(W_{1L}, W_{2U}) \geq \beta_0 \right].,  \label{eq:p.alt.is.less}
\end{eqnarray}  
and if that p-value equals $\alpha/2$, then the upper limit of equation~\ref{eq:meldedCI} would equal $\beta_0$.  The one-sided p-values are compatible with the one-sided confidence intervals (and the $100(1-\alpha)\%$ central intervals are compatible with testing both of the one-sided tests at level $\alpha/2$).
Those p-values and values $q(a,X)$ in equation~\ref{eq:meldedCI} can be estimated by Monte Carlo simulation or sometimes by numeric integration (see Section~\ref{sec-alg}). 

When the confidence intervals of equation~(\ref{eq:thetajCIs}) are  valid 
and  nested (i.e., if $1-\alpha_1 > 1-\alpha_2$ then
$L_{S_i(t)}(1-\alpha_1) \leq L_{S_i(t)}(1-\alpha_2) <  U_{S_i(t)}(1-\alpha_2) \leq U_{S_i(t)}(1-\alpha_1)$),  then
Fay, {\it et al}\cite{FayP:2015} conjectured that the melded intervals
of equation~(\ref{eq:meldedCI}) are valid and nested also. This conjecture is not true in general.
Frey and Zhang\cite{Frey:2023} showed that it does not hold for the difference in medians case where the confidence intervals on the single sample medians invert the sign test.  Nevertheless there are some important special cases. As previously mentioned, melding two exact binomial intervals for binomial parameters $\theta_1$ and $\theta_2$  gives confidence intervals on any function $b(\theta_1,\theta_2)$ in Table~\ref{tab:betas} and all are compatible with the central Fisher's exact test.
Further, extensive numerical calculations could find no two-sample binomial parameter example with less than nominal coverage for one-sided 95\% confidence intervals \citep{FayP:2015}.
Melding on normally distributed parameters gives the usual normal confidence intervals.
For example, if we assume normally of $h \left\{ \hat{S}_i(t) \right\}$ and treat $\hat{\Sigma}_i(t)$ as $\Sigma_i(t)$,
then the melding using the normal confidence intervals for $h \left\{ \hat{S}_i(t) \right\}$ for $i=1,2$ returns
the confidence interval for $\beta$  of equation~(\ref{eq:deltaMethodCIs}).
A melding method better for small sample coverage is to meld the BPCP intervals for $S_i(t)$, which is what we propose.

The original melding was designed to guarantee coverage. If we instead aim to achieve coverage that is closer to the nominal level,  a natural idea is to use a mid-p-like confidence distribution random variable. Thus, we propose using the mid-p version of the BPCP\cite{FayB:2016} to get such a CDRV.  Details are in the next two sections.  

\subsection{One-Sample Confidence Intervals using the Beta Product Confidence Procedure}
\label{sec-bpcp}

Now we define the CDRVs associated with the BPCP and its mid-p version.
Let the ordered times to the event or censoring for group $i$ be $t_{i1} < t_{i2} < \cdots < t_{ij_i}$
for $i=1,2$. Let $r_{ij}$ be the number at risk just before $t_{ij}$ and let $d_{ij}$ be the number of events at $t_{ij}$.
If $d_{ij}=0$ then $t_{ij}$ only has censored observations.

The BPCP method uses CDRVs that are products of independent beta random variables.
Let $B(a,b)$ be a beta random variable with parameters $a>0$ and $b>0$, and define $B(a,0) = \lim_{\epsilon \rightarrow 0} B(a,\epsilon)$ to be a point mass at $1$.
Then the upper and lower confidence distribution random variables for $S_i(t)$ using the BPCP at $t \in [t_{ij}, t_{i,j+1})$ are, respectively,
\begin{eqnarray}
W_{iU}(t) & = &  \prod_{\ell=1}^{j} B(r_{i\ell} - d_{i\ell} + 1, d_{i\ell}),  \nonumber  \\
\mbox{  and  } & &    \label{eq:Ws}  \\
W_{iL}(t) & = & W_{iU}(t) B(r_{i,j+1},1),   \nonumber
\end{eqnarray}
where all beta random variables are independent.
The $100(1-\alpha)\%$ BPCP for $S_i(t)$ is
\begin{eqnarray}
\left( q(\alpha/2, W_{iL}(t)), q(1-\alpha/2, W_{iU}(t) \right). \label{eq:BPCPinterval}
\end{eqnarray}
The melded BPCP interval for $\beta=b(S_1(t),S_2(t))$ is then given by  equation~(\ref{eq:meldedCI}).

Fay and Brittain\cite{FayB:2016} proposed a mid-p version of the BPCP by defining
a random variable halfway between $W_{iL}(t)$ and $W_{iU}(t)$ as $W^*_{i}(t) = U W_{iL}(t)
+(1-U) W_{iU}(t)$, where $U \sim \mathrm{Bernoulli}(0.5)$. The $100(1-\alpha)\%$ mid-p BPCP for
$\theta_i$ is
\begin{eqnarray*}
\left( q(\alpha/2, W_{i}^*(t)), q(1-\alpha/2, W_{i}^*(t) \right).
\end{eqnarray*}
For melding on the mid-p BPCP, we replace each of $W_{iL}$ and $W_{iU}$ in equation~(\ref{eq:meldedCI}) with $W_i^*(t)$ for $i=1,2$.
Section~\ref{sec-alg} gives algorithms and approximations for estimating those intervals.

When there is no censoring, the beta product random variables are all distributed
either beta or a point mass at $0$ or $1$, and the BPCP confidence interval and its mid-p version reduce to the exact binomial
confidence interval and its mid-p version\citep{FayB:2016}.
In general a beta product random variable is not distributed beta (although in the special case with no
censoring, they are).

\subsection{Algorithms and Approximations for Melding on the BPCP}
\label{sec-alg}

The estimation of the confidence intervals and p-values for melding can be broken into two parts. 
Part 1 is estimation of the confidence distributions from the one-sample parameters. 
Part 2 is taking those one-sample confidence distributions and melding them together to get two-sample inferences.

A straightforward way to estimate the melding confidence intervals and p-values is by  Monte Carlo simulation for both Parts (and we call this type of algorithm P1:MC,P2:MC). 
In Part~1, 
let the first  Monte Carlo draw of  $W_{iU}(t)$ be  $W_{iU}^{*(1)}$, which   may be simulated by multiplying $j$ independent pseudo-random beta variables with parameters given in expression~\ref{eq:Ws}. The 
associated first random draw of $W_{iL}(t)$, say $W_{iL}^{*(1)}$, is  simply $W_{iU}^{*(1)}$ multiplied by an independent pseudo-random beta variable with parameters $r_{i,j+1}$ and $1$. 
Repeat that process $N_{mc}$ times to get 
\begin{eqnarray}
W_{iU}^{*(1)}, W_{iU}^{*(2)}, \ldots, W_{iU}^{*(N_{mc})},  \mbox{ and }   W_{iL}^{*(1)}, W_{iL}^{*(2)}, \ldots, W_{iL}^{*(N_{mc})}. \label{eq:WULstar}
\end{eqnarray}
Then in Part~2, use the $b(\cdot,\cdot)$ function to get a sample from the lower (or upper)  confidence distribution random variable for $\beta$,  where the $k$th sample is $\beta^{*(k)}_L = b\left\{ W_{1U}^{*(k)}, W_{2L}^{*(k)} \right\}$ $\left( \mbox{or }  \beta^{*(k)}_U = b\left\{ W_{1L}^{*(k)}, W_{2U}^{*(k)} \right\} \right).$ 
This is a Monte Carlo draw from the lower (or upper) CDRV for $\beta$.
The quantiles $q(p,W)$ as in expression~\ref{eq:BPCPinterval} can be estimated by each set of $N_{mc}$ simulated CDRV and the {\sf quantile} function in R.
The p-value of equation~\ref{eq:p.alt.is.greater} may be estimated by 
\begin{eqnarray*}
p & = & \sum_{k=1}^{N_{mc}}  \frac{ I( \beta^{*(k)}_L \leq \beta_0 ) }{N_{mc}}
\end{eqnarray*}
where $I(A)=1$ if $A$ is true and $0$ otherwise,
and if $p=0$ we set it to $0.5/N_{mc}$. 
The p-value estimate for equation~\ref{eq:p.alt.is.less} is analogous.

An alternative appoximate implementation is to use method of moments to estimate the confidence distributions in Part 1,
and then Monte Carlo estimation for Part 2 (say P1:MM,P2:MC).
In Part~1, we  assume that the beta product random variables are beta, and estimate the beta parameters by
method of moments \cite{FayB:2016}.
Specifically,  for  $W= \prod_{i=1}^{j} B(a_i,b_i)$, we approximate its distribution with a beta with parameters $a^*=\frac{u_1(u_1-u_2)}{u_2-u_1^2}$ and 
$b^*=\frac{(1-u_1)(u_1-u_2)}{u_2-u_1^2}$, where 
$u_1= \prod_{i=1}^{j}  \frac{a_i}{a_i+b_i}$ and $u_2=\prod_{i=1}^{j} \frac{ a_i(a_i+1)}{(a_i+b_i)(a_i+b_i+1)}$.
This approximate implementation has
been shown by simulation to work well in the one-sample case \cite{FayB:2013,FayB:2016}.
Then for Part~2, we can take Monte Carlo draws  from beta distributions that are based on those method of moments beta parameters,
and these can be used instead of expression~\ref{eq:WULstar}. The Part~2 follows in the same manner as the previous paragraph.

The third alternative is to use method of moments for Part~1, and numeric integration for Part~2 (say P1:MM,P2:NI).  
After approximating  single sample CDRVs by beta distributions with parameters estimated by method of moments, we estimate the p-values by using numeric integration and cumulative probability functions and density functions for those beta distributions.
For example, suppose $\beta=S_2(t)- S_1(t)$. Let   $(a_{iL},b_{iL})$ and $(a_{iU}.b_{iU})$ be, respectively,  the method of moments parameter estimates for the beta shape parameters for the lower and upper confidence distribution 
for $S_i(t)$.  Then the p-value for testing $H_0: \beta \leq \beta_0$ (see equation~\ref{eq:p.alt.is.greater}) is $p_{NI}(\beta_0)$, given by 
\begin{eqnarray}
Pr \left[ b \left\{ B(a_{1U},b_{1U}), B(a_{2L},b_{2L}  \right\}    \leq \beta_0 \right] & = &  Pr \left[ B(a_{2L},b_{2L})  \leq   B(a_{1U},b_{1U}) + \beta_0 \right]   \nonumber \\
& \approx & \int_{0}^{1}  F_{beta}( s + \beta_0;  a_{2L},b_{2L}) f_{beta}(s; a_{1U},b_{1U} ) ds = p_{NI}(\beta_0) \label{eq:numeric.integration}
\end{eqnarray}
where $F_{beta}(\cdot; a,b)$ and $f_{beta}(\cdot; a,b)$ are the cumulative distribution and probability density function of $B(a,b)$.   
Expression~\ref{eq:numeric.integration} is solved by numeric integration. (Some simplication occurs and numeric integration may not be needed if the lower or upper confidence distribution is a point mass at $0$ or $1$.) Then the lower limit of equation~\ref{eq:meldedCI} is found by a root solving function as the value of 
$\beta_0$ such that $p_{NI}(\beta_0) = \alpha/2$. 
In an analogous way, we can estimate the  p-value for testing $H_0: \beta \leq \beta_0$ (see equation~\ref{eq:p.alt.is.less}) and the upper limit of equation~\ref{eq:meldedCI}.
For other parameters $\beta$ (see Supplementary Table~\ref{tab:betas}) different algebra is done inside the probability expression to get different expressions for numeric integration.
For example, if $\beta=S_2(t)/S_1(t)$ then the numeric integration of expression~\ref{eq:numeric.integration} would instead become 
$ \int_{0}^{1}  F_{beta}( s  \beta_0;  a_{2L},b_{2L}) f_{beta}(s; a_{1U},b_{1U} ) ds.$

For melding on BPCP, we used the algorithm P1:MM,P2:NI  in the simulations and it is the default in the {\sf bpcp2samp } function in the {\sf bpcp} R package. 
For melding on mid-p BPCP we used the algorithm P1:MM,P2:MC, which is the default when midp=TRUE. 
This means that the distributions of $W_{iL}(t)$ and $W_{iU}(t)$ are estimated by method of moments, and then the distributions of $W_i^*(t)$ for $i=1,2$ are 
estimated by Monte Carlo simulation, which is used in Part 2.

For melding on the mid-p BPCP, the calculations use Monte Carlo simulation so that there is variability.  The default number of Monte Carlo replications in the {\sf bpcp2samp} function in the current version of the {\sf bpcp} R package
 is $10^6$ so the Monte Carlo variability in the confidence intervals in negligible. (That was the number of replications used in the simulation of Section~\ref{sec-simulations}).
 For applications, the {\sf bpcp2samp} function by default sets the random number seed, so that a reanalysis of the same data set is replicated. In contrast, for simulations the seed is not set for each data set, since setting a seed in that situation would interfere with the simulation process for the simulated data sets. 

To test the negligibleness, we compared replicates. 
For each of 10 data sets from Scenario~3  with $n_1=n_2=30$ and heavy censoring (see Section~\ref{sec-simulations} for definition of Scenario 3 and other details),
we repeatedly calculated the 95\% confidence intervals (100 on each data set) on $\beta= S_2(1.2)-S_1(1.2)$  using melding on the mid-p BPCP (after setting the seed control option to NULL so that each repeat calculation would take a different Monte Carlo sample).  The 1000 confidence interval calculations  averaged less than 1 second each,
and the mean absolute difference (MAD) overall was $0.00037$, where the MAD was calculated within each confidence limit, within  each set of 100 (99 pairs),  within each of  the 10 data sets, and then averaged over all of those cases.
The analogous calculation with $10^5$ replications gives an overall MAD of $0.0012$.
Since the range of $\beta$ for the difference is $-1$ to $1$, this Monte Carlo variability is negligible compared to the confidence interval widths (see Figure~\ref{fig-RD01} for widths). 

For a rough  comparison of calculation times we timed calculations using the {\sf bpcp} R package twice and give the range of times. We calculated confidence intervals at one time point using standard delta method with zero-one adjustment on 1,000 data sets with $N_1=n_2=300$ which took about 13-18 seconds on a PC (1.8 GHz processor, 16 GB RAM).
For comparison,  melding on the BPCP (P1MM:p2:NI)  took about the same amount of time  (0.92 - 1.03 times as long), and the melding on the BPCP mid-p (P1:MM,P2:MC)  using $10^6$ Monte Carlo replicates took longer (47-73 times as long as the standard method)  (but for each data set averaged less than a second).
Using $10^5$ Monte Carlo replicates the melding on the mid-p BPCP took about 11-13 times as long as the standard.

\section{No Censoring Cases: Numeric Evaluation}
\label{sec-no-censoring}

Without censoring, the problem simplifies to comparing two binomials via $\beta$ parameters.
Because of this  simplification,  we first turn to the literature on the two-sample binomial problem (see e.g., \cite{FayH:2021, Lyde:2009}).
Among valid tests for this problem (i.e., those that guarantee type I error rate is less than the alpha-level), those that tend to be more powerful are unconditional exact tests, and they 
have not been generalized to handle censored data.  When looking at the two-sample binomial literature recall that the p-values for the standard 
delta method reduces to the Wald-test version of the binomial delta method application, the p-values for the BPCP are equivalent to the one-sided or central Fisher's exact test p-values, and hence the melding on mid-p BPCP version is expected to be similar to a 
mid-p version of Fisher's exact test.  
In the two-sample binomial case, Fisher's exact test tends to be conservative, but  its mid-p version can be a good approximation to the 
unconditional exact tests \cite{FayH:2021,Lyde:2009}, and that suggests the mid-p BPCP may be a good approximation as well.
Because the zero-one adjustments for the delta method are new to this paper, and Borkowf's adjustments have not been studied with the two-sample binomial problem, 
we do some brief calculations to compare the methods of this paper.  Studying the no censoring case additionally allows us to see type I error rates issues that are due only to small sample sizes,
 whereas if we left out this section  we might mistakenly attribute those  issues primarily due to censoring.

\begin{figure}
\includegraphics[width=6.0in]{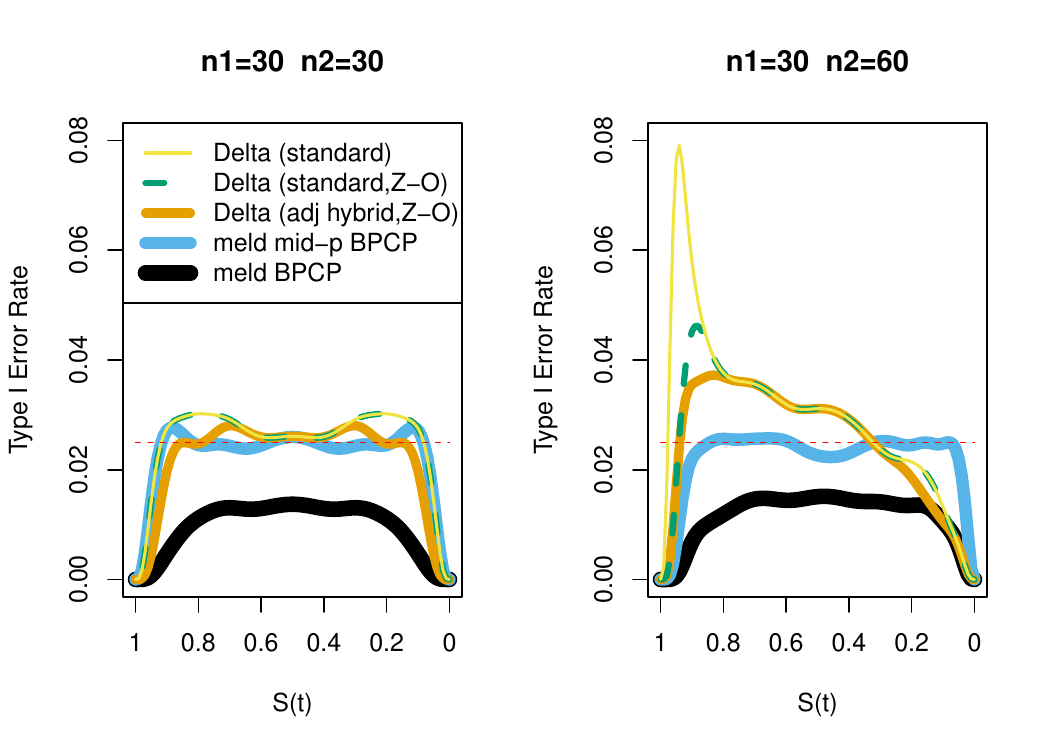}

\caption{
Type I error rate for testing when the null hypothesis is $S_1(t)=S_2(t)=S(t)$ with no censoring. 
Because the survival function goes from $1$ to $0$ as time progesses, we create the horizontal axes with $1$ on the left and $0$ on the right. 
The left panel has $n_1=n_2=30$ testing the alternative  
alternative that $S_2(t) - S_1(t) > 0$  (or testing the alternative that $S_2(t) - S_1(t) < 0$)  at the one-sided 2.5\% alpha-level. 
The right panel has $n_1=30$ and $n_2=60$ testing the alternative  
 that $S_2(t) - S_1(t) < 0$.  Type I error rates are lines through the fixed time points such that $S(t)=1,0.99,0.98,\ldots,0$. We test  3 ``delta'' methods applied to the difference, the standard method 
without the zero-one adjustment (Wald test on the difference in proportions with no censoring), the standard method with the zero-one adjustment,
and the adjusted hybrid with the zero-one adjustment, and two BPCP methods, the usual BPCP (Fisher's exact test with no censoring) and mid-p BPCP (similar to mid-p Fisher's exact test with no censoring). 
\label{fig.calculation_no_censor}
}

\end{figure}

When there is no censoring, all that matters is the values of the two binomial parameters (which in this case are $S_1(t)$ and $S_2(t)$), and the sample size. 
So to compare type I error rates, we need only plot the size of the different methods for all values of $S_1(t)=S_2(t)=S(t)  \in (0,1)$.  
We plot two cases of type I error rates for 5 methods  in Figure~\ref{fig.calculation_no_censor}.  
For small sample sizes, we can calculate the confidence limits for the entire sample 
space, then use those limits to quickly calculate the type I error rates of the associated tests under different values of $S_1(t)=S_2(t)=S(t) \in \left[ 1, 0.99, 0.98,\ldots, 0 \right]$. 
In the left panel we consider $n_1=n_2=30$ testing the alternative that $S_2(t) - S_1(t) > 0$ at the one-sided 2.5\% alpha-level. 
This test might be comparing time to a bad event (e.g., death), and we want to show that Group 2 (experimental arm) is better than Group 1 (control arm).
Because of symmetry, we would get the same plot if the event were a good event (e.g., time to cure) and we were testing the 
alternative that   $S_2(t) - S_1(t) < 0$.  In the right panel, we are testing only the latter alternative (e.g., showing that Group 2 has faster cure),
in a case where twice as many are randomized to Group 2.
Both panels show that the BPCP is conservative with type I error rate much less than the 2.5\% alpha-level, while the mid-p BPCP is closest to the
nominal alpha-level.  The left panels show that the delta methods are slightly more anti-conservative than the mid-p BPCP, with the standard method  the most anti-conservative (but not by much). 
The right panel gives an interesting insight. First, counter-intuitively, doubling the sample size of Group 2  can in some cases increase the type I error rate for the  delta methods. 
This is especially true for the standard delta method without the zero-one adjustment, where  for large survival times the type I error rate can be over 3 times its target (e.g., when $S_1(t)-S_2(t)=S(t) = 0.94$).
If we  instead test for a time to a bad event to show     $S_2(t) - S_1(t) > 0$, then a mirror image would result with the standard delta method having size over 3 times the alpha-level when $S(t)$ is very small. 
The difference between the standard delta method with and without the zero-one adjustment is due to the group 1 reaching an extreme value ($\hat{S}_1(t)=1$ in the right panel, or $\hat{S}_1(t)=0$ in its unshown mirror image). 
We see that even with the adjusted hybrid modifications and the  zero-one adjustment there can still be cases with type I error rate slightly less than $4\%$ (for  the $2.5\%$ target). 

If we consider testing $S_2(t)-S_1(t)<0$ for both panels, then the compatible 97.5\% one-sided  upper confidence limits are smaller than the BPCP method by on average (over the entire sample space) by: (left panel=0.0295, right panel=0.0224)  for standard method, (left=0.0284, right=0.0215) for standard method with zero-one adjustment, 
(left=0.0250, right=0.0190) for adjusted hybrid method with zero-one adjustment,  and (left=0.0268,right=0.0203) for BPCP mid-p.    
 
The above alpha-level type I error rate issue is a small sample one and as the sample size increases with the same proportional allocation the differences between the methods are expected to disappear. 
 The delta methods will have different type I error rates for different transformations. For example, in Supplementary Figure~\ref{sfig.calculation_no_censor_efflogs} we show the type I error rates for the 
 delta methods with  $\beta=1- \log(S_2(t))/\log(S_1(t)))$.  The standard method gives large type I error rates for very small or large $S(t)$ since there is no adjustment for  $\hat{S}(t)=0$ or $1$.  
Because of this, we will always perform the zero-one asdjustments in the simulations of the next section.

For small imbalanced sample sizes (e.g., $n_1=12$ and $n_2=24$),  there can be striking differences in the one-sided coverage.  Supplementary Figures~\ref{sfig-diff.coverage} ( for $\beta = S_2(t)-S_1(t)$)
and \ref{sfig-ratiolo.coverage} (for $\beta=S_2(t)/S_1(t)$)  show that the melding on the BPCP appears to have guaranteed coverage, melding on the mid-p BPCP has coverage closer to 
nominal but not guaranteed, and both delta methods with the zero-one adjustments can in some cases have coverage lower than 85\% for a nominal 97.5\% confidence.
In contrast,  Supplementary Figure~\ref{sfig-efflogs.coverage} (for $\beta=1-\log(S_2(t))/\log(S_1(t))$) shows that the delta method has one-sided coverage almost as good as melding on the mid-p BPCP.  

\section{Censoring Cases: Simulations}
\label{sec-simulations}

In Section~\ref{sec-no-censoring} we could calculate coverage for a grid of the entire parameter
space of $S_1(t)$ and $S_2(t)$. When there is censoring, the shape of the survival distributions
matter before $t$ and there are infinitely many ways of modeling censoring.

\begin{figure}
\includegraphics[width=6.0in]{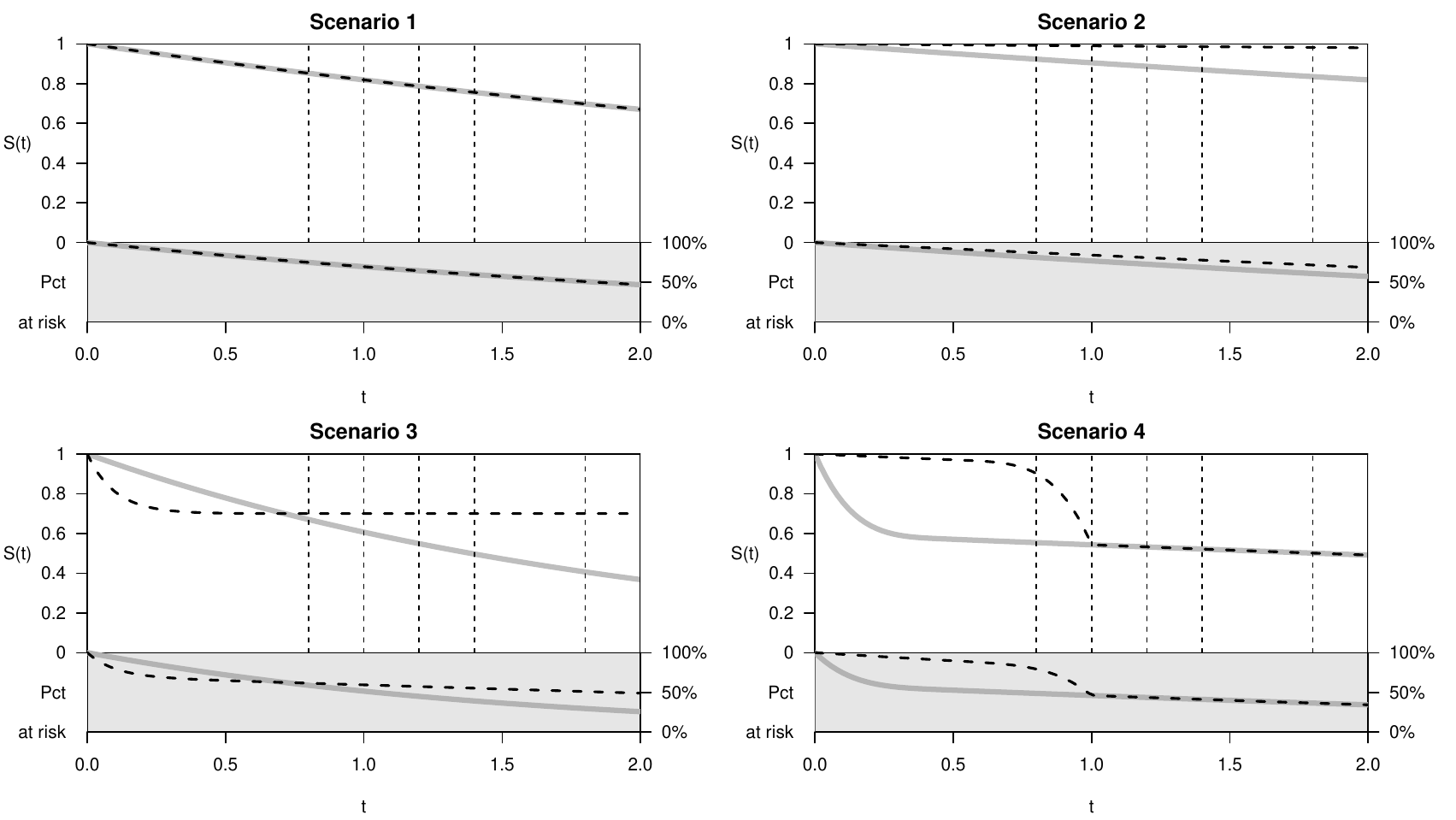}

\caption{
Simulation scenarios for Group 1 (gray lines) and Group 2 (dashed thick black lines). The upper part of each panel gives $S_1(t)$ and $S_2(t)$ by $t$,
and the lower part give the expected percent at risk at $t$ under heavy censoring.  Vertical dotted lines are the 5 values of $t$ tested in the simulations. 
\label{fig:simScenarios}
}

\end{figure}

We first consider 4 failure time generating scenarios (see Figure~\ref{fig:simScenarios}), under either heavy or light censoring, with 9 different sample sizes. 
The sample sizes have different total sample size ($n$), with 1:1, 1:2, and 2:1 allocations.  Specifically, the sample sizes are  $n=60$  with ($n_1$:$n_2$)=(30:30), (20:40), or  (40:20); 
$n=180$ with ($n_1$:$n_2$)=(90:90), (60:120), or (120:60); or 
$n=300$ with ($n_1$:$n_2$)=(150:150), (100:200), or  (200:100).  
For each of those $72=4 \times 2 \times 9$  data generating cases, we try $5$ different milestone times (0.8,1,1.2,1.4,1.8).
We run simulations for $\beta=S_2(t)-S_1(t)$ and repeat the whole group of simulations for $\beta=1-\log(S_2(t))/\log(S_1(t))$.
For each case we simulate 10,000 data sets, so that for example if the simulated error rate was 2.5\%, we are 95\% confident that the 
the true error rate is roughly between $2.2$\% and $2.8$\% (by binomial exact test).  Loosely speaking, the simulated error should be within about $\pm 0.3\%$ of its true value.

The censoring is simulated as independent of the failure times. For heavy censoring  $p_C=40\%$ of the study population has uniform censoring on the interval $(0,2)$, while the other $1-p_C=95\%$ is 
administratively censored just after 2, denoting the end of the study.  Light censoring is analogous with $p_C=10\%$. 

Scenario~1 has both groups exponential with  the same rate of $\lambda_1=\lambda_2=0.20$ to represent  the null case where there is no difference between the two arms. 
Scenario~2  has different exponential distributions,  $\lambda_1=0.1$ for Group~1 and $\lambda_2=0.01$ for Group~2, representing a highly effective treatment compared to control.
The third and fourth event time scenarios consist of a mixture of two populations that cannot be identified at baseline. 
In Scenario~3, we call the populations $A$ ($70\%$)  and $B$ ($30\%$). In Group~1 (control arm), both populations have the same the failure time distribution; it is exponential with rate $\lambda_1=0.5$.  
In Group~2, the experimental treatment  is harsh and it either kills you more quickly (those in population $B$ have exponential failures with rate $\lambda_B=10$) or cures you (those in population~$A$ all survive to at least 2 years). 
Scenario~4  is also a mixture of two populations, now $A$ ($60\%$) and $B$ ($40\%$).  In this scenario, the treatment does not affect population~$A$, 
and regardless of the arm a type A individual has failure that is exponential with rate $0.10$. 
Those in population $B$ have under the control arm (Group~1) have  a random failure time of   $T=B(1,4)/2$ (recall $B(a,b) \sim Beta(a,b)$), so that all fail by $t=0.5$,
while for the experimental treatment that have $T= 0.5 + B(4,1)/2$, so that that treatment delays failures by at least 0.5 years  for those in population~B, but by 1 year all those in population $B$ fail.  
So for the fourth scenario, there is no difference between the arms after 1 year.

\begin{figure}
\includegraphics[width=6.0in]{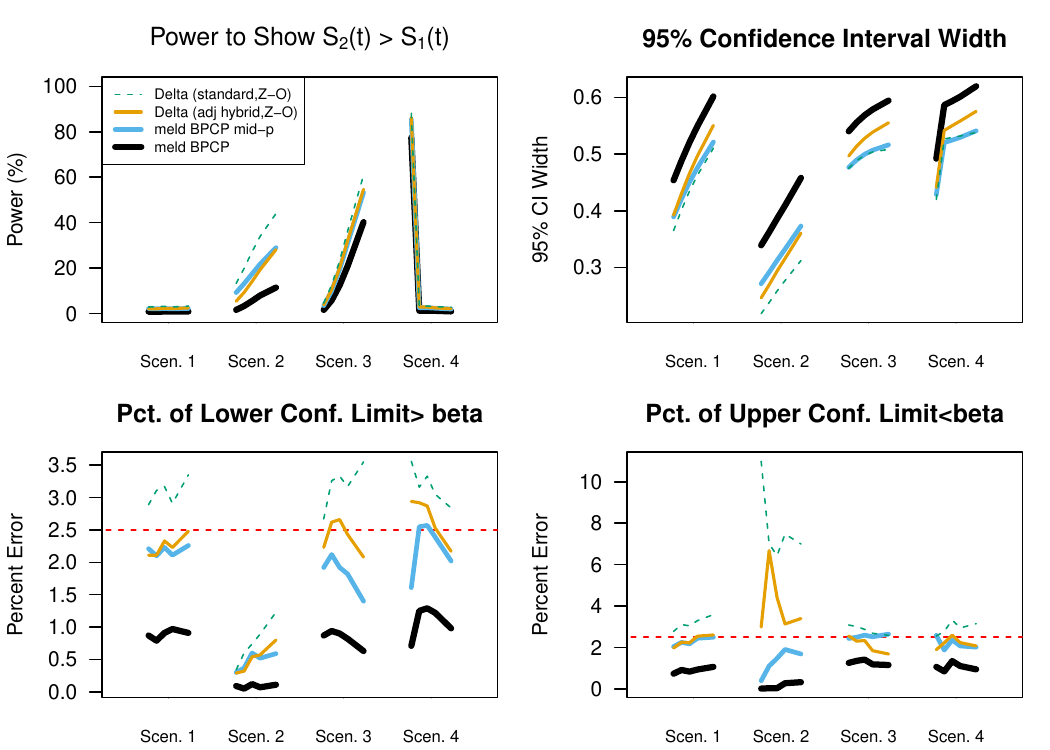}

\caption{
{\bf Simulation Results for $\beta=S_2(t)-S_1(t)$ with $n_1=n_2=30$ with heavy censoring.}
In each panel the four lines connect the 5 milestone times for each of the 4 methods.
 Upper left panel is the power to show $S_2(t) >  S_1(t)$ at the one-sided 2.5\% level.
The upper right panel is the 95\% central confidence interval width (which is the intersection of the two 
one-sided 97.5\% intervals).
The lower panels represent the one-sided error rates of the one-sided 97.5\% confidence intervals.
Scenario~1 and the last 4 times of Scenario~4 represent the null case when $S_2(t)=S_1(t)$.  The power for those null case points may be read off of the lower left panel, since in the null case the power is error in the one-sided test.
\label{fig-RD01}
}

\end{figure}

In Figure~\ref{fig-RD01} we plot the simulation results for $\beta=\beta_d= S_2(t)-S_1(t)$ on all 4 scenarios when $n_1=n_2=30$ with heavy censoring. The simulation results are in 
4 panels, with 4 line segments within each scenario of each panel, representing the 5 times that are tested on each of the 4 methods (standard delta method with zero-one adjustment [standard], 
adjusted hybrid method with zero-one adjustment [adjusted hybrid], melding on the BPCP [mBPCP], and melding on the mid-p  BPCP  [mBPCP mid-p]).
We plot the power and confidence interval width in the top panels, and the one-sided confidence limit errors on the bottom panels. For Scenario~1 and the last 4 test times in Scenario~4,
the bottom panels represent the type I error rate for the corresponding one-sided test at the 2.5\% level.  
We give analgous figures for the rest of the simulations of this section in the Supplement. 
 Supplementary Figures~\ref{sfig-RD02} to  \ref{sfig-RD18} give other sample size and censoring results when $\beta$ is the difference.
Supplementary Figures ~\ref{sfig-RE01} to  \ref{sfig-RE18} give analogous results when $\beta=\beta_{els} =1- \left\{ \log(S_2(t))/\log(S_1(t)) \right\} $.  
As in Figure~\ref{fig:B5panels} we use a transformation on the confidence intervals when $\beta=\beta_{els}$ so that the width to extreme values (e.g., $-\infty$) may be calculated 
and compared to the widths for the confidence intervals for $\beta_d$ in a fairer way.

For simulations on both  $\beta_d$ and $\beta_{els}$, mBPCP has confidence limit error less than the nomial 2.5\% in all cases, most of the time substantially less. 
The other 3 methods have at least some situations where the one-sided confidence interval error is greater than the 2.5\% nominal level.  Unsurprisingly, mBPCP generally has the lowest power
and longest width confidence intervals, although with Scenario~2 on $\beta_{els}$ it usually has better power and smaller width intervals than the adjusted hybrid method.
By design, mBPCP mid-p has better power than mBPCP. In almost all cases, the mBPCP mid-p has the one-sided confidence interval error closest to the nominal level; however in some cases 
(see Figure~\ref{fig-RD01}, lower limit error Scenario 3 ) adjusted hybrid is closest. Adjusted hybrid has some situations where the confidence interval error is clearly above the nominal (see upper limit error for $\beta_d$  Scenario~2 except in cases where $n_1=2*n_2$). 
Although less often, there are situations (see upper limit error for $\beta_d$ when  $n_1=2*n_2$)  where mBPCP mid-p has error above the nomial. The standard method and less often the adjusted hybrid method can have one-sided confidence limit error substantially greater than 
nominal (see $\beta_d$ Scenario~2 when $n_2=2*n_1$).  In agreement with the asymptotic theory, all methods have error closer to nominal levels as the sample sizes gets larger.

It appears that for $\beta_{els}$ the adjusted hybrid has less than nominal error rates in almost all cases, except the first time point in Scenario~4. 
We did another simulation with details presented in 
Supplemental Section~\ref{sec-simDiscrete} that shows the adjusted hybrid method can have error over 11\% for $\beta_{els}$ with large sample sizes ($n_1=n_2=300$) but very heavy censoring.  See Supplementary Figure~\ref{sfig-simDiscreteResults} which shows that, mBPCP always bounds the type I error rate with rates often substantially less than the nominal 2.5\%, the adjusted hybrid performs the worst, with type I error rates above 11\%, while the standard and mBPCP mid-p perform better but with type I error rate sometimes close to 5 or 6\%.

Overall, if validity is a top priority, then mBPCP is preferred, otherwise mBPCP mid-p is perferred since it is generally closest to the nominal error rate, often has the highest power (especially with $\beta_{els}$) and has only a few situations where it has error rates clearly higher than the nominal.

AI was occasionally used to assist with debugging code or with suggestions of small pieces of R
code.

\section{Applications}
\label{sec-applications}

In Table~\ref{tab:applications1}
we apply all the methods studied to the three examples of 
Swennes {\it et al}\cite{Swen:2007}  presented in Table~\ref{tab:delta.method.no.mods}. For these extreme examples, the
standard (Greenwood variance) delta method and Borkowf's adjusted hybrid delta method
are nearly identical, perhaps because both are using the same zero-one adjustments.
The shrunken adjusted hybrid method differs because it gives $\beta$ estimates based on the shruken survival curves,. Since the melding BPCP method appears valid (see Sections~\ref{sec-no-censoring} and \ref{sec-simulations}), we expect it to have the largest
intervals; however, this is not always so (see the last column for the efficacy on log(S)). The melding on the mid-p BPCP often gives the smallest width intervals.

In Table~\ref{tab:applications2} we apply the methods to data from Moertel {\it et al}\cite{Moer:1995}.
From Figure~\ref{lev3} we see little difference between the methods at time $t=6$ years when there
are still large numbers at risk, so it is unsurprising that the methods give similar confidence intervals.
By $t=8$ years, there are fewer at risk and the methods differ more, and for the subset with small sample sizes we see the biggest difference between the methods.

Overall, the melding BCPC method tends to give larger width intervals than the delta method ones (although not always), we suspect from Sections~\ref{sec-no-censoring} and \ref{sec-simulations} is much more likely to be valid.
Melding on the  mid-p BPCP has smaller width intervals than melding on the BPCP,  but may be smaller or larger than the delta method intervals.

\section{Discussion}
\label{sec-discussion}

In this paper we have modified the standard delta method confidence intervals by (1) substituting
the Greenwood variance for ones developed by Borkowf\cite{Bork:2005} which have been shown by simulation to be less biased,
and (2) developed modifications that give more reasonable intervals when either of the Kaplan-Meier estimates is $0$ or $1$.   We have compared these confidence intervals to melded confidence intervals using either beta product confidence procedure for the Kaplan-Meier or its mid-p version. These
melded intervals automatically adjust and produce sensible intervals for small sample sizes or Kaplan-Meier estimates on the boundary. Calculations and simulations suggest that the melded BPCP gives valid intervals, while the other methods are shown to have coverage less than nominal in some situations.

When there is no censoring, the problem is a two-sample binomial one, and the BPCP gives confidence intervals for the entire class of estimands
studied (see Supplementary Table~\ref{tab:betas}) that are compatible with the central Fisher's exact test. Unconditional exact test are more powerful valid tests for the two-sample binomial problem \citep{FayH:2021}; however, those tests have not been
generalized to allow for censoring. The melding on the mid-p BPCP can approximate the unconditional exact tests with no censoring, and provides an option with censoring. 
Compared to melding on the BPCP, melding on the mid-p BPCP is more powerful,  gives type I error rates closer to the nominal, but it is known to sometimes give type I error rates greater than nominal.
Thus, if validity is of primary concern, melding on the BPCP is recommended, otherwise melding the on mid-p BPCP is recommended.

The {\sf bpcp} R package (version 1.5.5 or later) has functions for all the methods of this paper. Importantly,
the melding on the BPCP (usual or mid-p version)  are quick (calculations take less than a second for a total sample size of 1,000),
so are practical for routine use.


\section*{Acknowledgements}

This work utilized the computational resources of the NIH HPC Biowulf cluster \verb@(https://hpc.nih.gov).@

Funding:
This research was supported in part by the Intramural Research Program of the National Institutes of
Health (NIH).
The contributions of the NIH author(s) are considered Works of the United States
Government. The findings and conclusions presented in this paper are those of the authors) and do not
necessarily reflect the views of the NIH or the U.S. Department of Health and Human Services.

This project has been funded in whole or in part with federal funds from the National Cancer Institute,
National Institutes of Health, under Contract No. 75N91019D00024. The content of this publication does
not necessarily reflect the views or policies of the Department of Health and Human Services, nor does
mention of trade names, commercial products, or organizations imply endorsement by the U.S.
Government.

Competing interests:
 All authors declare no financial or non-financial competing interests.



\bibliographystyle{unsrtnat}
\bibliography{refs}

\clearpage

\setcounter{section}{0}
\renewcommand*{\thesection}{S\arabic{section}}

\setcounter{table}{0}
\renewcommand*{\thetable}{S\arabic{table}}

\setcounter{figure}{0}
\renewcommand*{\thefigure}{S\arabic{figure}}

\setcounter{equation}{0}
\renewcommand*{\theequation}{S\arabic{equation}}

\begin{center}
{\LARGE \bf Supplement to:Comparing Two Survival Functions at a Fixed Timepoint}
\end{center}

Michael P. Fay, Allyson Mateja, and Megan C. Grieco

\section*{Summary}

We provide supplementary tables, figures and extra simulation details referenced in the main paper.
References are given in the main paper.

\section{Supplemental Tables}
\label{app-tables}

\begin{table}[H]
\begin{center}
\caption{Estimands for comparing $S_2(t)$ and $S_1(t)$. All ratios define $0/0$ or $\infty/\infty$ as $1$.
For asymptotic normality transformations, we write $\beta = g \left\{ h \left( S_2(t) \right) - h \left( S_1(t) \right) \right\} = g(D)$ for transformation functions $g()$ and $h()$.
The derivative of $h()$, $h'()$, is used in the delta method formula.
\label{tab:betas}}

\begin{tabular}{llcccc} \hline
&  & Equality  & \multicolumn{3}{c}{ Transformation Functions } \\
Name   &  Definition & Value ($\beta_e$) &  $g(D)$ & $h(S)$  & $h'(S)$ \\ \hline
 Difference &  $\beta_d = S_2(t) - S_1(t)$ & 0 & $D$ & $S$ & 1 \\
 Ratio &  $\beta_r = S_2(t)/ S_1(t)$ & 1 &  $\exp(D)$ & $\log(S)$ & $\frac{1}{S}$ \\
Odds Ratio &  $\beta_{or} = \frac{ S_2(t) (1-S_1(t)) }{ S_1(t) (1-S_2(t)) }$ & 1 &  $\exp(D)$ & $\log \left\{ \frac{S}{1-S} \right\}$ & $\frac{1}{S(1-S)}$  \\
 Effect using &  $\beta_{ecd} = 1 - \frac{ 1- S_2(t) }{1-S_1(t)}$ & 0   &  $1 - \exp(D)$ & $\log(1-S)$ &  $\frac{-1}{1-S}$ \\
  Cumulative Distribution &   \\
 Effect using & $\beta_{els} = 1 - \frac{ \log\left\{ S_2(t) \right\} }{ \log\left\{ S_1(t) \right\}}$ & 0 & $1-\exp(D)$ &  $\log \left\{ - \log( S ) \right\}$ & $\frac{1}{S \log(S)}$ \\
 log Survival  &  \\ \hline
\end{tabular}
\end{center}
\end{table}

\begin{table}[H]
\begin{center}
\caption{Three examples from
Swennes, et al, 2007
of delta method without modifications. All examples have $n=20$ in each arm, with no censoring.
NA values are due to for example, $0/0$ or $\log$ of a negative number.
\label{tab:delta.method.no.mods}}

\begin{tabular}{lcccccccl} \hline
Estimand   &   $h(S)$  & $\hat{S}_1(1)$ & $\hat{S}_2(1)$ & $h(\hat{S}_1(1))$ & $h(\hat{S}_2(1))$ & $\hat{\beta}$ & 95\% CI & p-value  \\ \hline
 $S_2(t) - S_1(t)$ &  $S$ & $0$  & $1$ & $0$ & $1$ & $1$ & $[1, 1]$ & $0$ \\
  $S_2(t)/ S_1(t)$ &  $\log(S)$ & $0$  & $1$ & $-\infty$ & $0$ & $\infty$ & [NA,NA] & NA \\
 $\frac{ S_2(t) (1-S_1(t)) }{ S_1(t) (1-S_2(t)) }$ &  $\log \left\{ \frac{S}{1-S} \right\}$ & $0$  & $1$ & $-\infty$ & $\infty$ & $\infty$ & [NA,NA] & NA\\
 $1 - \frac{ 1- S_2(t) }{1-S_1(t)}$  & $\log(1-S)$ &   $0$  & $1$ & $0$ &  $-\infty$ & $1$ & [NA,NA] & NA \\
 $1 - \frac{ \log\left\{ S_2(t) \right\} }{ \log\left\{ S_1(t) \right\}}$ &  $\log \left\{ - \log( S ) \right\}$ & $0$  & $1$ & $\infty$ & $-\infty$ & $1$ & [NA,NA] & NA \\
 \hline
  $S_2(t) - S_1(t)$ &  $S$ & $0$ &  $0.80$ & $0$ & $0.80$ & $0.80$ & $[0.625, 0.975]$ & $<0.0001$ \\
  $S_2(t)/ S_1(t)$ &  $\log(S)$ & $0$  & $0.80$ & $-\infty$ & $-0.223$ & $\infty$ & [NA, NA] & NA \\
 $\frac{ S_2(t) (1-S_1(t)) }{ S_1(t) (1-S_2(t)) }$ &  $\log \left\{ \frac{S}{1-S} \right\}$ & $0$ &  $0.80$ & $-\infty$ & $1.39$  & $\infty$ & [NA,NA] & NA\\
 $1 - \frac{ 1- S_2(t) }{1-S_1(t)}$  & $\log(1-S)$ &  $0$ &  $0.80$  &  $0$ & $-1.61$ & $0.8$ & [0.519,0.917] & $0.0003$ \\
 $1 - \frac{ \log\left\{ S_2(t) \right\} }{ \log\left\{ S_1(t) \right\}}$ &  $\log \left\{ - \log( S ) \right\}$ & $0$ & $0.80$ & $\infty$  & $-1.50$ & $1$ & [NA,NA] & NA \\ \hline
 $ S_2(t) - S_1(t)$ &  $S$ & $0.80$  & $1$ & $0.80$ & $1$ & $0.20$ & $[0.025, 0.375]$ & $0.025$ \\
  $S_2(t)/ S_1(t)$ &  $\log(S)$ & $0.80$  & $1$ & $-0.223$ & $0$ & $1.25$ & [1.004,1.556] & 0.046 \\
 $ \frac{ S_2(t) (1-S_1(t)) }{ S_1(t) (1-S_2(t)) }$ &  $\log \left\{ \frac{S}{1-S} \right\}$ & $0.80$  & $1$ & $1.39$ & $\infty$ & $\infty$ & [NA,NA] & NA\\
 $1 - \frac{ 1- S_2(t) }{1-S_1(t)}$  & $\log(1-S)$ &   $0.80$  & $1$ &   $-1.61$ & $-\infty$ & $1$ & [NA,NA] & NA \\
 $1 - \frac{ \log\left\{ S_2(t) \right\} }{ \log\left\{ S_1(t) \right\}}$ &  $\log \left\{ - \log( S ) \right\}$ & $0.80$  & $1$ & $-1.50$ & $-\infty$ & $1$ & [NA,NA] & NA \\ \hline
\end{tabular}
\end{center}
\end{table}

\begin{table}[H]
\begin{center}
\caption{Estimates (95\% confidence intervals) for $\beta$ estimands using proposed methods
on three examples from
Swennes, et al, 2007.
All Delta methods use the zero-one adjustment.
All examples have $n=20$ in each arm with $t=26$ days.
\label{tab:applications1}}

\begin{tabular}{lccc} \hline
 & \multicolumn{3}{c}{Difference: $S_2(t)-S_1(t)$ } \\
Method & $\hat{S}_2(t)=1$ vs. $\hat{S}_1(t)=0$ & $\hat{S}_2(t)=0.80$ vs. $\hat{S}_1(t)=0$ &  $\hat{S}_2(t)=1$ vs. $\hat{S}_1(t)=0.80$ \\ \hline
Standard Delta & 1.00 [0.90, 1.00] & 0.80 [0.61, 0.99] & 0.20 [0.01, 0.39] \\
Reg. Hybrid Delta & 1.00 [0.90, 1.00] & 0.80 [0.61, 0.99] & 0.20 [0.01, 0.39] \\
Adj. Hybrid Delta & 1.00 [0.90, 1.00] & 0.80 [0.61,0.99] & 0.20 [0.01, 0.39] \\
Shr. Adj. Hybrid Delta & 0.95 [0.85, 1.05] & 0.76 [0.57, 0.95] & 0.19 [0.00, 0.38] \\
Melding BPCP & 1.00 [0.75, 1.00] & 0.80 [0.49, 0.94] & 0.20 [-0.03, 0.44] \\
Melding mid-p BPCP & 1.00 [0.80, 1.00] & 0.80 [0.54, 0.92] & 0.20 [0.01, 0.40] \\ \hline
 &  \multicolumn{3}{c}{Ratio: $S_2(t)/S_1(t)$ } \\
 Method & $\hat{S}_2(t)=1$ vs. $\hat{S}_1(t)=0$ & $\hat{S}_2(t)=0.80$ vs. $\hat{S}_1(t)=0$ &  $\hat{S}_2(t)=1$ vs. $\hat{S}_1(t)=0.80$ \\ \hline
Standard Delta & $\infty$ [2.59, $\infty$] & $\infty$ [2.05, $\infty$] & 1.25 [0.99, 1.57] \\
Reg. Hybrid Delta & $\infty$ [2.59 $\infty$] & $\infty$ [2.05, $\infty$] & 1.25 [0.99, 1.57] \\
Adj. Hybrid Delta & $\infty$ [2.59, $\infty$] & $\infty$ [2.05, $\infty$] & 1.25 [0.99, 1.58] \\
Shr. Adj. Hybrid Delta & 39 [0.00, $\infty$] & 31.4 [0.00, $\infty$] & 1.24 [0.98, 1.57] \\
Melding BPCP & $\infty$ [5.63, $\infty$] & $\infty$ [4.39, $\infty$] & 1.25 [0.96, 1.77] \\
Melding mid-p BPCP & $\infty$ [7.00, $\infty$] & $\infty$ [5.52, $\infty$] & 1.25 [1.01, 1.68] \\ \hline
 & \multicolumn{3}{c}{Efficacy on log(S): $1 - \log \left\{ S_2(t) \right\}/ \log \left\{ S_1(t) \right\}$ } \\
 Method & $\hat{S}_2(t)=1$ vs. $\hat{S}_1(t)=0$ & $\hat{S}_2(t)=0.80$ vs. $\hat{S}_1(t)=0$ &  $\hat{S}_2(t)=1$ vs. $\hat{S}_1(t)=0.80$ \\ \hline
Standard Delta & 1.00 [0.88, 1.00] & 1.00 [0.79, 1.00] & 1.00 [-1.15, 1.00] \\
Reg. Hybrid Delta & 1.00 [0.88, 1.00] & 1.00 [0.79, 1.00] & 1.00 [-1.15, 1.00] \\
Adj. Hybrid Delta & 1.00 [0.88, 1.00] & 1.00 [0.79, 1.00] & 1.00 [-1.17, 1.00] \\
Shr. Adj. Hybrid Delta & 0.99 [-$\infty$, 1.00] & 0.93 [-$\infty$, 1.00] & 0.90 [-$\infty$, 1.00] \\
Melding BPCP & 1.00 [0.94, 1.00] & 1.00 [0.78, 1.00] & 1.00 [-0.40, 1.00] \\
Melding mid-p BPCP & 1.00 [0.96, 1.00] & 1.00 [0.83, 1.00] & 1.00 [0.11, 1.00] \\ \hline
\end{tabular}
\end{center}
\end{table}

\begin{table}[H]
\begin{center}
\caption{Estimates (95\% confidence intervals) for $\beta$ estimands using proposed methods on three examples from
Moertel, {\it et al.} (1995)
comparing Lev (group 1) to Lev+5FU (group 2).
Sample sizes are $n=310$ for Lev arm,
and $n=304$ for Lev+5FU arm, and the subset with perforated colon at $t=0$ has
$n=10$ for Lev arm,
and $n=8$ for Lev+5FU arm.
\label{tab:applications2}}

\begin{tabular}{lccc} \hline
 & \multicolumn{3}{c}{Difference: $S_2(t)-S_1(t)$ } \\
Method & Lev vs. Lev+5FU (all, $t=6$)  & Lev vs. Lev+5FU (all, $t=8$) &  Lev vs. Lev+5FU (perf. colon, $t=6$)  \\ \hline
Standard Delta & 0.11 [0.04, 0.19] & 0.17 [0.04, 0.30] & 0.38 [-0.06, 0.81] \\
Reg. Hybrid Delta & 0.11 [0.03, 0.20] & 0.17 [0.06, 0.28] & 0.38 [-0.08, 0.83] \\
Adj. Hybrid Delta & 0.11 [0.03, 0.20] & 0.17 [0.06, 0.28] & 0.38 [-0.09, 0.84] \\
Shr. Adj. Hybrid Delta & 0.11 [0.03, 0.20] & 0.17 [0.05, 0.28] & 0.33 [-0.13, 0.79] \\
Melding BPCP & 0.11 [0.03, 0.20] & 0.17 [-0.02, 0.35] & 0.38 [-0.23, 0.78] \\
Melding mid-p BPCP & 0.11 [0.03, 0.19] & 0.17 [0.01, 0.32] & 0.38 [-0.13, 0.73] \\ \hline
 &  \multicolumn{3}{c}{Ratio: $S_2(t)/S_1(t)$ } \\
Method & Lev vs. Lev+5FU (all, $t=6$)  & Lev vs. Lev+5FU (all, $t=8$) &  Lev vs. Lev+5FU (perf. colon, $t=6$)  \\ \hline
Standard Delta & 1.23 [1.06, 1.43] & 1.43 [1.05, 1.94] & 2.00 [0.79, 5.07] \\
Reg. Hybrid Delta & 1.23 [1.05, 1.44] & 1.43 [1.12, 1.82] & 2.00 [0.77, 5.20] \\
Adj. Hybrid Delta & 1.23 [1.05, 1.44] & 1.43 [1.12, 1.82] & 2.00 [0.76, 5.25] \\
Shr. Adj. Hybrid Delta & 1.23 [1.05, 1.44] & 1.43 [1.12, 1.82] & 1.85 [0.71, 4.87] \\
Melding BPCP & 1.23 [1.06, 1.44] & 1.43 [0.95, 2.46] & 2.00 [0.63, 8.87] \\
Melding mid-p BPCP & 1.23 [1.06, 1.43] & 1.43 [1.02, 2.24] & 2.00 [0.77, 6.72] \\ \hline
 & \multicolumn{3}{c}{Efficacy on log(S): $1 - \log \left\{ S_2(t) \right\}/ \log \left\{ S_1(t) \right\}$ } \\
Method & Lev vs. Lev+5FU (all, $t=6$)  & Lev vs. Lev+5FU (all, $t=8$) &  Lev vs. Lev+5FU (perf. colon, $t=6$)  \\ \hline
Standard Delta & 0.30 [0.10, 0.45] & 0.38 [0.11, 0.57] & 0.71 [-0.50, 0.94] \\
Reg. Hybrid Delta & 0.30 [0.08, 0.46] & 0.38 [0.14, 0.56] & 0.71 [-0.64, 0.95]  \\
Adj. Hybrid Delta & 0.30 [0.08, 0.46] & 0.38 [0.14, 0.56] & 0.71 [-0.73, 0.95] \\
Shr. Adj. Hybrid Delta & 0.29 [0.08, 0.46] & 0.38 [0.14, 0.55] & 0.65 [-1.05, 0.94] \\
Melding BPCP & 0.30 [0.09, 0.46] & 0.38 [-0.07, 0.63] & 0.71 [-1.03, 0.98] \\
Melding mid-p BPCP & 0.30 [0.10, 0.45] & 0.38 [0.02, 0.60] & 0.71 [-0.50, 0.96] \\ \hline
\end{tabular}
\end{center}
\end{table}

\section{Figures of Moertel, et al Data}
\label{app-figures}

\begin{figure}[H]
\includegraphics[width=5.0in]{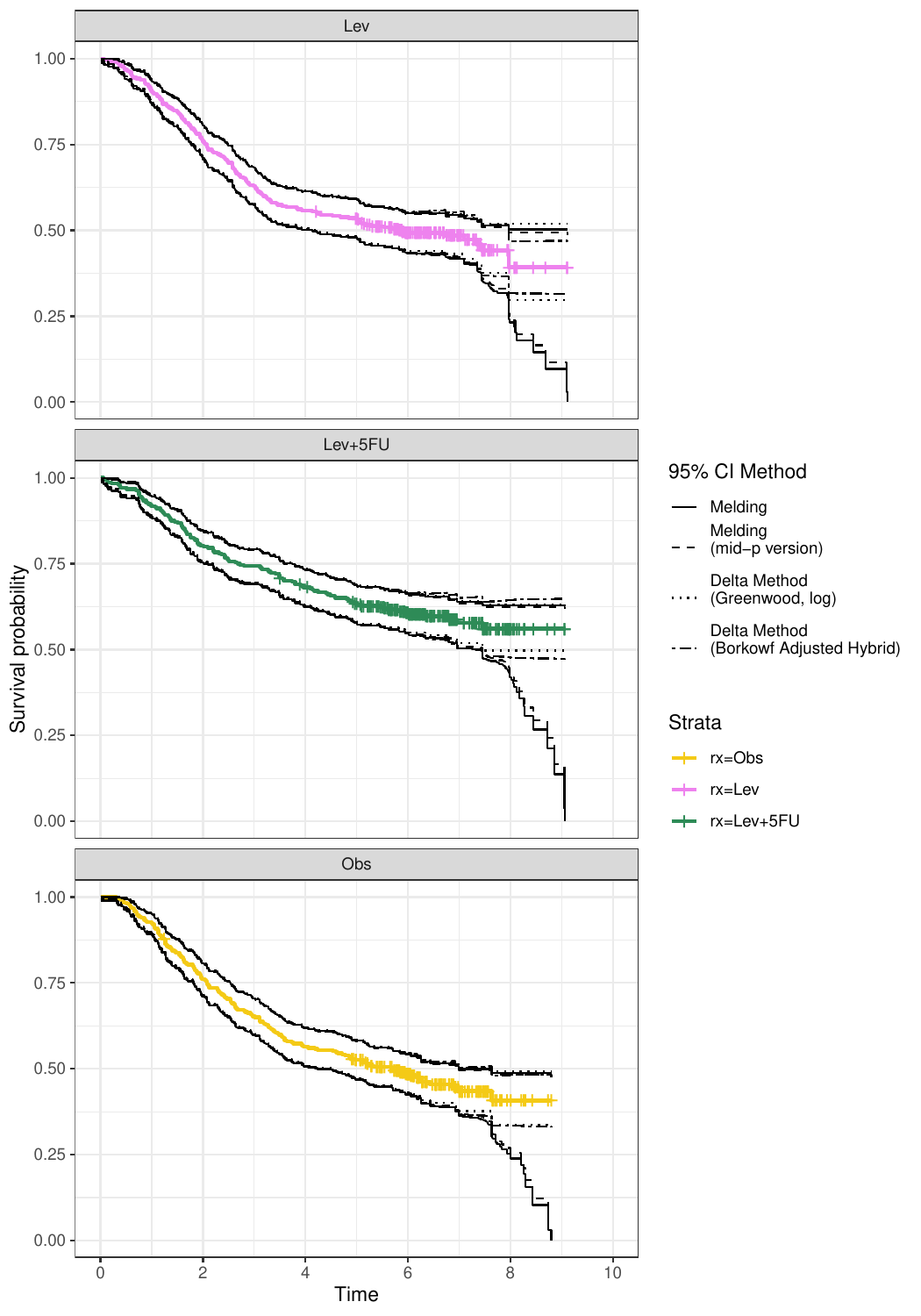}

\caption{
Kaplan-Meier curves of years from randomization to death in three arms of a
 randomized study of patients with stage III colon cancer recovered from surgery
Moertel, et al (1995).
 Vertical bars represent censored observations. The three arms are: levamisole alone (Lev) [$n=310$], flourouracil plus levamisole (Lev+5FU) [$n=304$], and control (Obs) [$n=315$].
95\% confidence intervals using melded with BPCP (solid), melded with mid-p version of BPCP (dashed),
delta method using Greenwood variance with a  log transformation (dotted),
or Borkowf's adjusted hybrid  method (dash-dot). The number at risk  in each of the groups  are: at 6 years ($n=108$ Lev, $n=128$ Lev+5FU, $n=101$ Obs) and at 8 years
($n=7$ Lev, $n=12$ Lev+5FU, $n=7$ Obs).
\label{lev3}
}
\end{figure}

\begin{figure}[bh!]
\includegraphics[width=6.5in]{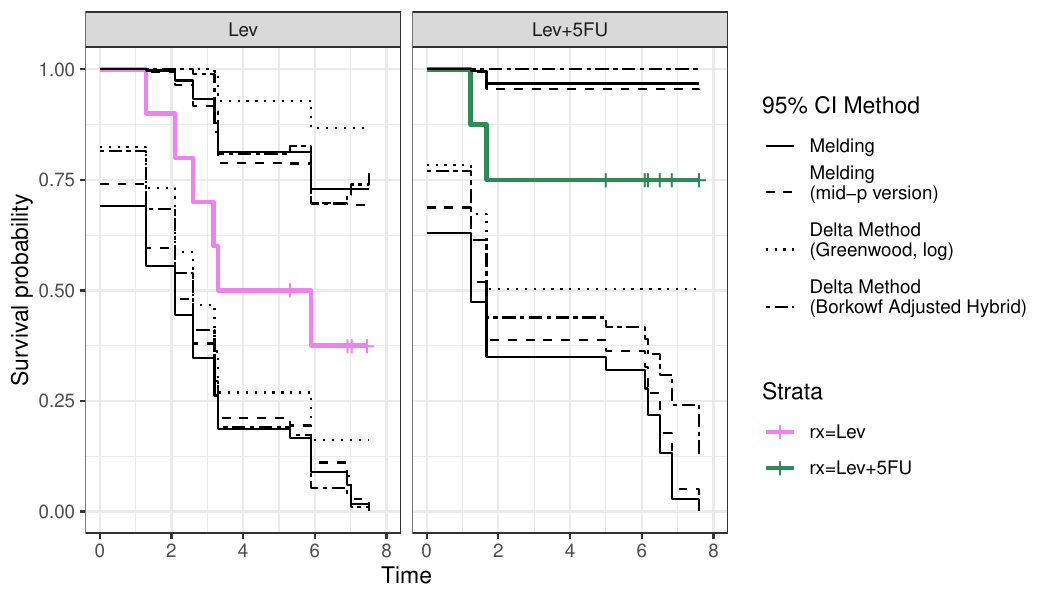}

\caption{Subset of
Moertel, {\it et al.} (1995)
data from Figure~\ref{lev3}: only those with perforated colon at
time 0, and only the two treatment arms. Sample sizes are: levamisole alone (Lev) [$n=10$], flourouracil plus levamisole (Lev+5FU) [$n=8$].
95\% confidence intervals using melded with BPCP (solid), melded with mid-p version of BPCP (dashed),
delta method using Greenwood variance with a  log transformation (dotted),
or Borkowf's adjusted hybrid method (dash-dot).
\label{Perf2panel}
}

\end{figure}

\clearpage

\section{Figures for Numeric Calculations with No Censoring}

\begin{figure}[H]
\includegraphics[width=6.0in]{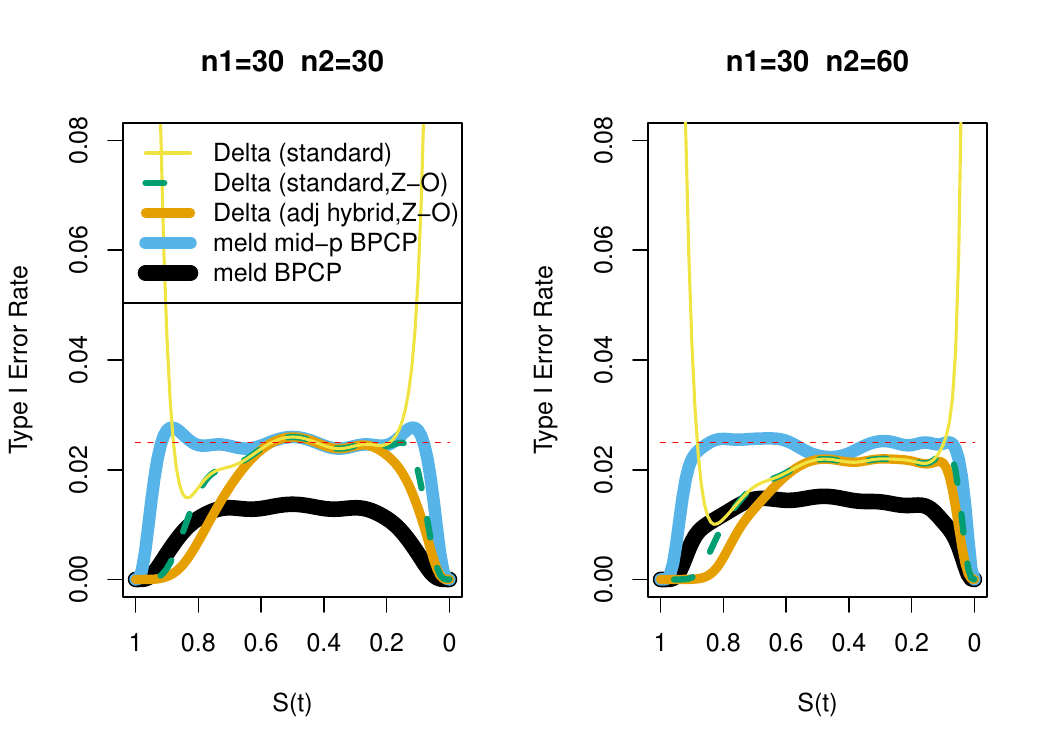}

\caption{
Type I error rate for testing when the null hypothesis is $S_1(t)=S_2(t)=S(t)$ with no censoring. For the delta methods we use the $\log \left\{ -\log(S) \right\}$ transformation and the 
parameter $\beta = 1-\log(S_2(t))/\log(S_1(t))$ (see Table~\ref{tab:betas}). 
Because the surival function goes from $1$ to $0$ as time progesses, we create the horizontal axes with $1$ on the left and $0$ on the right. 
The left panel has $n_1=n_2=30$ testing the alternative  
alternative that $S_2(t) - S_1(t) > 0$  (or testing the alternative that $S_2(t) - S_1(t) < 0$)  at the one-sided 2.5\% alpha-level. 
The right panel has $n_1=30$ and $n_2=60$ testing the alternative  
 that $S_2(t) - S_1(t) > 0$.  Type I error rates are lines through the fixed time points such that $S(t)=1,0.99,0.98,\ldots,0$. We test  3 delta methods applied to the efficacy using $\log(S)$, the standard method 
without the zero-one adjustment (Wald test on the difference in proportions with no censoring), the standard method with the zero-one adjustment,
and the adjusted hybrid with the zero-one adjustmet. and two BPCP methods, the usual BPCP (Fisher's exact test with no censoring) and BPCP mid-p version (mid-p Fisher's exact test with no censoring). 
\label{sfig.calculation_no_censor_efflogs}
}

\end{figure}




\begin{figure}
\includegraphics[width=6.0in]{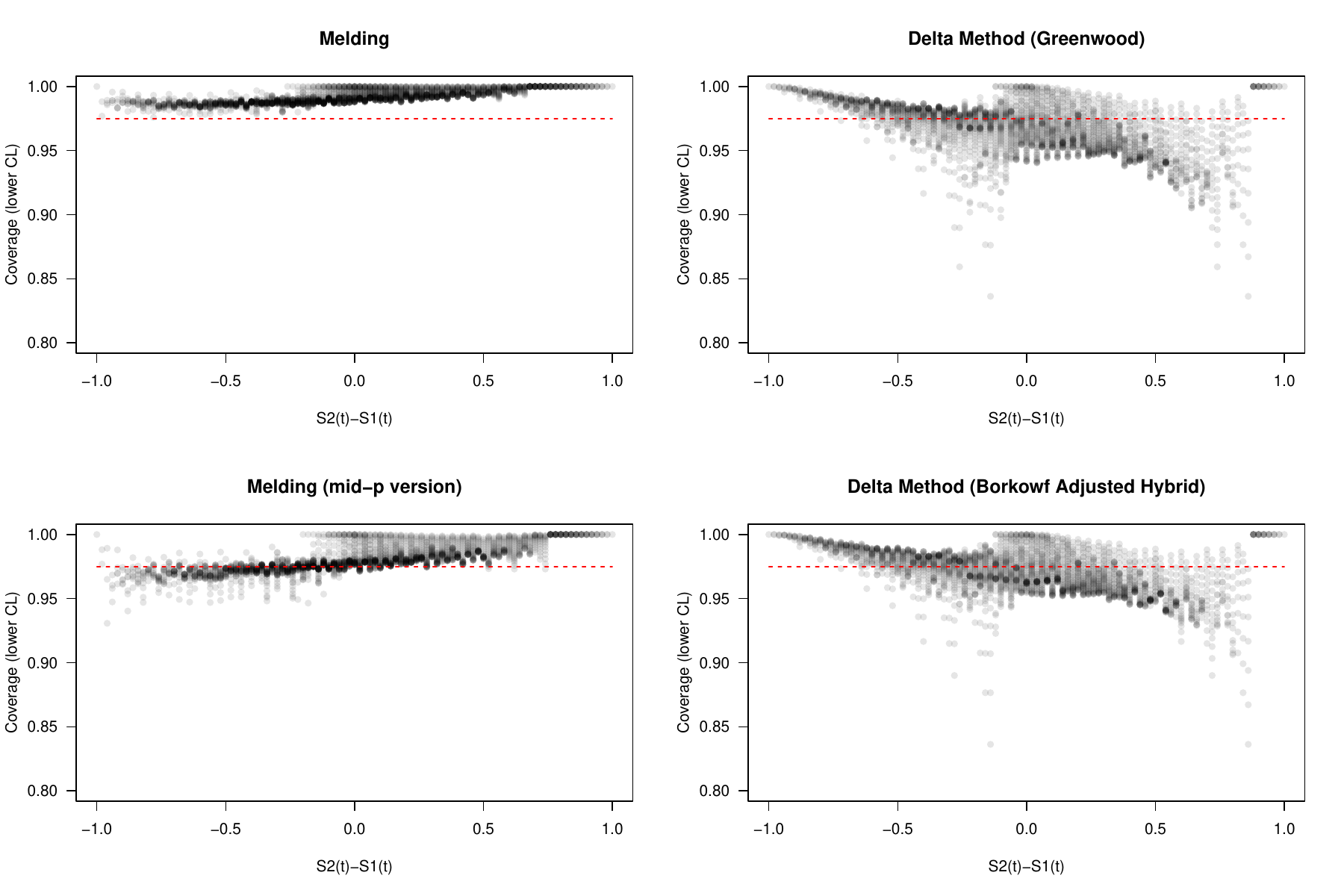}

\caption{
Coverage of the 97.5\%  one-sided confidence intervals of $S_2(t) - S_1(t)$ using the lower limit
 ($n_1=12$ and $n_2=24$).
The left panels are the melding methods (original on top, and mid-p version on the bottom).
 The right panels are the delta methods using  the zero-one modifications
 (Greenwood [i.e., standard] on top, Borkowf adjusted hybrid on bottom).
  Each point represents the coverage at
 one of the $51 \times 51 = 2601$ values of $S_1(t)$ and $S_2(t)$ on a grid.
 A valid confidence interval would have no points below the dotted red line at 97.5\%. Points are graphed semi-transparently,
 so that darker points represent more overlapping points.
\label{sfig-diff.coverage}
}

\end{figure}

\begin{figure}[htb!]
\includegraphics[width=6.0in]{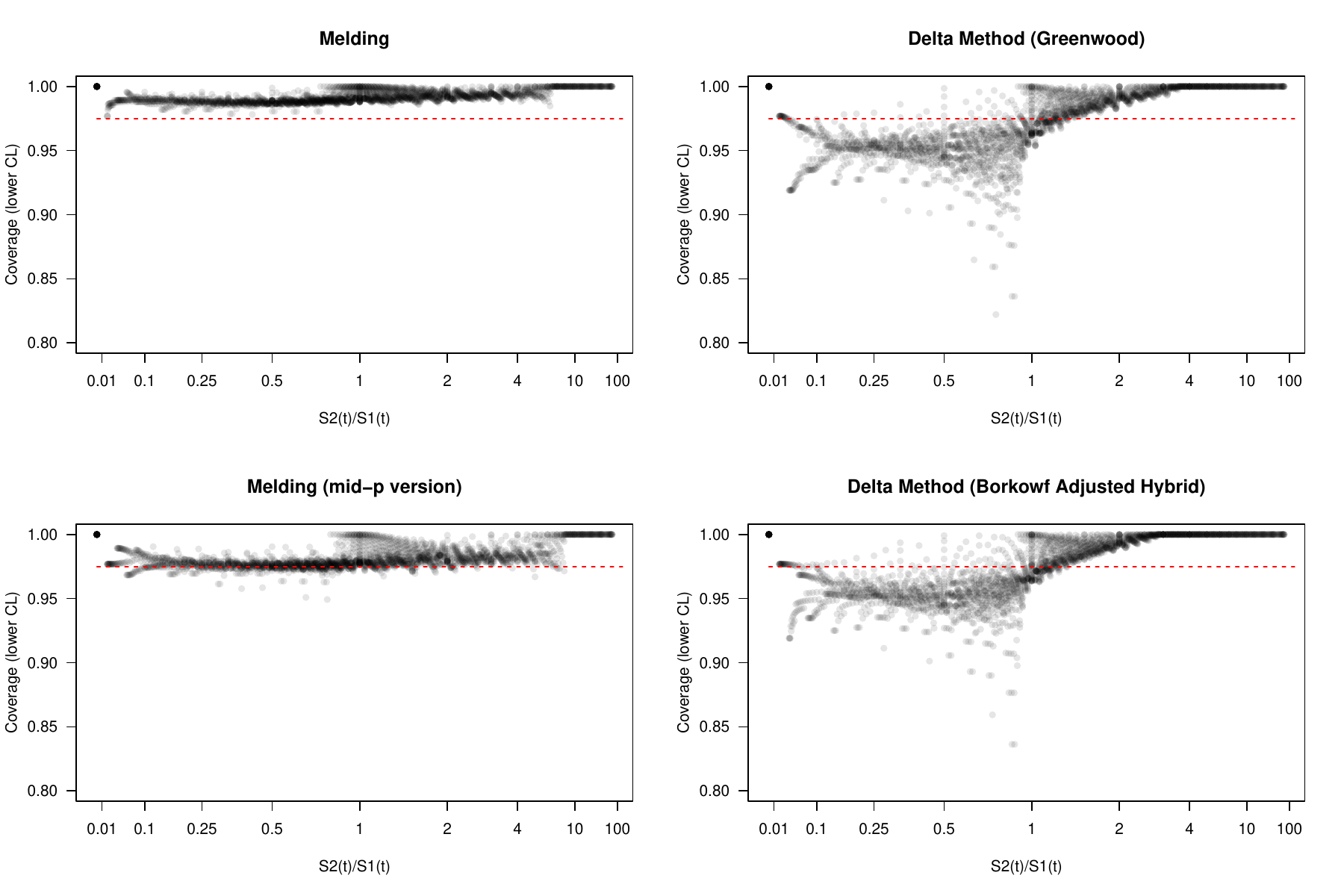}

\caption{ 
Coverage of the 97.5\%  one-sided confidence intervals of $S_2(t)/S_1(t)$ using the lower limit ($n_1=12$ and $n_2=24$).
The left panels are the melding methods (original on top, and mid-p version on the bottom).
 The right panels are the delta methods using  the zero-one modifications
 (Greenwood [i.e., standard] on top, Borkowf adjusted hybrid  on bottom).
  Each point represents the coverage at
 one of the $51 \times 51 = 2601$ values of $S_1(t)$ and $S_2(t)$ on a grid.
 A valid confidence interval would have no points below the dotted red line at 97.5\%. Points are graphed semi-transparently,
 so that darker points represent more overlapping points.
\label{sfig-ratiolo.coverage}
}

\end{figure}

\begin{figure}
\includegraphics[width=6.0in]{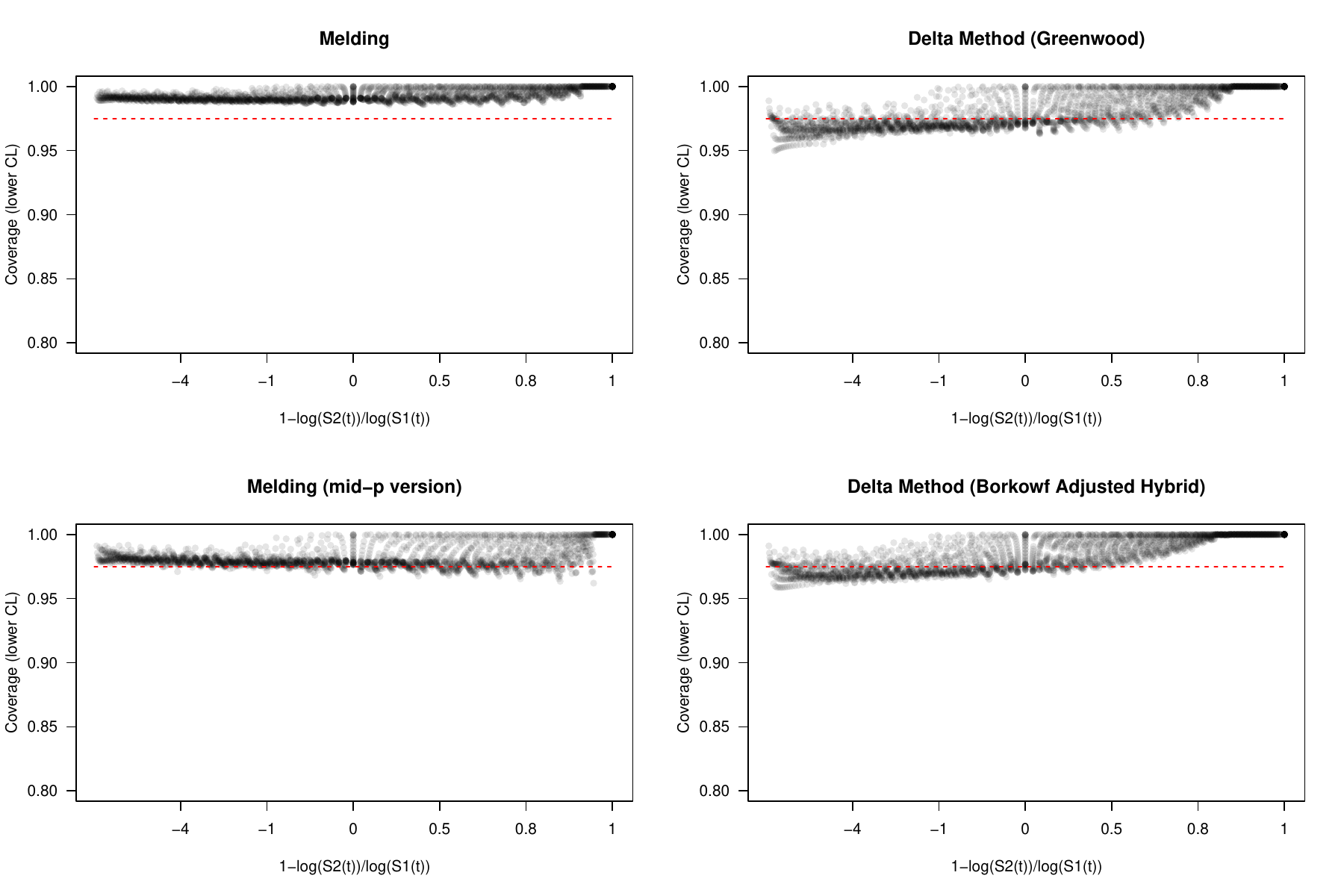}

\caption{
Coverage of the 97.5\%  one-sided confidence intervals of $1- \log(S_2(t))/\log(S_1(t))$ using the lower limit
 ($n_1=12$ and $n_2=24$).
The left panels are the melding methods (original on top, and mid-p version on the bottom).
 The right panels are the delta methods using  the zero-one modifications
 (Greenwood [i.e., standard] on top, Borkowf adjusted hybrid on bottom).
 Each point represents the coverage at
 one of the $51 \times 51 = 2601$ values of $S_1(t)$ and $S_2(t)$ on a grid.
 A valid confidence interval would have no points below the dotted red line at 97.5\%. Points are graphed semi-transparently,
 so that darker points represent more overlapping points.
\label{sfig-efflogs.coverage}
}

\end{figure}




\clearpage

\section{Figures for Simulation with Continuous Responses and Censoring}

\begin{figure}[H]
\includegraphics[width=6.0in]{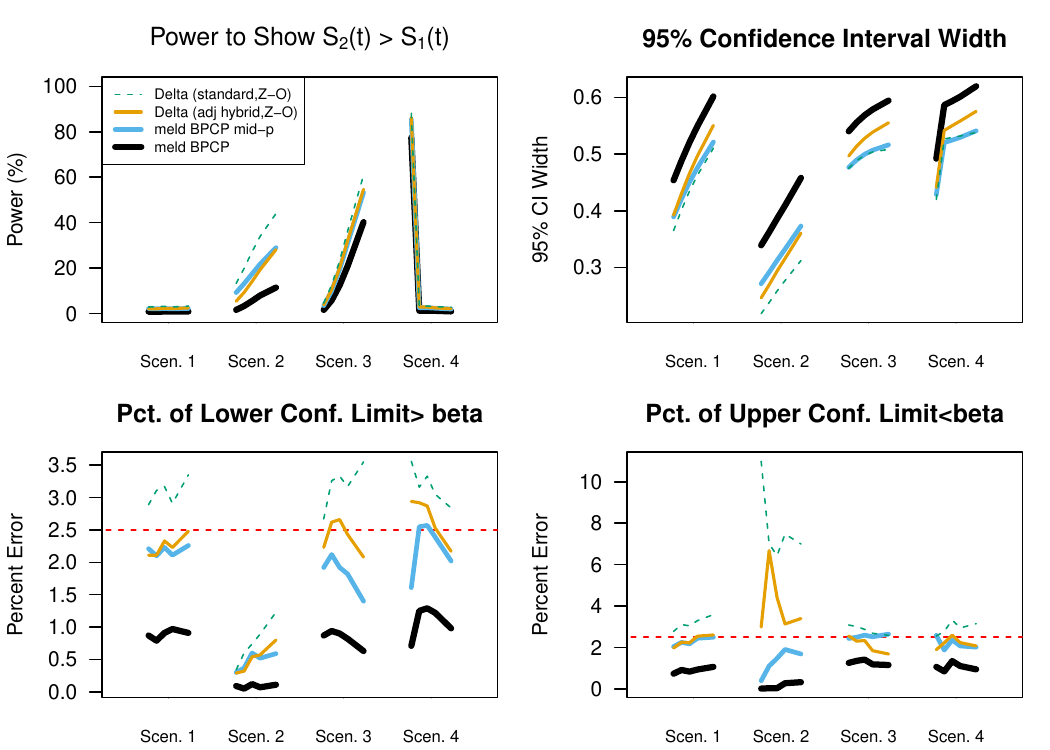}

\caption{
(Same as Figure 4) {\bf Simulation Results for $\beta=S_2(t)-S_1(t)$ with $n_1=n_2=30$ with heavy censoring.}
In each panel the four lines connect the 5 milestone times for each of the 4 methods.
 Upper left panel is the power to show $S_2(t) >  S_1(t)$ at the one-sided 2.5\% level.
The upper right panel is the 95\% central confidence interval width (which is the intersection of the two 
one-sided 97.5\% intervals).
The lower panels represent the one-sided error rates of the one-sided 97.5\% confidence intervals.
Scenario~1 and the last 4 times of Scenario~4 represent the null case when $S_2(t)=S_1(t)$.  The power for those null case points may be read off of the lower left panel, since in the null case the power is error in the one-sided test.
\label{sfig-RD01}
}

\end{figure}

\begin{figure}
\includegraphics[width=6.0in]{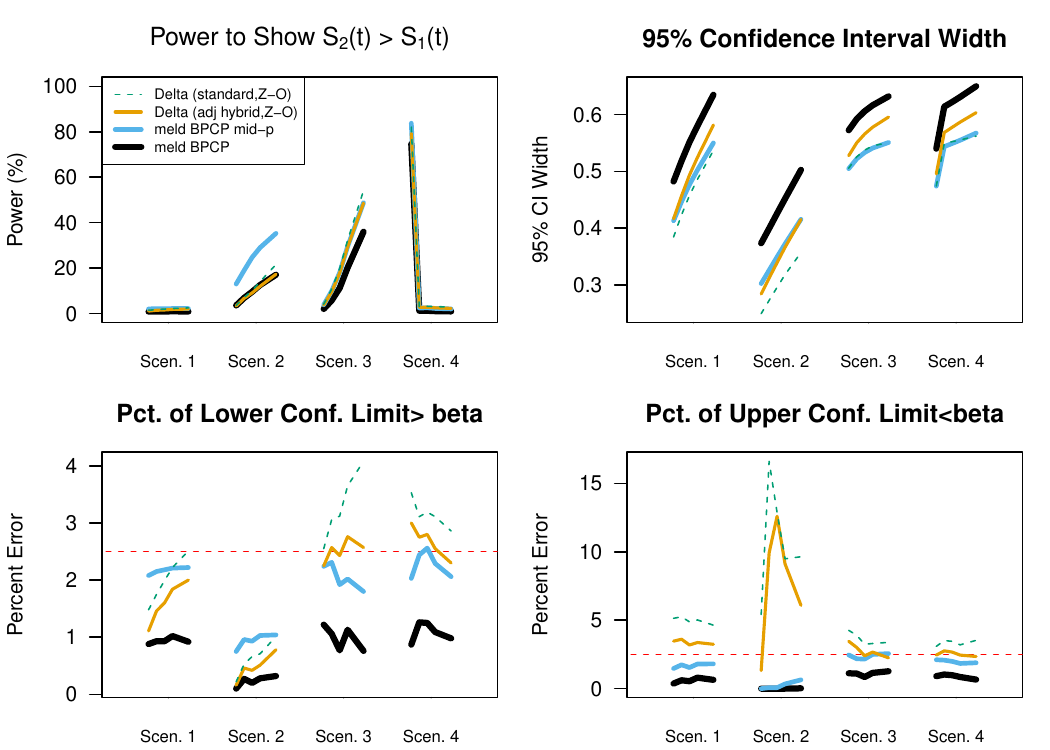}

\caption{
{\bf Simulation Results for $\beta=S_2(t)-S_1(t)$ with $n_1=20, n_2=40$ with heavy censoring.}
In each panel the four lines connect the 5 milestone times for each of the 4 methods.
 Upper left panel is the power to show $S_2(t) >  S_1(t)$ at the one-sided 2.5\% level.
The upper right panel is the 95\% central confidence interval width (which is the intersection of the two 
one-sided 97.5\% intervals).
The lower panels represent the one-sided error rates of the one-sided 97.5\% confidence intervals.
Scenario~1 and the last 4 times of Scenario~4 represent the null case when $S_2(t)=S_1(t)$.  The power for those null case points may be read off of the lower left panel, since in the null case the power is error in the one-sided test.
\label{sfig-RD02}
}

\end{figure}

\begin{figure}
\includegraphics[width=6.0in]{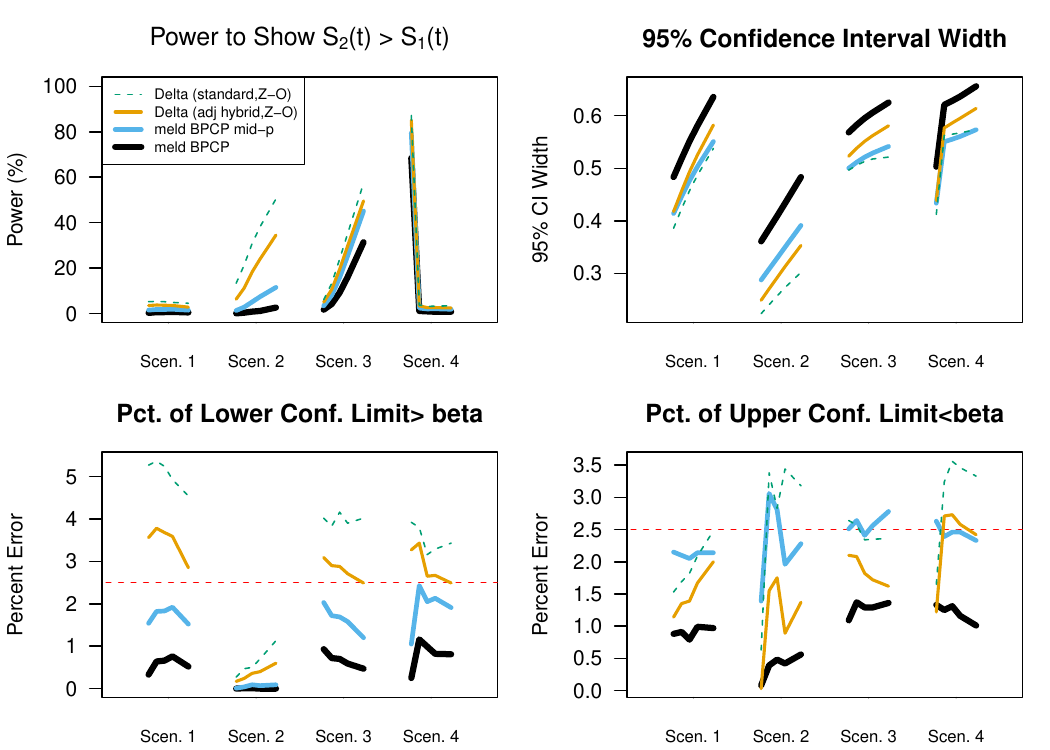}

\caption{
{\bf Simulation Results for $\beta=S_2(t)-S_1(t)$ with $n_1=40, n_2=20$ with heavy censoring.}
In each panel the four lines connect the 5 milestone times for each of the 4 methods.
 Upper left panel is the power to show $S_2(t) >  S_1(t)$ at the one-sided 2.5\% level.
The upper right panel is the 95\% central confidence interval width (which is the intersection of the two 
one-sided 97.5\% intervals).
The lower panels represent the one-sided error rates of the one-sided 97.5\% confidence intervals.
Scenario~1 and the last 4 times of Scenario~4 represent the null case when $S_2(t)=S_1(t)$.  The power for those null case points may be read off of the lower left panel, since in the null case the power is error in the one-sided test.
\label{sfig-RD03}
}

\end{figure}

\begin{figure}
\includegraphics[width=6.0in]{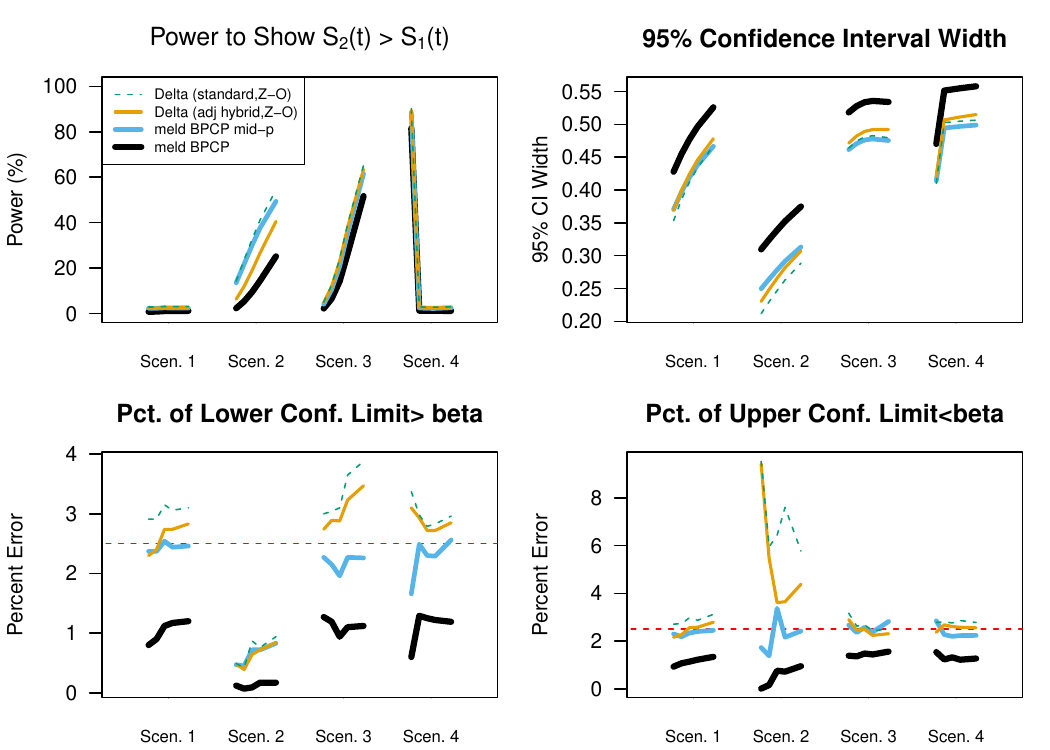}

\caption{
{\bf Simulation Results for $\beta=S_2(t)-S_1(t)$ with $n_1=n_2=30$ with light censoring.}
In each panel the four lines connect the 5 milestone times for each of the 4 methods.
 Upper left panel is the power to show $S_2(t) >  S_1(t)$ at the one-sided 2.5\% level.
The upper right panel is the 95\% central confidence interval width (which is the intersection of the two 
one-sided 97.5\% intervals).
The lower panels represent the one-sided error rates of the one-sided 97.5\% confidence intervals.
Scenario~1 and the last 4 times of Scenario~4 represent the null case when $S_2(t)=S_1(t)$.  The power for those null case points may be read off of the lower left panel, since in the null case the power is error in the one-sided test.
\label{sfig-RD04}
}

\end{figure}

\begin{figure}
\includegraphics[width=6.0in]{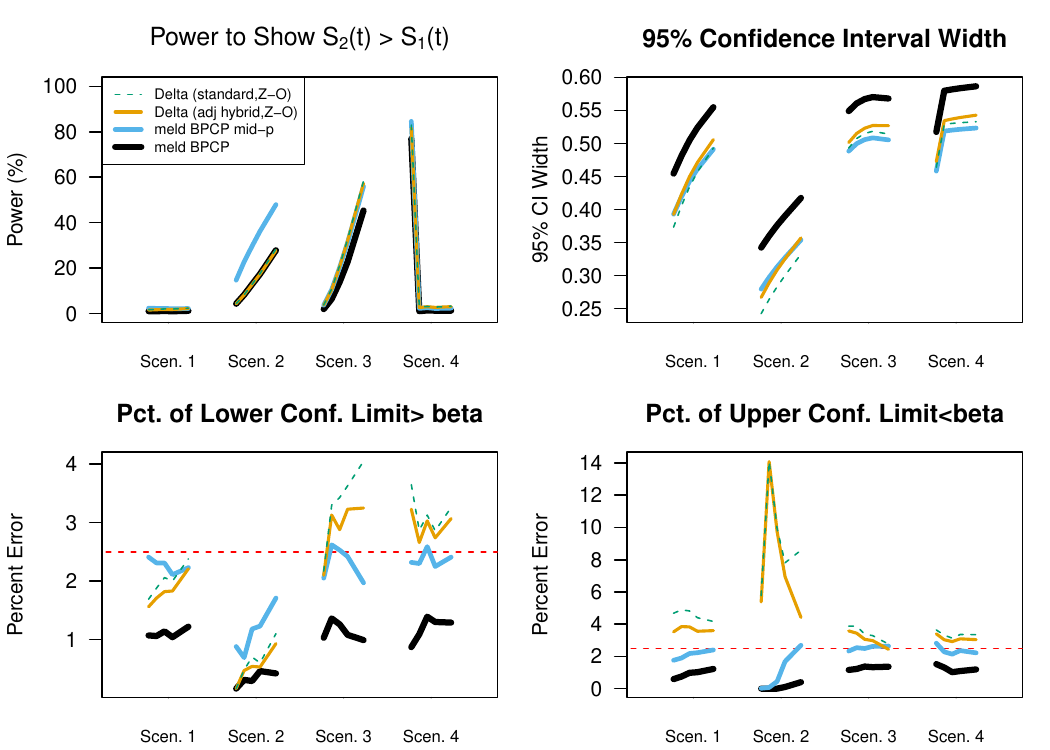}

\caption{
{\bf Simulation Results for $\beta=S_2(t)-S_1(t)$ with $n_1=20, n_2=40$ with light censoring.}
In each panel the four lines connect the 5 milestone times for each of the 4 methods.
 Upper left panel is the power to show $S_2(t) >  S_1(t)$ at the one-sided 2.5\% level.
The upper right panel is the 95\% central confidence interval width (which is the intersection of the two 
one-sided 97.5\% intervals).
The lower panels represent the one-sided error rates of the one-sided 97.5\% confidence intervals.
Scenario~1 and the last 4 times of Scenario~4 represent the null case when $S_2(t)=S_1(t)$.  The power for those null case points may be read off of the lower left panel, since in the null case the power is error in the one-sided test.
\label{sfig-RD05}
}

\end{figure}

\begin{figure}
\includegraphics[width=6.0in]{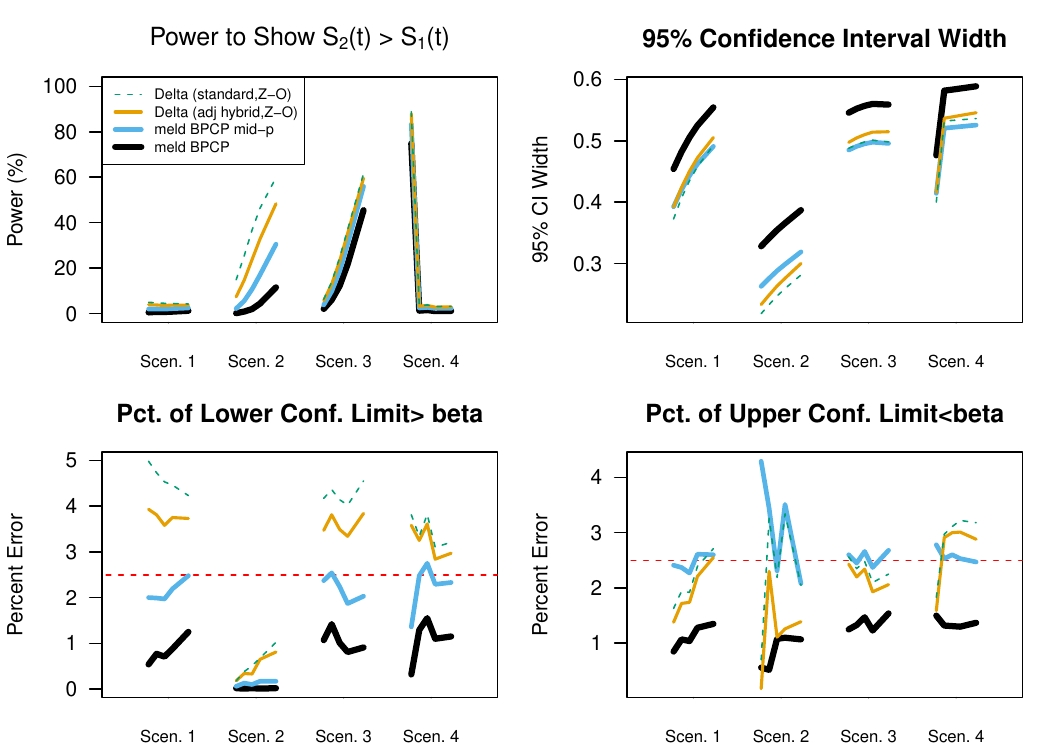}

\caption{
{\bf Simulation Results for $\beta=S_2(t)-S_1(t)$ with $n_1=40, n_2=20$ with light censoring.}
In each panel the four lines connect the 5 milestone times for each of the 4 methods.
 Upper left panel is the power to show $S_2(t) >  S_1(t)$ at the one-sided 2.5\% level.
The upper right panel is the 95\% central confidence interval width (which is the intersection of the two 
one-sided 97.5\% intervals).
The lower panels represent the one-sided error rates of the one-sided 97.5\% confidence intervals.
Scenario~1 and the last 4 times of Scenario~4 represent the null case when $S_2(t)=S_1(t)$.  The power for those null case points may be read off of the lower left panel, since in the null case the power is error in the one-sided test.
\label{sfig-RD06}
}

\end{figure}

\begin{figure}
\includegraphics[width=6.0in]{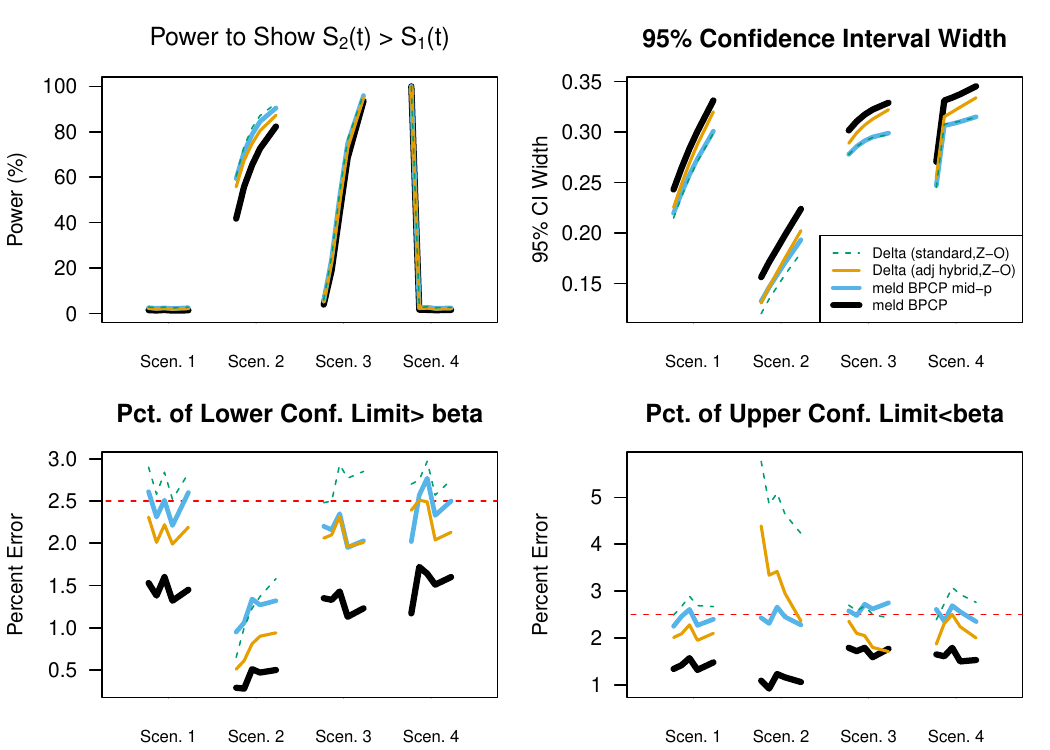}

\caption{
{\bf Simulation Results for $\beta=S_2(t)-S_1(t)$ with $n_1=n_2=90$ with heavy censoring.}
In each panel the four lines connect the 5 milestone times for each of the 4 methods.
 Upper left panel is the power to show $S_2(t) >  S_1(t)$ at the one-sided 2.5\% level.
The upper right panel is the 95\% central confidence interval width (which is the intersection of the two 
one-sided 97.5\% intervals).
The lower panels represent the one-sided error rates of the one-sided 97.5\% confidence intervals.
Scenario~1 and the last 4 times of Scenario~4 represent the null case when $S_2(t)=S_1(t)$.  The power for those null case points may be read off of the lower left panel, since in the null case the power is error in the one-sided test.
\label{sfig-RD07}
}

\end{figure}

\begin{figure}
\includegraphics[width=6.0in]{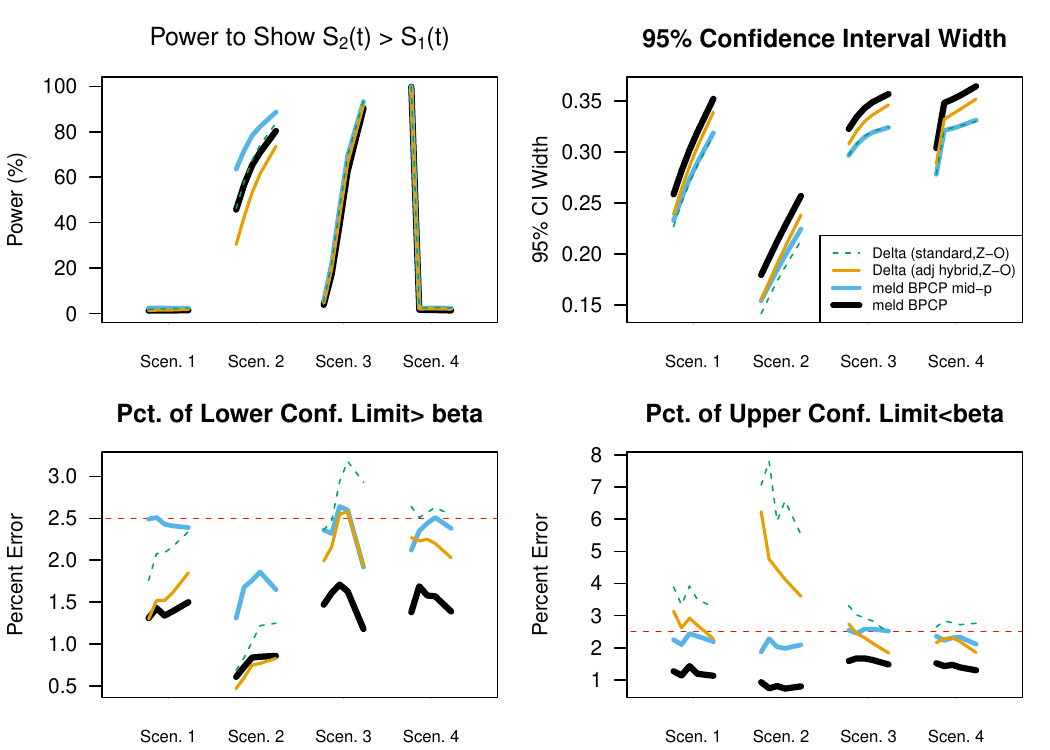}

\caption{
{\bf Simulation Results for $\beta=S_2(t)-S_1(t)$ with $n_1=60, n_2=120$ with heavy censoring.}
In each panel the four lines connect the 5 milestone times for each of the 4 methods.
 Upper left panel is the power to show $S_2(t) >  S_1(t)$ at the one-sided 2.5\% level.
The upper right panel is the 95\% central confidence interval width (which is the intersection of the two 
one-sided 97.5\% intervals).
The lower panels represent the one-sided error rates of the one-sided 97.5\% confidence intervals.
Scenario~1 and the last 4 times of Scenario~4 represent the null case when $S_2(t)=S_1(t)$.  The power for those null case points may be read off of the lower left panel, since in the null case the power is error in the one-sided test.
\label{sfig-RD08}
}

\end{figure}

\begin{figure}
\includegraphics[width=6.0in]{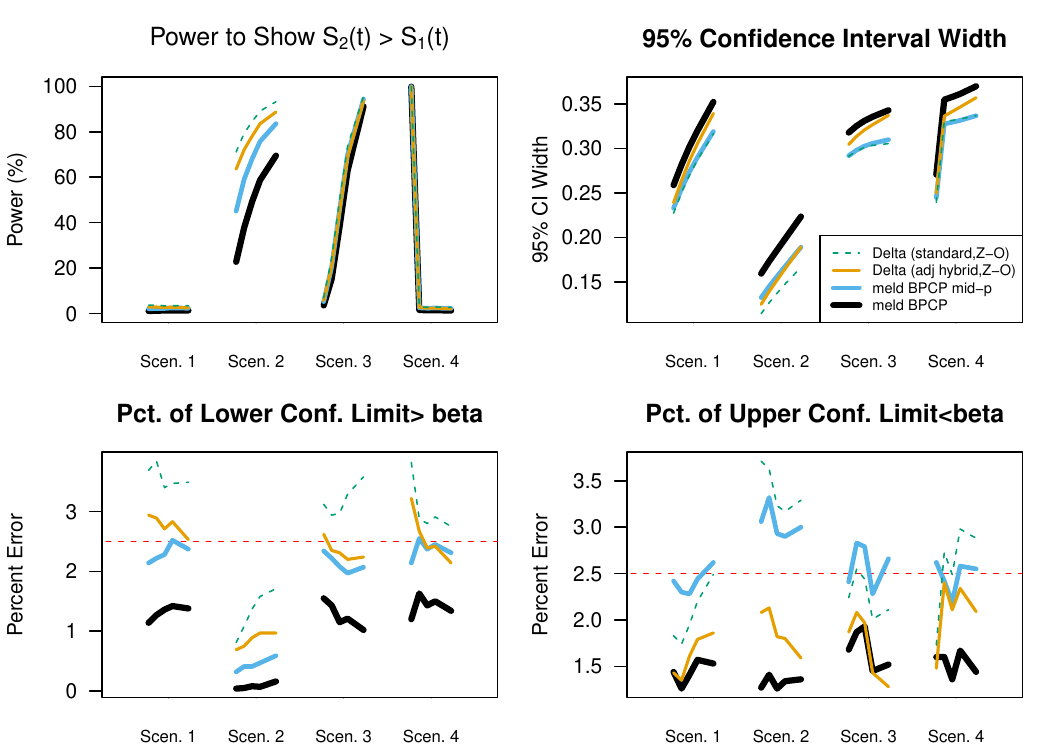}

\caption{
{\bf Simulation Results for $\beta=S_2(t)-S_1(t)$ with $n_1=120, n_2=60$ with heavy censoring.}
In each panel the four lines connect the 5 milestone times for each of the 4 methods.
 Upper left panel is the power to show $S_2(t) >  S_1(t)$ at the one-sided 2.5\% level.
The upper right panel is the 95\% central confidence interval width (which is the intersection of the two 
one-sided 97.5\% intervals).
The lower panels represent the one-sided error rates of the one-sided 97.5\% confidence intervals.
Scenario~1 and the last 4 times of Scenario~4 represent the null case when $S_2(t)=S_1(t)$.  The power for those null case points may be read off of the lower left panel, since in the null case the power is error in the one-sided test.
\label{sfig-RD09}
}

\end{figure}

\begin{figure}
\includegraphics[width=6.0in]{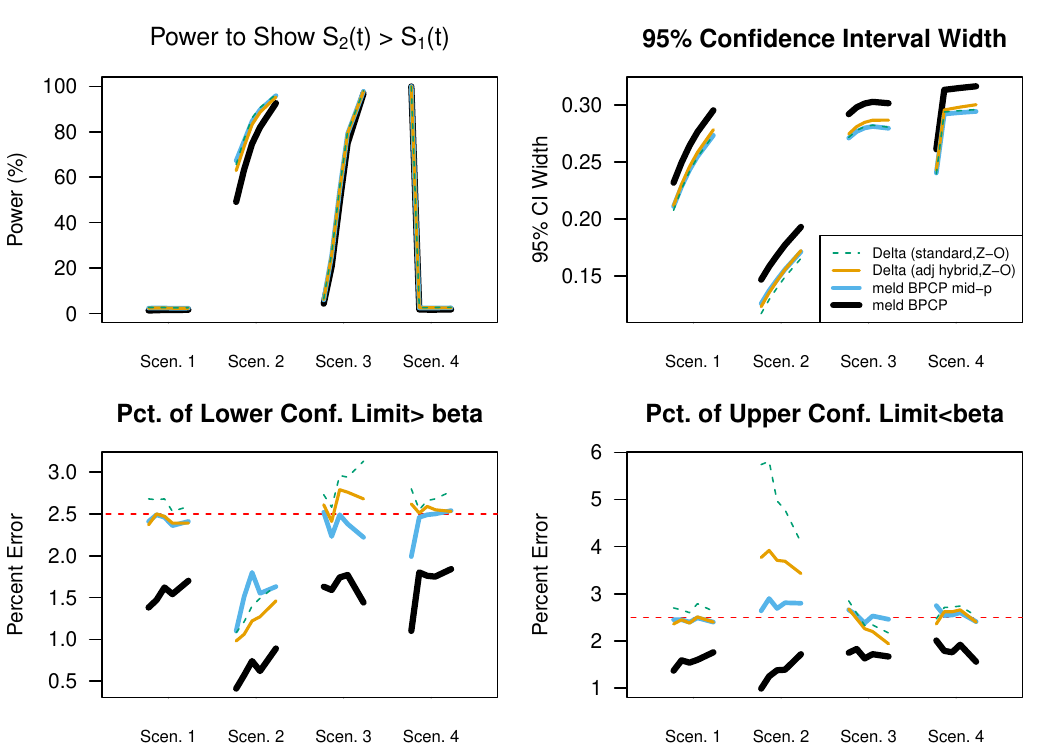}

\caption{
{\bf Simulation Results for $\beta=S_2(t)-S_1(t)$ with $n_1=n_2=90$ with light censoring.}
In each panel the four lines connect the 5 milestone times for each of the 4 methods.
 Upper left panel is the power to show $S_2(t) >  S_1(t)$ at the one-sided 2.5\% level.
The upper right panel is the 95\% central confidence interval width (which is the intersection of the two 
one-sided 97.5\% intervals).
The lower panels represent the one-sided error rates of the one-sided 97.5\% confidence intervals.
Scenario~1 and the last 4 times of Scenario~4 represent the null case when $S_2(t)=S_1(t)$.  The power for those null case points may be read off of the lower left panel, since in the null case the power is error in the one-sided test.
\label{sfig-RD10}
}

\end{figure}

\begin{figure}
\includegraphics[width=6.0in]{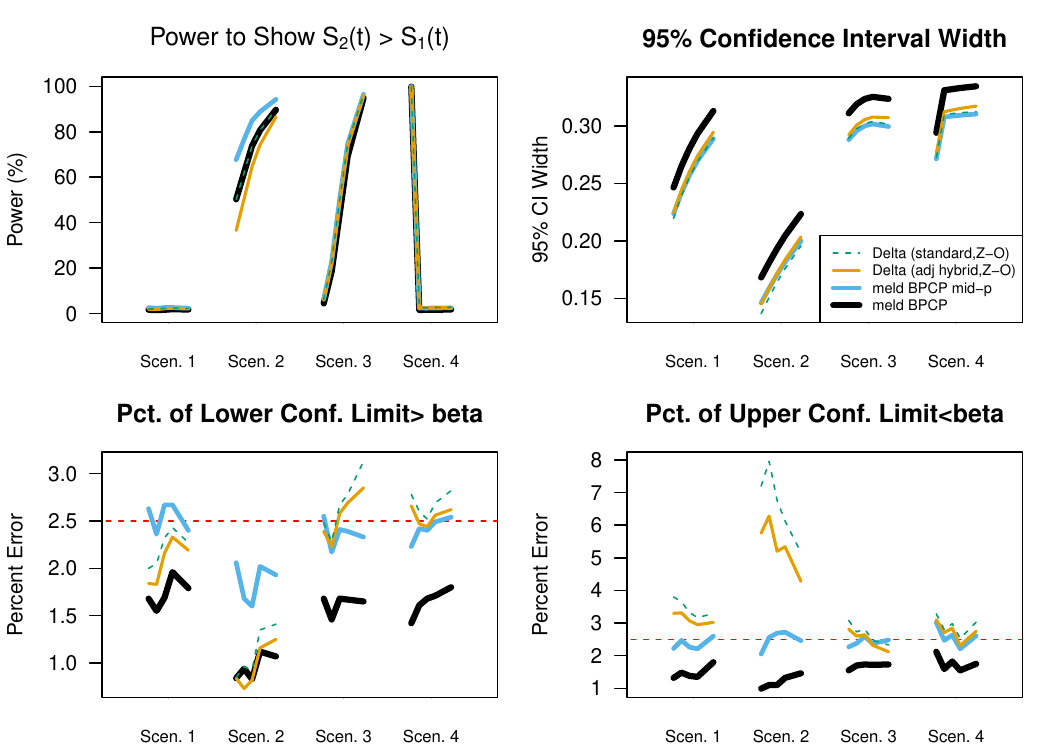}

\caption{
{\bf Simulation Results for $\beta=S_2(t)-S_1(t)$ with $n_1=60, n_2=120$ with light censoring.}
In each panel the four lines connect the 5 milestone times for each of the 4 methods.
 Upper left panel is the power to show $S_2(t) >  S_1(t)$ at the one-sided 2.5\% level.
The upper right panel is the 95\% central confidence interval width (which is the intersection of the two 
one-sided 97.5\% intervals).
The lower panels represent the one-sided error rates of the one-sided 97.5\% confidence intervals.
Scenario~1 and the last 4 times of Scenario~4 represent the null case when $S_2(t)=S_1(t)$.  The power for those null case points may be read off of the lower left panel, since in the null case the power is error in the one-sided test.
\label{sfig-RD11}
}

\end{figure}

\begin{figure}
\includegraphics[width=6.0in]{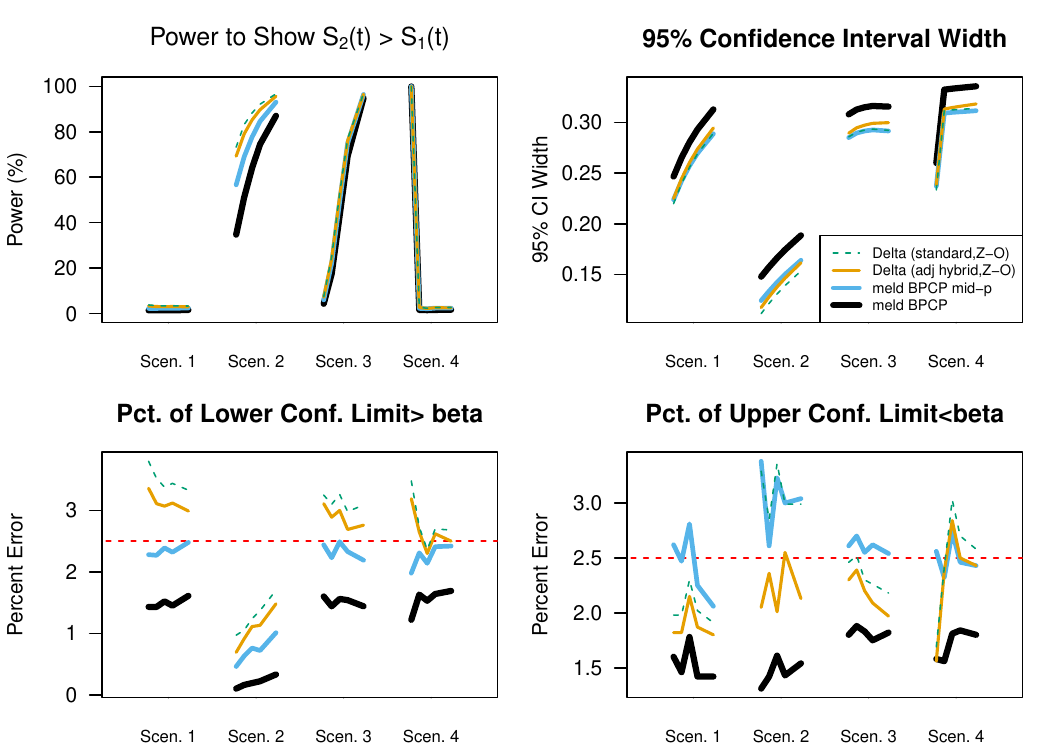}

\caption{
{\bf Simulation Results for $\beta=S_2(t)-S_1(t)$ with $n_1=120, n_2=60$ with light censoring.}
In each panel the four lines connect the 5 milestone times for each of the 4 methods.
 Upper left panel is the power to show $S_2(t) >  S_1(t)$ at the one-sided 2.5\% level.
The upper right panel is the 95\% central confidence interval width (which is the intersection of the two 
one-sided 97.5\% intervals).
The lower panels represent the one-sided error rates of the one-sided 97.5\% confidence intervals.
Scenario~1 and the last 4 times of Scenario~4 represent the null case when $S_2(t)=S_1(t)$.  The power for those null case points may be read off of the lower left panel, since in the null case the power is error in the one-sided test.
\label{sfig-RD12}
}

\end{figure}

\begin{figure}
\includegraphics[width=6.0in]{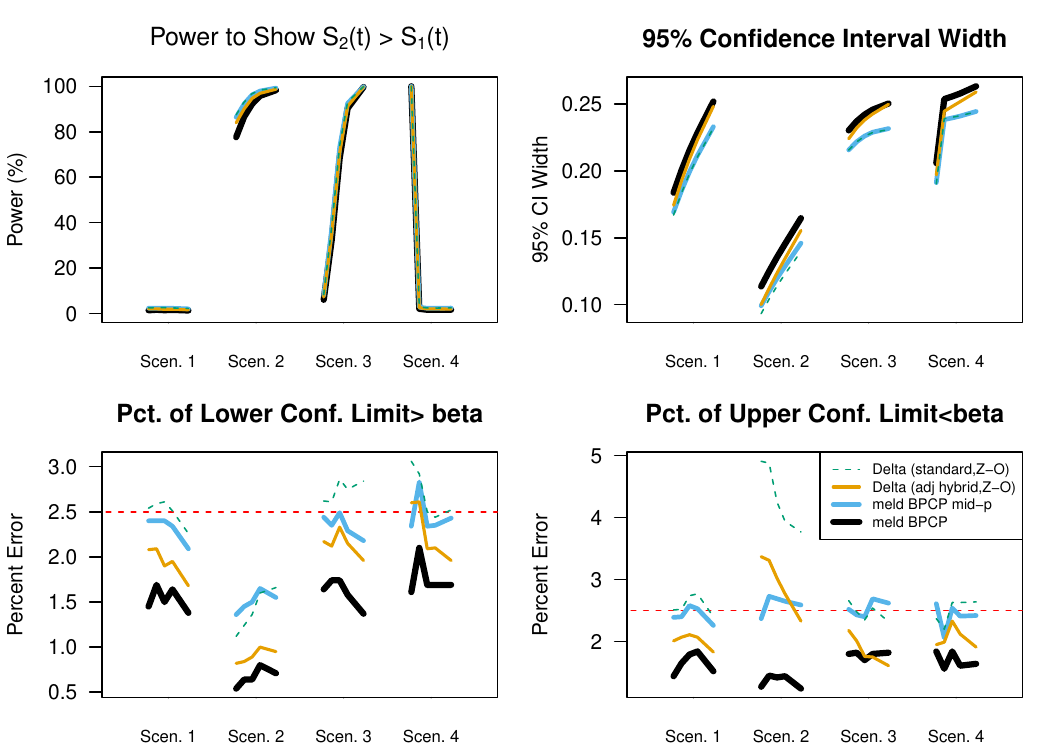}

\caption{
{\bf Simulation Results for $\beta=S_2(t)-S_1(t)$ with $n_1=n_2=150$ with heavy censoring.}
In each panel the four lines connect the 5 milestone times for each of the 4 methods.
 Upper left panel is the power to show $S_2(t) >  S_1(t)$ at the one-sided 2.5\% level.
The upper right panel is the 95\% central confidence interval width (which is the intersection of the two 
one-sided 97.5\% intervals).
The lower panels represent the one-sided error rates of the one-sided 97.5\% confidence intervals.
Scenario~1 and the last 4 times of Scenario~4 represent the null case when $S_2(t)=S_1(t)$.  The power for those null case points may be read off of the lower left panel, since in the null case the power is error in the one-sided test.
\label{sfig-RD13}
}

\end{figure}

\begin{figure}
\includegraphics[width=6.0in]{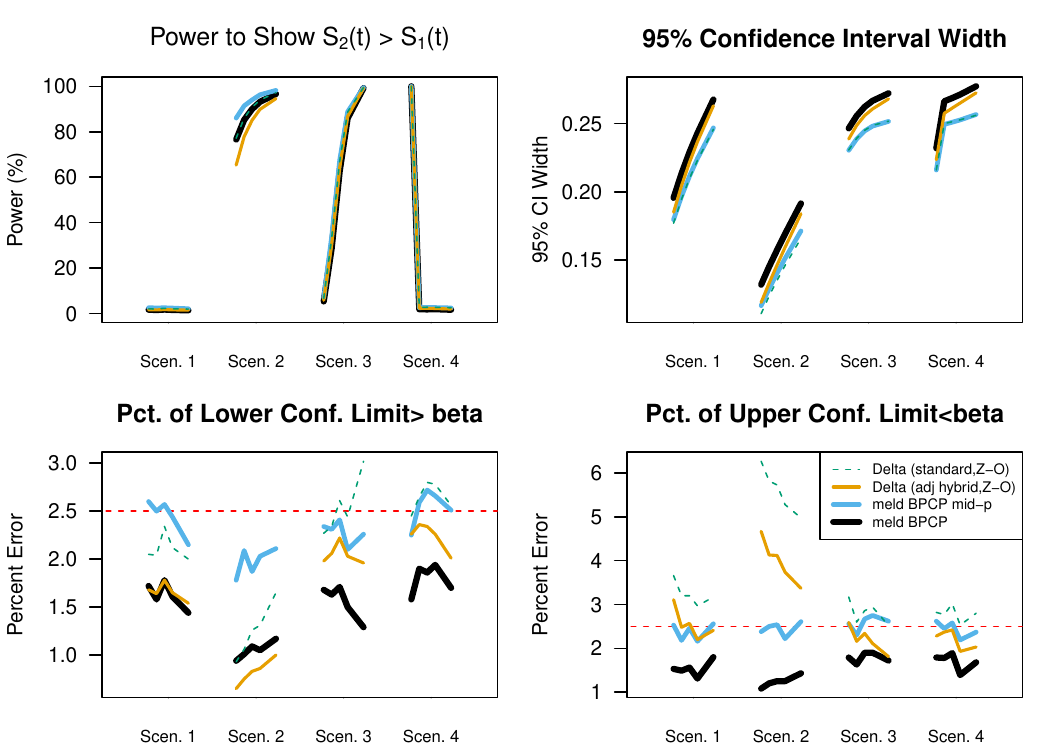}

\caption{
{\bf Simulation Results for $\beta=S_2(t)-S_1(t)$ with $n_1=100, n_2=200$ with heavy censoring.}
In each panel the four lines connect the 5 milestone times for each of the 4 methods.
 Upper left panel is the power to show $S_2(t) >  S_1(t)$ at the one-sided 2.5\% level.
The upper right panel is the 95\% central confidence interval width (which is the intersection of the two 
one-sided 97.5\% intervals).
The lower panels represent the one-sided error rates of the one-sided 97.5\% confidence intervals.
Scenario~1 and the last 4 times of Scenario~4 represent the null case when $S_2(t)=S_1(t)$.  The power for those null case points may be read off of the lower left panel, since in the null case the power is error in the one-sided test.
\label{sfig-RD14}
}

\end{figure}

\begin{figure}
\includegraphics[width=6.0in]{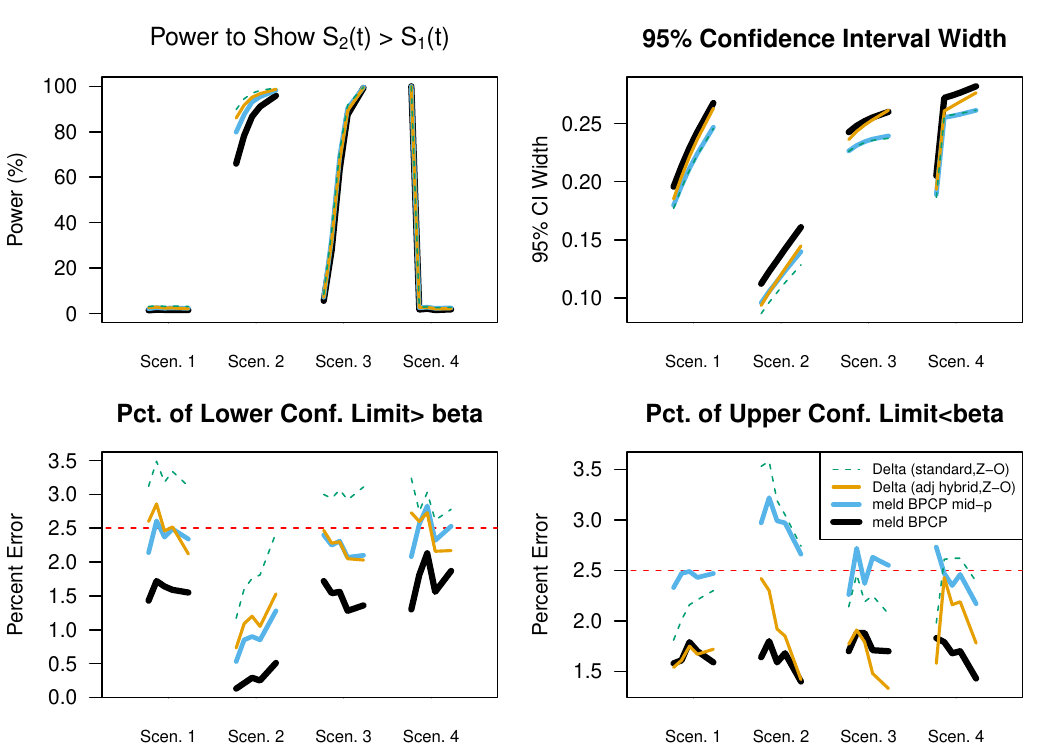}

\caption{
{\bf Simulation Results for $\beta=S_2(t)-S_1(t)$ with $n_1=200, n_2=100$ with heavy censoring.}
In each panel the four lines connect the 5 milestone times for each of the 4 methods.
 Upper left panel is the power to show $S_2(t) >  S_1(t)$ at the one-sided 2.5\% level.
The upper right panel is the 95\% central confidence interval width (which is the intersection of the two 
one-sided 97.5\% intervals).
The lower panels represent the one-sided error rates of the one-sided 97.5\% confidence intervals.
Scenario~1 and the last 4 times of Scenario~4 represent the null case when $S_2(t)=S_1(t)$.  The power for those null case points may be read off of the lower left panel, since in the null case the power is error in the one-sided test.
\label{sfig-RD15}
}

\end{figure}

\begin{figure}
\includegraphics[width=6.0in]{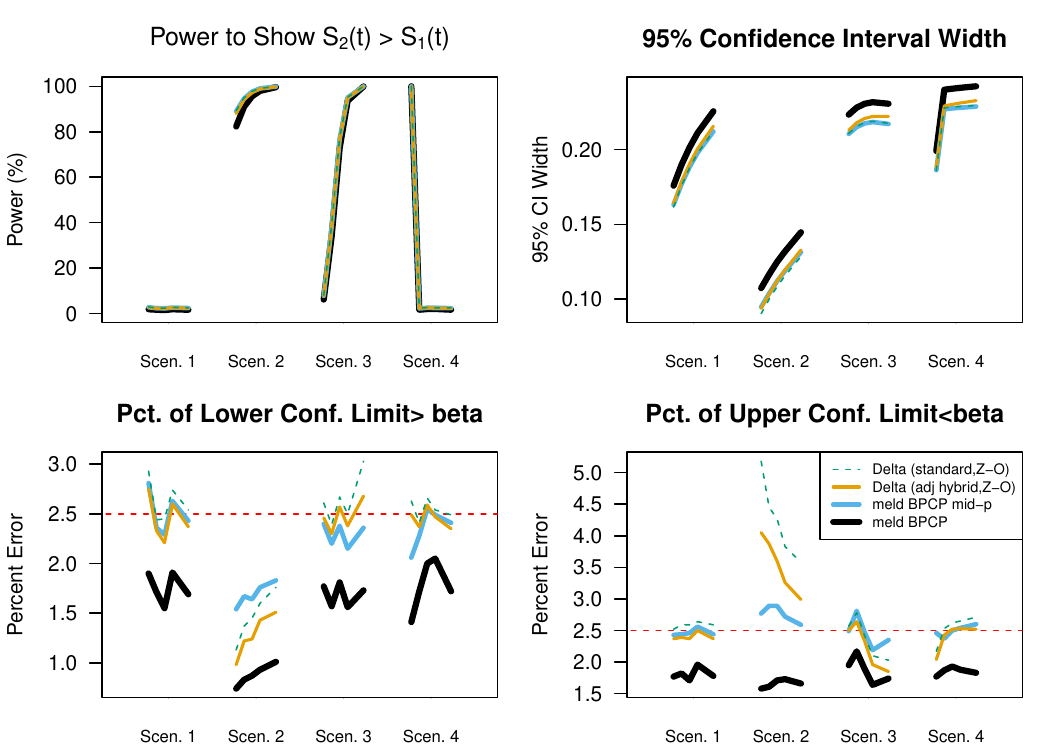}

\caption{
{\bf Simulation Results for $\beta=S_2(t)-S_1(t)$ with $n_1=n_2=150$ with light censoring.}
In each panel the four lines connect the 5 milestone times for each of the 4 methods.
 Upper left panel is the power to show $S_2(t) >  S_1(t)$ at the one-sided 2.5\% level.
The upper right panel is the 95\% central confidence interval width (which is the intersection of the two 
one-sided 97.5\% intervals).
The lower panels represent the one-sided error rates of the one-sided 97.5\% confidence intervals.
Scenario~1 and the last 4 times of Scenario~4 represent the null case when $S_2(t)=S_1(t)$.  The power for those null case points may be read off of the lower left panel, since in the null case the power is error in the one-sided test.
\label{sfig-RD16}
}

\end{figure}

\begin{figure}
\includegraphics[width=6.0in]{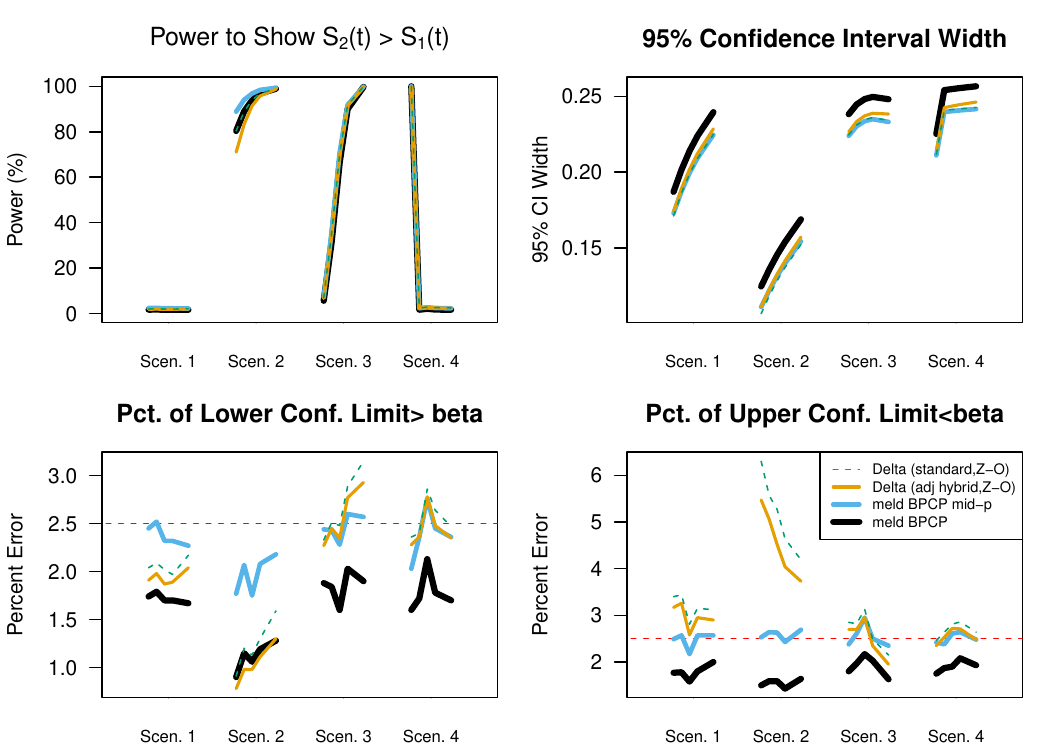}

\caption{
{\bf Simulation Results for $\beta=S_2(t)-S_1(t)$ with $n_1=100, n_2=200$ with light censoring.}
In each panel the four lines connect the 5 milestone times for each of the 4 methods.
 Upper left panel is the power to show $S_2(t) >  S_1(t)$ at the one-sided 2.5\% level.
The upper right panel is the 95\% central confidence interval width (which is the intersection of the two 
one-sided 97.5\% intervals).
The lower panels represent the one-sided error rates of the one-sided 97.5\% confidence intervals.
Scenario~1 and the last 4 times of Scenario~4 represent the null case when $S_2(t)=S_1(t)$.  The power for those null case points may be read off of the lower left panel, since in the null case the power is error in the one-sided test.
\label{sfig-RD17}
}

\end{figure}

\begin{figure}
\includegraphics[width=6.0in]{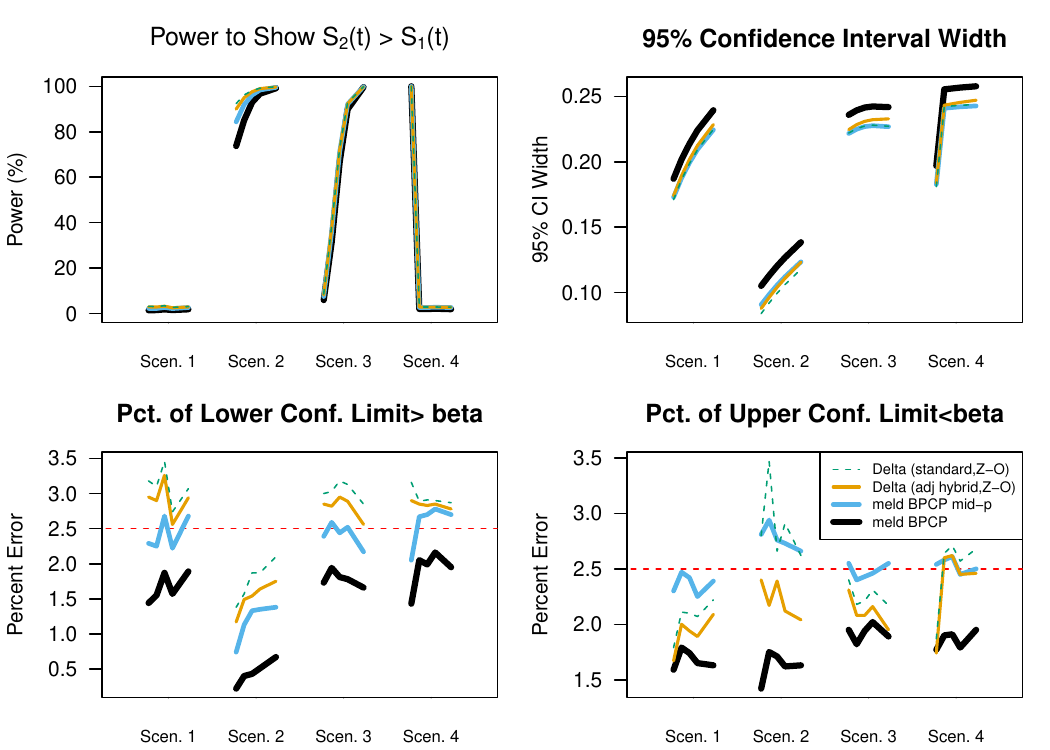}

\caption{
{\bf Simulation Results for $\beta=S_2(t)-S_1(t)$ with $n_1=200, n_2=100$ with light censoring.}
In each panel the four lines connect the 5 milestone times for each of the 4 methods.
 Upper left panel is the power to show $S_2(t) >  S_1(t)$ at the one-sided 2.5\% level.
The upper right panel is the 95\% central confidence interval width (which is the intersection of the two 
one-sided 97.5\% intervals).
The lower panels represent the one-sided error rates of the one-sided 97.5\% confidence intervals.
Scenario~1 and the last 4 times of Scenario~4 represent the null case when $S_2(t)=S_1(t)$.  The power for those null case points may be read off of the lower left panel, since in the null case the power is error in the one-sided test.
\label{sfig-RD18}
}

\end{figure}

\begin{figure}
\includegraphics[width=6.0in]{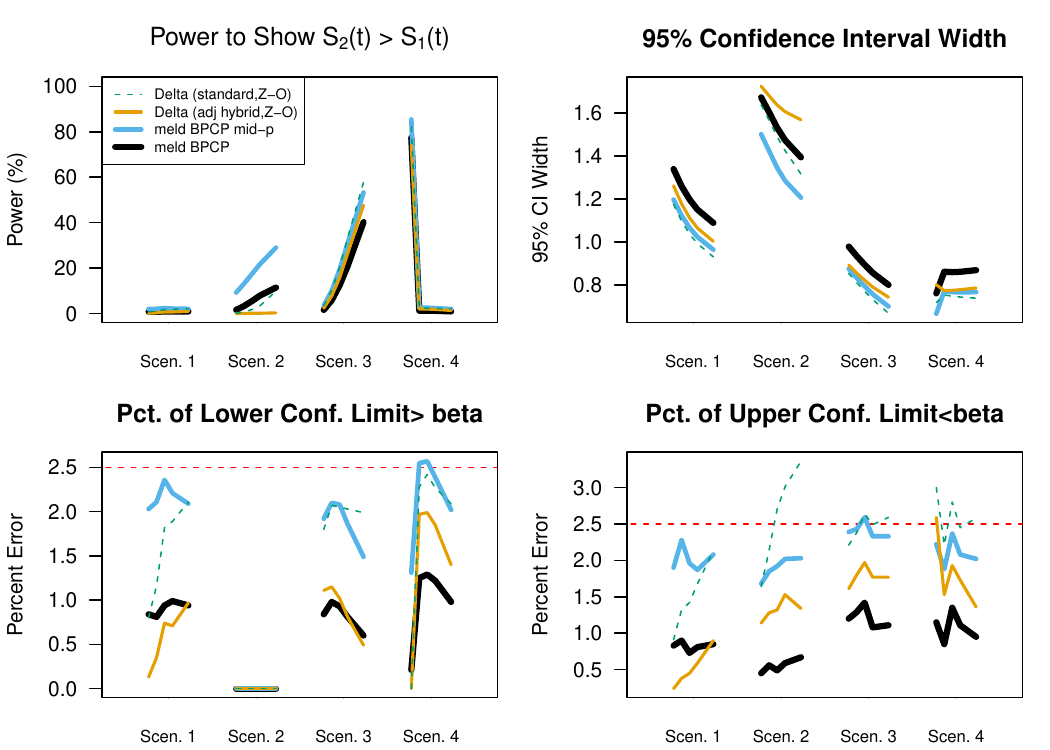}

\caption{
{\bf Simulation Results for $\beta=1-\log(S_2(t))/\log(S_1(t))$ with $n_1=n_2=30$ with heavy censoring.}
In each panel the four lines connect the 5 milestone times for each of the 4 methods.
 Upper left panel is the power to show $S_2(t) >  S_1(t)$ at the one-sided 2.5\% level.
The upper right panel is the 95\% central confidence interval width for the transformed limits (transformation is $\beta/(2-\beta)$ and allows fair comparisons with finite limits).
The lower panels represent the one-sided error rates of the one-sided 97.5\% confidence intervals.
Scenario~1 and the last 4 times of Scenario~4 represent the null case when $S_2(t)=S_1(t)$.  The power for those null case points may be read off of the lower left panel, since in the null case the power is error in the one-sided test.
\label{sfig-RE01}
}

\end{figure}

\begin{figure}
\includegraphics[width=6.0in]{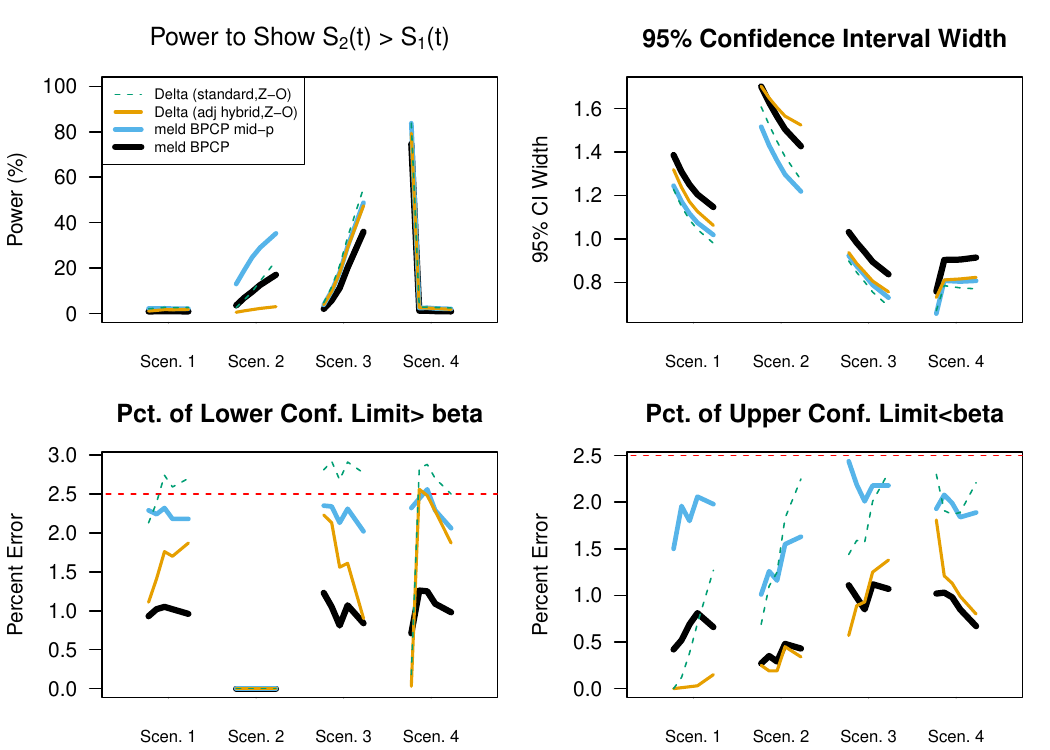}

\caption{
{\bf Simulation Results for $\beta=1-\log(S_2(t))/\log(S_1(t))$ with $n_1=20, n_2=40$ with heavy censoring.}
In each panel the four lines connect the 5 milestone times for each of the 4 methods.
 Upper left panel is the power to show $S_2(t) >  S_1(t)$ at the one-sided 2.5\% level.
The upper right panel is the 95\% central confidence interval width for the transformed limits (transformation is $\beta/(2-\beta)$ and allows fair comparisons with finite limits).
The lower panels represent the one-sided error rates of the one-sided 97.5\% confidence intervals.
Scenario~1 and the last 4 times of Scenario~4 represent the null case when $S_2(t)=S_1(t)$.  The power for those null case points may be read off of the lower left panel, since in the null case the power is error in the one-sided test.
\label{sfig-RE02}
}

\end{figure}

\begin{figure}
\includegraphics[width=6.0in]{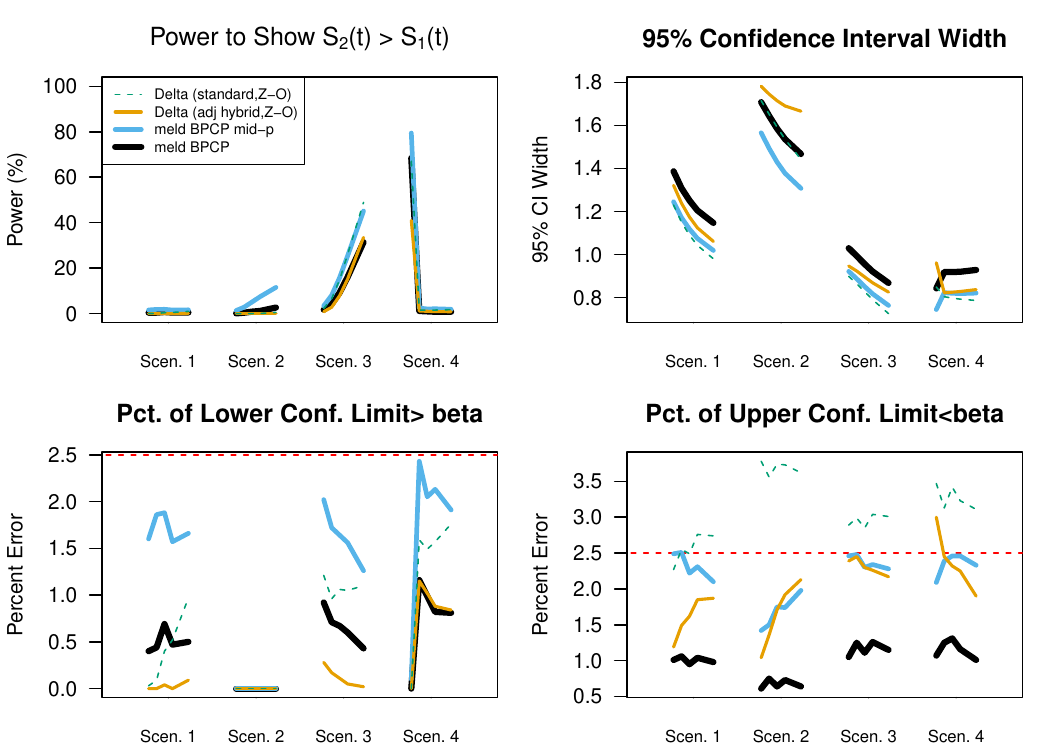}

\caption{
{\bf Simulation Results for $\beta=1-\log(S_2(t))/\log(S_1(t))$ with $n_1=40, n_2=20$ with heavy censoring.}
In each panel the four lines connect the 5 milestone times for each of the 4 methods.
 Upper left panel is the power to show $S_2(t) >  S_1(t)$ at the one-sided 2.5\% level.
The upper right panel is the 95\% central confidence interval width for the transformed limits (transformation is $\beta/(2-\beta)$ and allows fair comparisons with finite limits).
The lower panels represent the one-sided error rates of the one-sided 97.5\% confidence intervals.
Scenario~1 and the last 4 times of Scenario~4 represent the null case when $S_2(t)=S_1(t)$.  The power for those null case points may be read off of the lower left panel, since in the null case the power is error in the one-sided test.
\label{sfig-RE03}
}

\end{figure}

\begin{figure}
\includegraphics[width=6.0in]{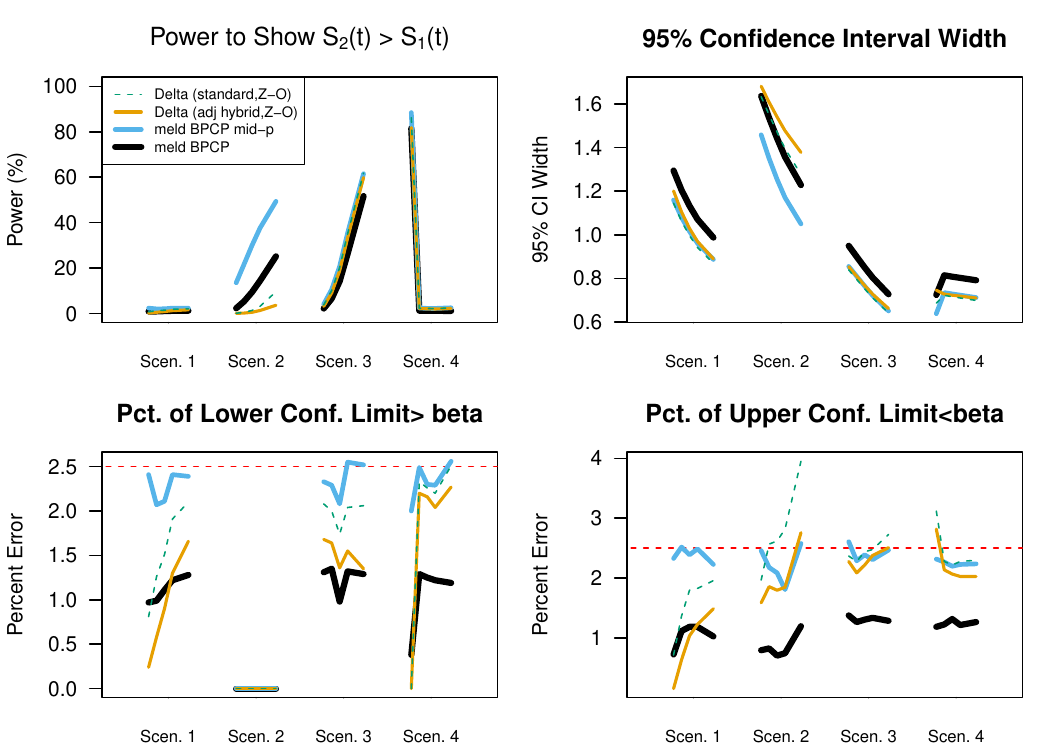}

\caption{
{\bf Simulation Results for $\beta=1-\log(S_2(t))/\log(S_1(t))$ with $n_1=n_2=30$ with light censoring.}
In each panel the four lines connect the 5 milestone times for each of the 4 methods.
 Upper left panel is the power to show $S_2(t) >  S_1(t)$ at the one-sided 2.5\% level.
The upper right panel is the 95\% central confidence interval width for the transformed limits (transformation is $\beta/(2-\beta)$ and allows fair comparisons with finite limits).
The lower panels represent the one-sided error rates of the one-sided 97.5\% confidence intervals.
Scenario~1 and the last 4 times of Scenario~4 represent the null case when $S_2(t)=S_1(t)$.  The power for those null case points may be read off of the lower left panel, since in the null case the power is error in the one-sided test.
\label{sfig-RE04}
}

\end{figure}

\begin{figure}
\includegraphics[width=6.0in]{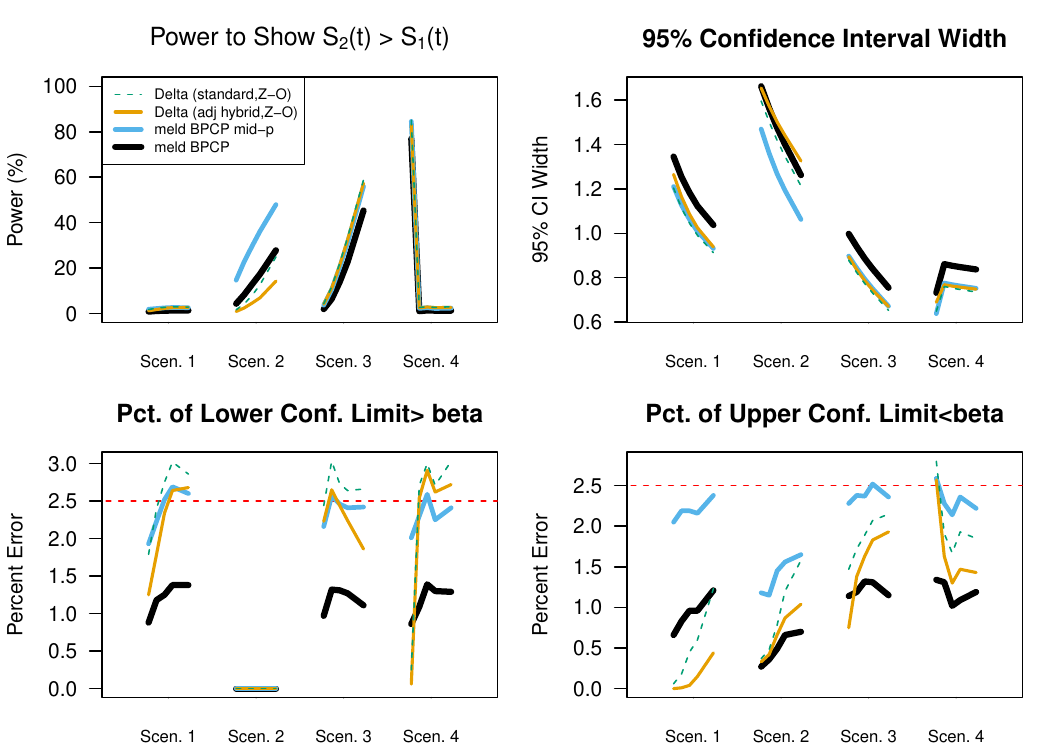}

\caption{
{\bf Simulation Results for $\beta=1-\log(S_2(t))/\log(S_1(t))$ with $n_1=20, n_2=40$ with light censoring.}
In each panel the four lines connect the 5 milestone times for each of the 4 methods.
 Upper left panel is the power to show $S_2(t) >  S_1(t)$ at the one-sided 2.5\% level.
The upper right panel is the 95\% central confidence interval width for the transformed limits (transformation is $\beta/(2-\beta)$ and allows fair comparisons with finite limits).
The lower panels represent the one-sided error rates of the one-sided 97.5\% confidence intervals.
Scenario~1 and the last 4 times of Scenario~4 represent the null case when $S_2(t)=S_1(t)$.  The power for those null case points may be read off of the lower left panel, since in the null case the power is error in the one-sided test.
\label{sfig-RE05}
}

\end{figure}

\begin{figure}
\includegraphics[width=6.0in]{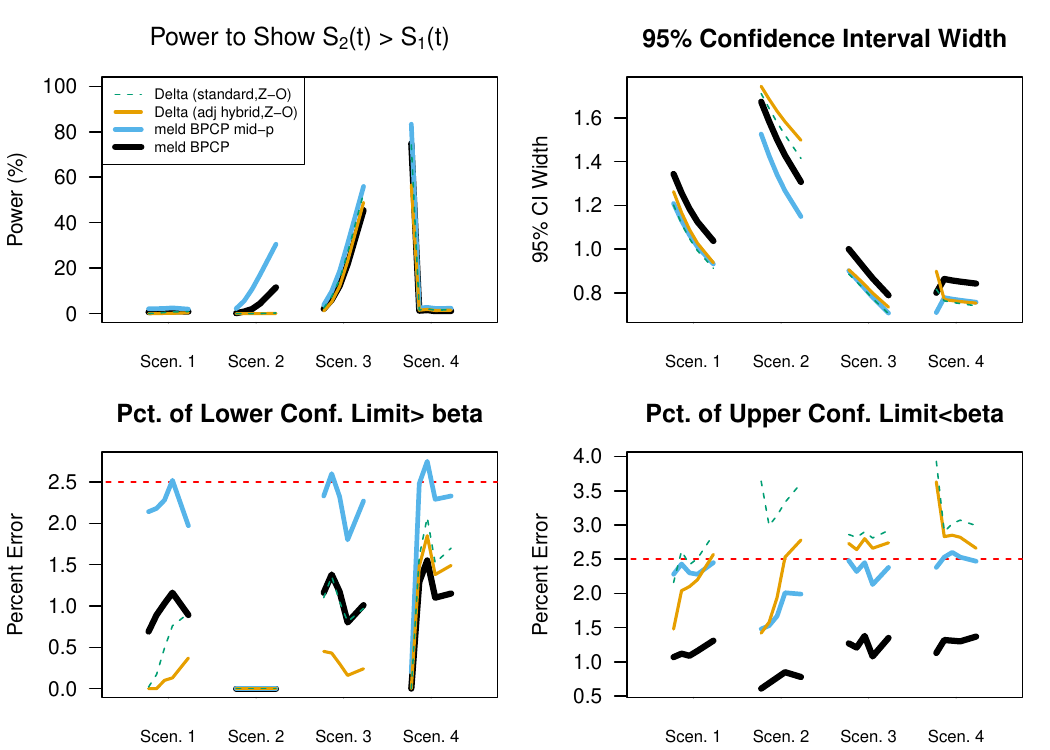}

\caption{
{\bf Simulation Results for $\beta=1-\log(S_2(t))/\log(S_1(t))$ with $n_1=40, n_2=20$ with light censoring.}
In each panel the four lines connect the 5 milestone times for each of the 4 methods.
 Upper left panel is the power to show $S_2(t) >  S_1(t)$ at the one-sided 2.5\% level.
The upper right panel is the 95\% central confidence interval width for the transformed limits (transformation is $\beta/(2-\beta)$ and allows fair comparisons with finite limits).
The lower panels represent the one-sided error rates of the one-sided 97.5\% confidence intervals.
Scenario~1 and the last 4 times of Scenario~4 represent the null case when $S_2(t)=S_1(t)$.  The power for those null case points may be read off of the lower left panel, since in the null case the power is error in the one-sided test.
\label{sfig-RE06}
}

\end{figure}

\begin{figure}
\includegraphics[width=6.0in]{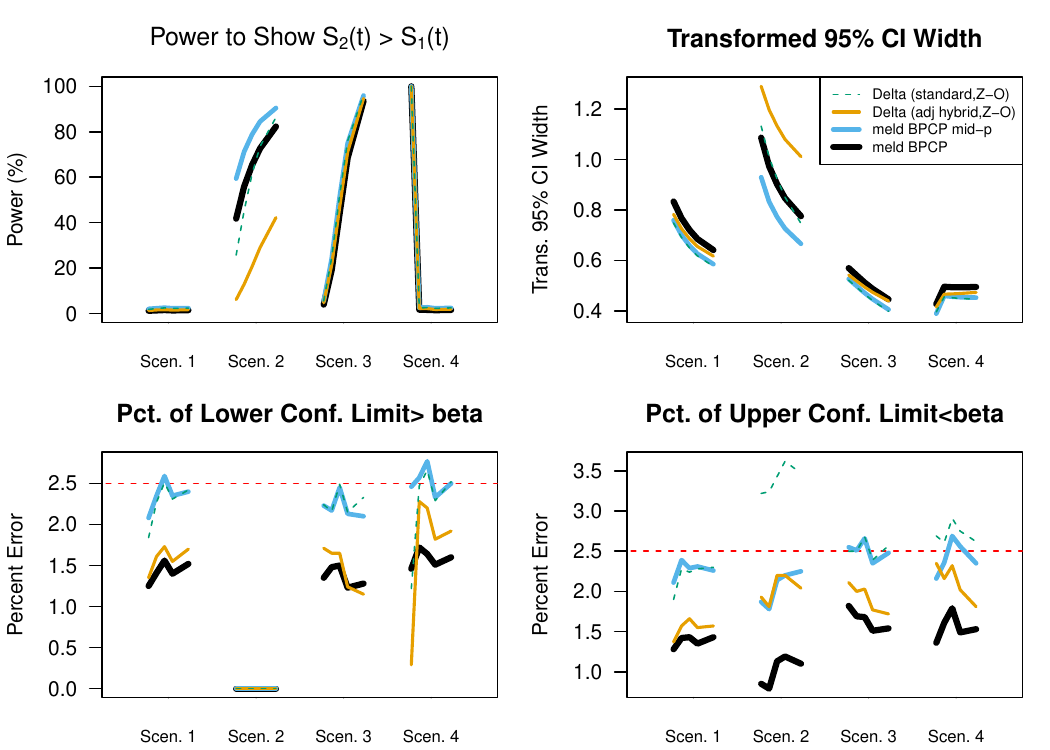}

\caption{
{\bf Simulation Results for $\beta=1-\log(S_2(t))/\log(S_1(t))$ with $n_1=n_2=90$ with heavy censoring.}
In each panel the four lines connect the 5 milestone times for each of the 4 methods.
 Upper left panel is the power to show $S_2(t) >  S_1(t)$ at the one-sided 2.5\% level.
The upper right panel is the 95\% central confidence interval width for the transformed limits (transformation is $\beta/(2-\beta)$ and allows fair comparisons with finite limits).
The lower panels represent the one-sided error rates of the one-sided 97.5\% confidence intervals.
Scenario~1 and the last 4 times of Scenario~4 represent the null case when $S_2(t)=S_1(t)$.  The power for those null case points may be read off of the lower left panel, since in the null case the power is error in the one-sided test.
\label{sfig-RE07}
}

\end{figure}

\begin{figure}
\includegraphics[width=6.0in]{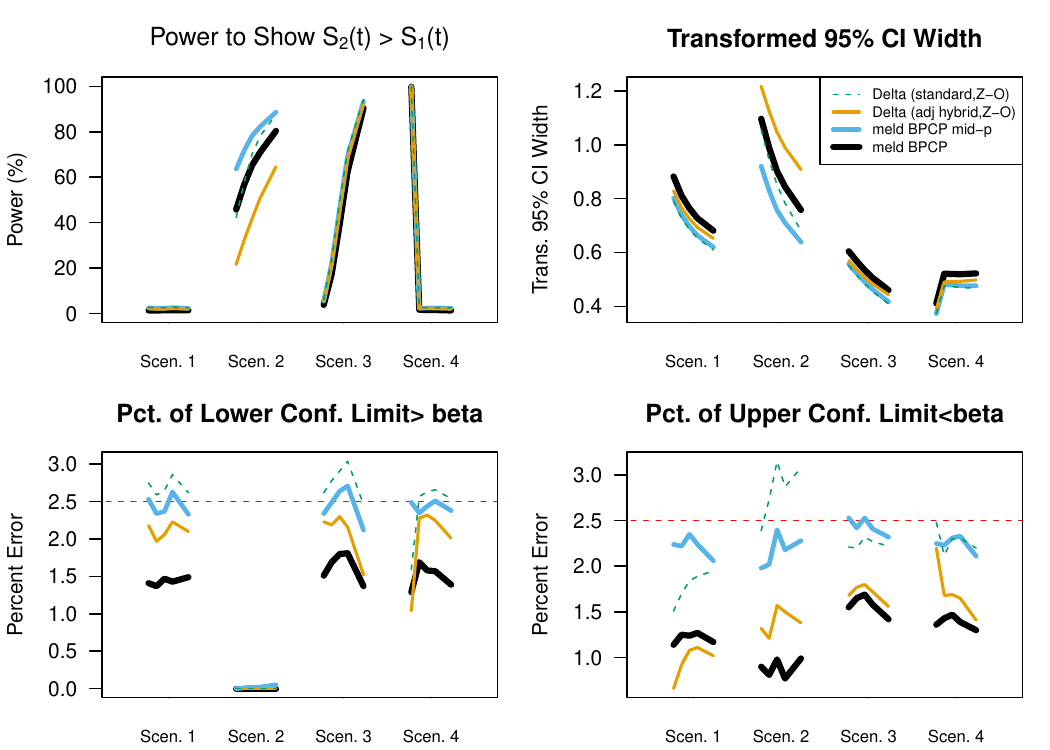}

\caption{
{\bf Simulation Results for $\beta=1-\log(S_2(t))/\log(S_1(t))$ with $n_1=60, n_2=120$ with heavy censoring.}
In each panel the four lines connect the 5 milestone times for each of the 4 methods.
 Upper left panel is the power to show $S_2(t) >  S_1(t)$ at the one-sided 2.5\% level.
The upper right panel is the 95\% central confidence interval width for the transformed limits (transformation is $\beta/(2-\beta)$ and allows fair comparisons with finite limits).
The lower panels represent the one-sided error rates of the one-sided 97.5\% confidence intervals.
Scenario~1 and the last 4 times of Scenario~4 represent the null case when $S_2(t)=S_1(t)$.  The power for those null case points may be read off of the lower left panel, since in the null case the power is error in the one-sided test.
\label{sfig-RE08}
}

\end{figure}

\begin{figure}
\includegraphics[width=6.0in]{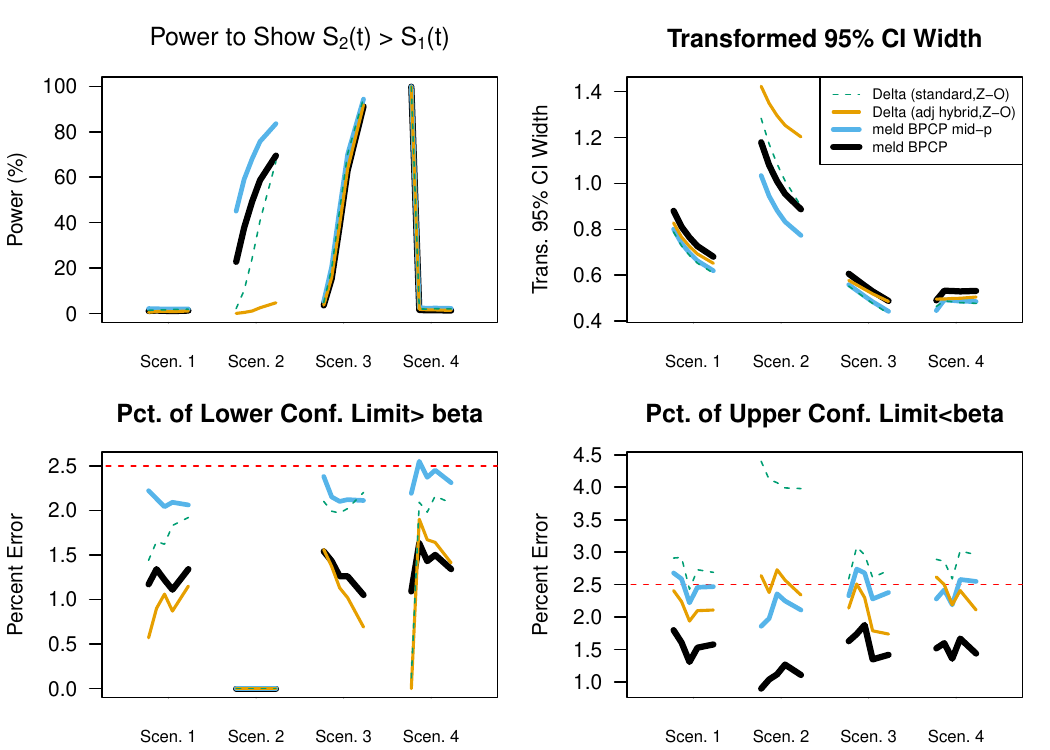}

\caption{
{\bf Simulation Results for $\beta=1-\log(S_2(t))/\log(S_1(t))$ with $n_1=120, n_2=60$ with heavy censoring.}
In each panel the four lines connect the 5 milestone times for each of the 4 methods.
 Upper left panel is the power to show $S_2(t) >  S_1(t)$ at the one-sided 2.5\% level.
The upper right panel is the 95\% central confidence interval width for the transformed limits (transformation is $\beta/(2-\beta)$ and allows fair comparisons with finite limits).
The lower panels represent the one-sided error rates of the one-sided 97.5\% confidence intervals.
Scenario~1 and the last 4 times of Scenario~4 represent the null case when $S_2(t)=S_1(t)$.  The power for those null case points may be read off of the lower left panel, since in the null case the power is error in the one-sided test.
\label{sfig-RE09}
}

\end{figure}

\begin{figure}
\includegraphics[width=6.0in]{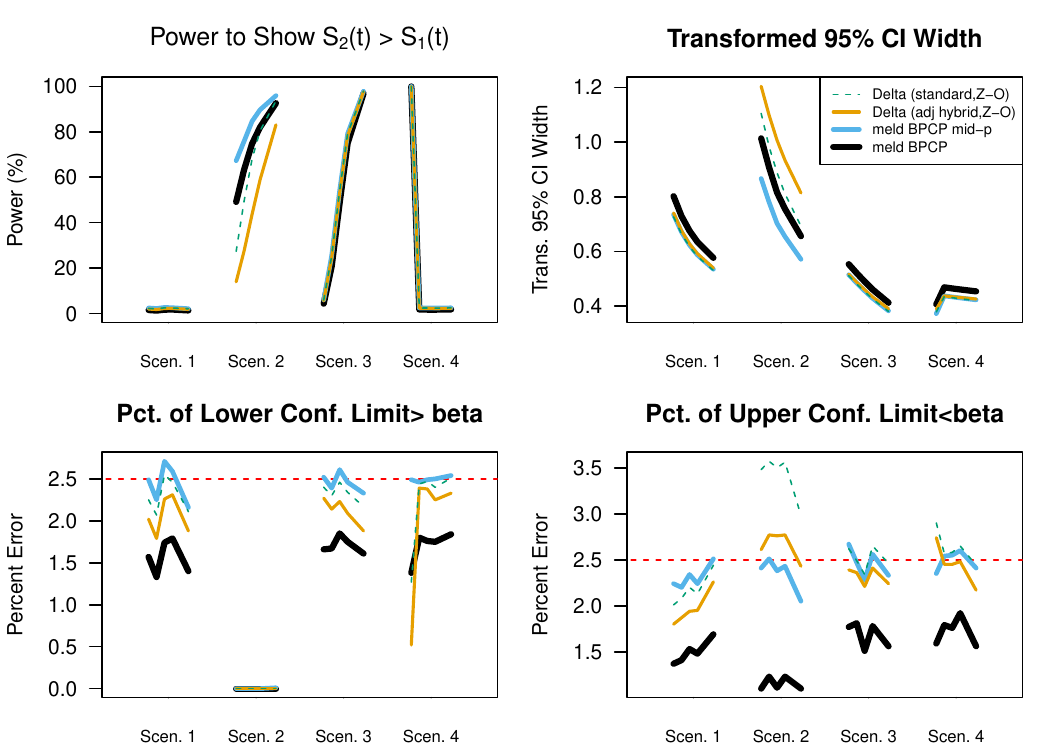}

\caption{
{\bf Simulation Results for $\beta=1-\log(S_2(t))/\log(S_1(t))$ with $n_1=n_2=90$ with light censoring.}
In each panel the four lines connect the 5 milestone times for each of the 4 methods.
 Upper left panel is the power to show $S_2(t) >  S_1(t)$ at the one-sided 2.5\% level.
The upper right panel is the 95\% central confidence interval width for the transformed limits (transformation is $\beta/(2-\beta)$ and allows fair comparisons with finite limits).
The lower panels represent the one-sided error rates of the one-sided 97.5\% confidence intervals.
Scenario~1 and the last 4 times of Scenario~4 represent the null case when $S_2(t)=S_1(t)$.  The power for those null case points may be read off of the lower left panel, since in the null case the power is error in the one-sided test.
\label{sfig-RE10}
}

\end{figure}

\begin{figure}
\includegraphics[width=6.0in]{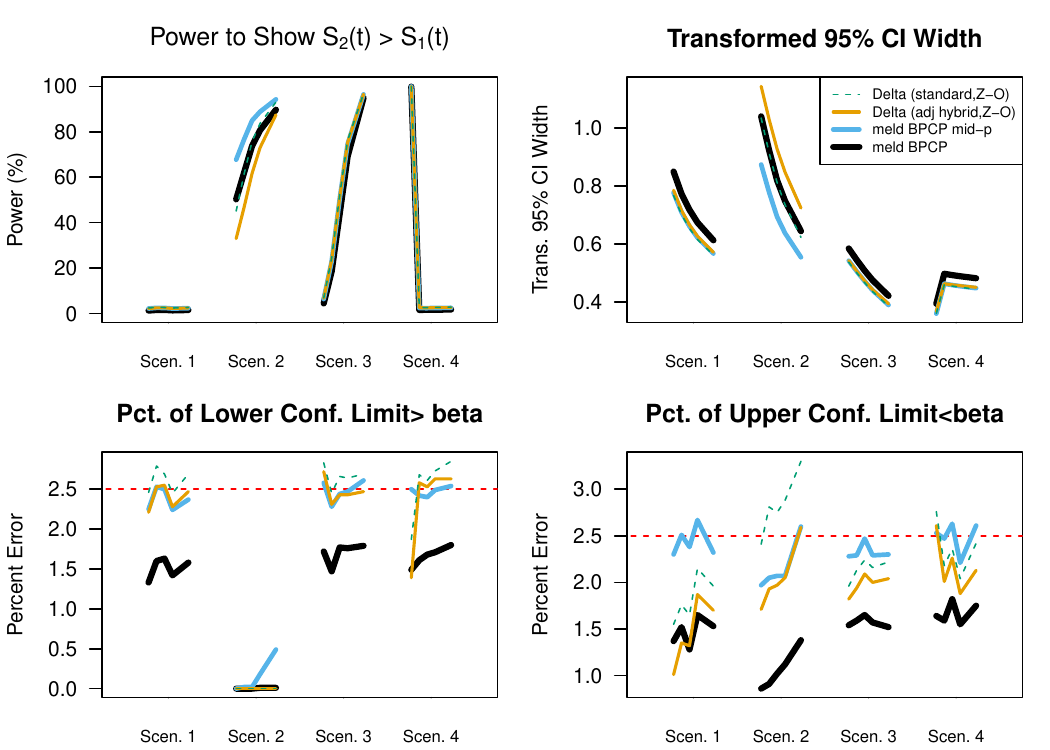}

\caption{
{\bf Simulation Results for $\beta=1-\log(S_2(t))/\log(S_1(t))$ with $n_1=60, n_2=120$ with light censoring.}
In each panel the four lines connect the 5 milestone times for each of the 4 methods.
 Upper left panel is the power to show $S_2(t) >  S_1(t)$ at the one-sided 2.5\% level.
The upper right panel is the 95\% central confidence interval width for the transformed limits (transformation is $\beta/(2-\beta)$ and allows fair comparisons with finite limits).
The lower panels represent the one-sided error rates of the one-sided 97.5\% confidence intervals.
Scenario~1 and the last 4 times of Scenario~4 represent the null case when $S_2(t)=S_1(t)$.  The power for those null case points may be read off of the lower left panel, since in the null case the power is error in the one-sided test.
\label{sfig-RE11}
}

\end{figure}

\begin{figure}
\includegraphics[width=6.0in]{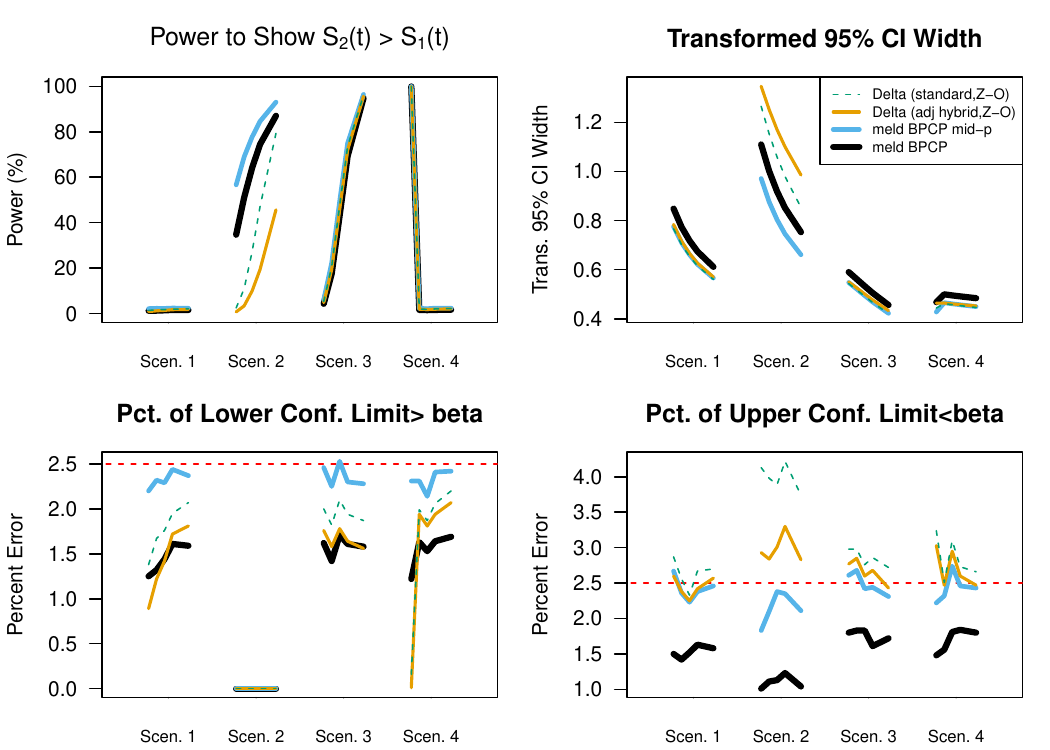}

\caption{
{\bf Simulation Results for $\beta=1-\log(S_2(t))/\log(S_1(t))$ with $n_1=120, n_2=60$ with light censoring.}
In each panel the four lines connect the 5 milestone times for each of the 4 methods.
 Upper left panel is the power to show $S_2(t) >  S_1(t)$ at the one-sided 2.5\% level.
The upper right panel is the 95\% central confidence interval width for the transformed limits (transformation is $\beta/(2-\beta)$ and allows fair comparisons with finite limits).
The lower panels represent the one-sided error rates of the one-sided 97.5\% confidence intervals.
Scenario~1 and the last 4 times of Scenario~4 represent the null case when $S_2(t)=S_1(t)$.  The power for those null case points may be read off of the lower left panel, since in the null case the power is error in the one-sided test.
\label{sfig-RE12}
}

\end{figure}

\begin{figure}
\includegraphics[width=6.0in]{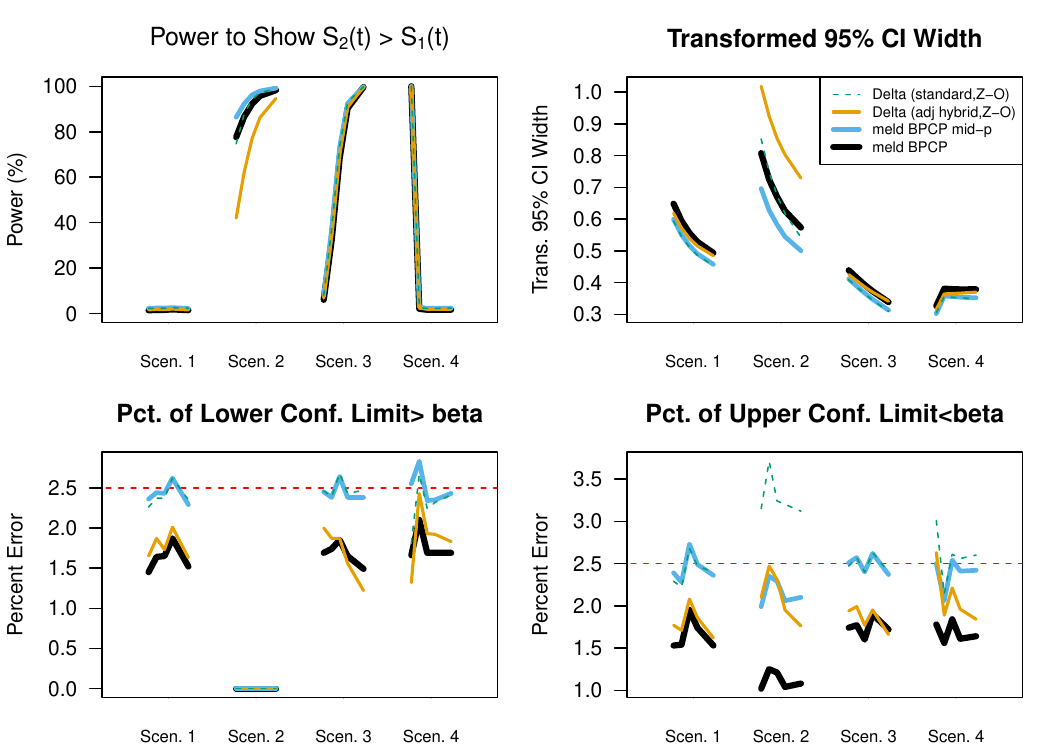}

\caption{
{\bf Simulation Results for $\beta=1-\log(S_2(t))/\log(S_1(t))$ with $n_1=n_2=150$ with heavy censoring.}
In each panel the four lines connect the 5 milestone times for each of the 4 methods.
 Upper left panel is the power to show $S_2(t) >  S_1(t)$ at the one-sided 2.5\% level.
The upper right panel is the 95\% central confidence interval width for the transformed limits (transformation is $\beta/(2-\beta)$ and allows fair comparisons with finite limits).
The lower panels represent the one-sided error rates of the one-sided 97.5\% confidence intervals.
Scenario~1 and the last 4 times of Scenario~4 represent the null case when $S_2(t)=S_1(t)$.  The power for those null case points may be read off of the lower left panel, since in the null case the power is error in the one-sided test.
\label{sfig-RE13}
}

\end{figure}

\begin{figure}
\includegraphics[width=6.0in]{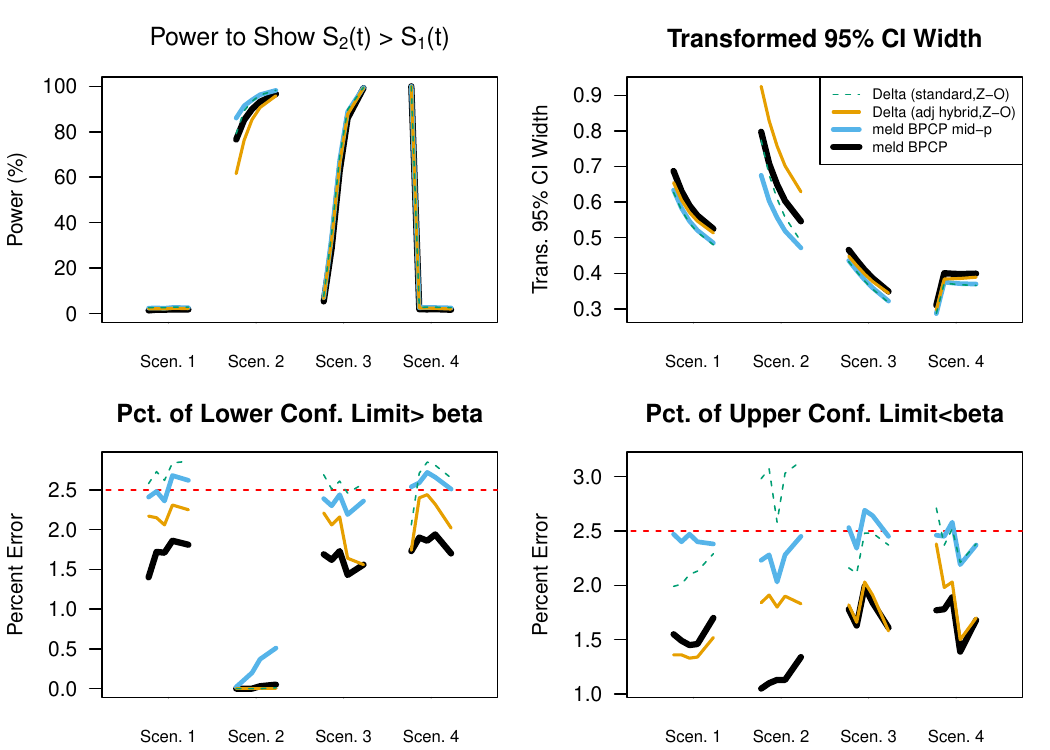}

\caption{
{\bf Simulation Results for $\beta=1-\log(S_2(t))/\log(S_1(t))$ with $n_1=100, n_2=200$ with heavy censoring.}
In each panel the four lines connect the 5 milestone times for each of the 4 methods.
 Upper left panel is the power to show $S_2(t) >  S_1(t)$ at the one-sided 2.5\% level.
The upper right panel is the 95\% central confidence interval width for the transformed limits (transformation is $\beta/(2-\beta)$ and allows fair comparisons with finite limits).
The lower panels represent the one-sided error rates of the one-sided 97.5\% confidence intervals.
Scenario~1 and the last 4 times of Scenario~4 represent the null case when $S_2(t)=S_1(t)$.  The power for those null case points may be read off of the lower left panel, since in the null case the power is error in the one-sided test.
\label{sfig-RE14}
}

\end{figure}

\begin{figure}
\includegraphics[width=6.0in]{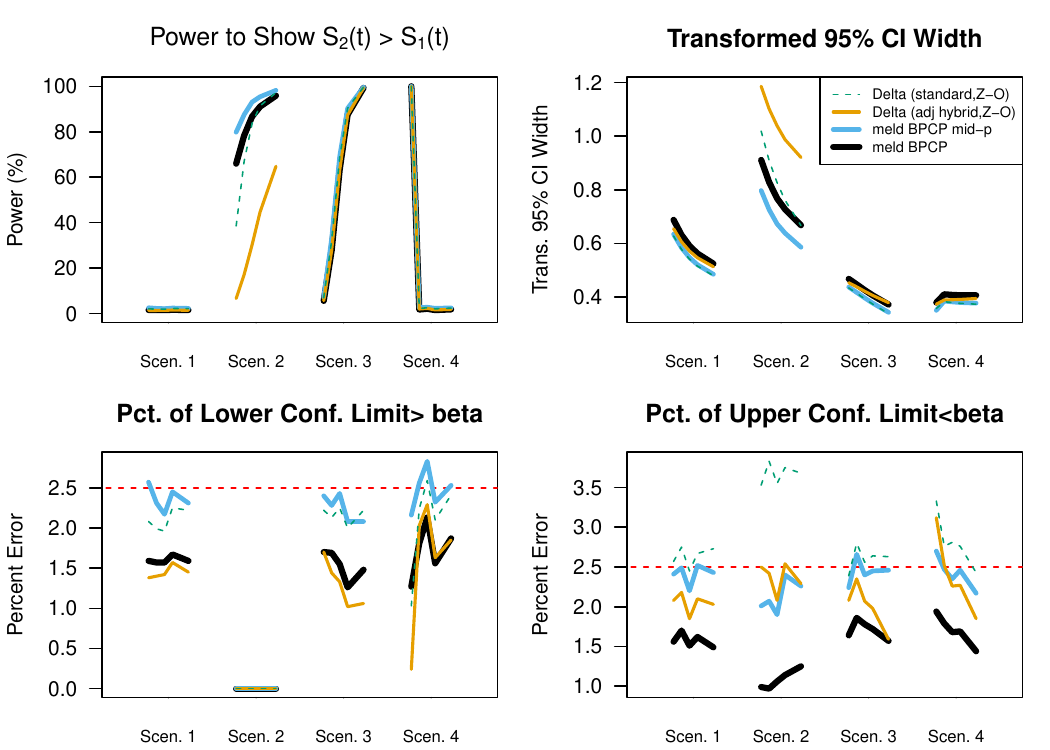}

\caption{
{\bf Simulation Results for $\beta=1-\log(S_2(t))/\log(S_1(t))$ with $n_1=200, n_2=100$ with heavy censoring.}
In each panel the four lines connect the 5 milestone times for each of the 4 methods.
 Upper left panel is the power to show $S_2(t) >  S_1(t)$ at the one-sided 2.5\% level.
The upper right panel is the 95\% central confidence interval width for the transformed limits (transformation is $\beta/(2-\beta)$ and allows fair comparisons with finite limits).
The lower panels represent the one-sided error rates of the one-sided 97.5\% confidence intervals.
Scenario~1 and the last 4 times of Scenario~4 represent the null case when $S_2(t)=S_1(t)$.  The power for those null case points may be read off of the lower left panel, since in the null case the power is error in the one-sided test.
\label{sfig-RE15}
}

\end{figure}

\begin{figure}
\includegraphics[width=6.0in]{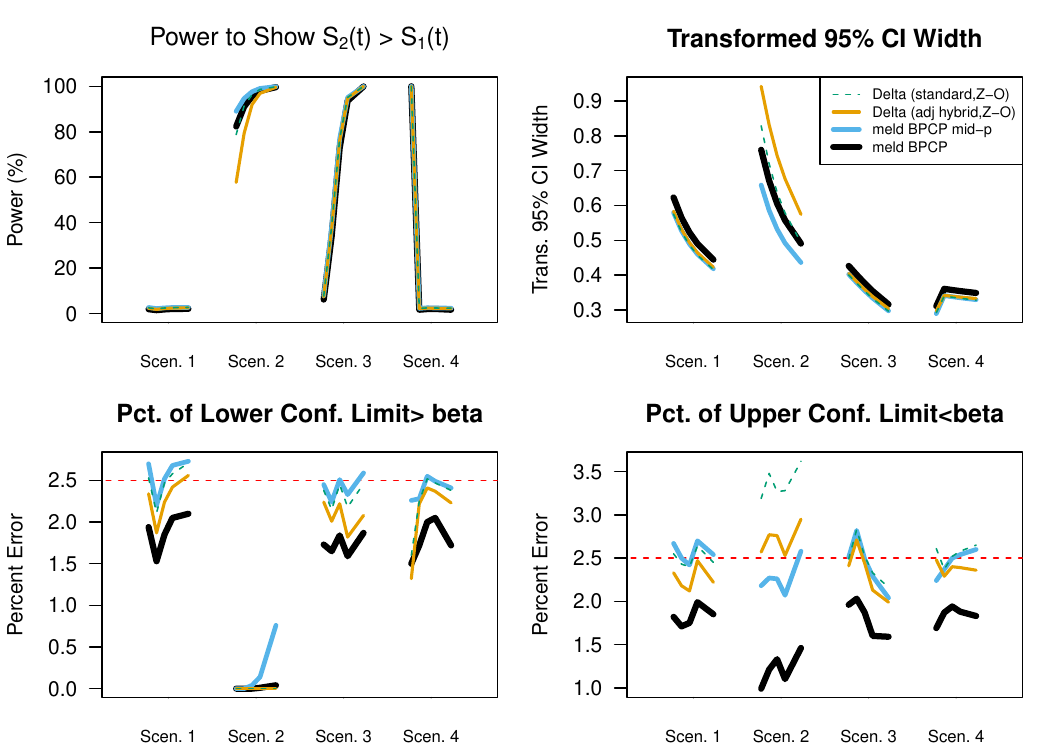}

\caption{
{\bf Simulation Results for $\beta=1-\log(S_2(t))/\log(S_1(t))$ with $n_1=n_2=150$ with light censoring.}
In each panel the four lines connect the 5 milestone times for each of the 4 methods.
 Upper left panel is the power to show $S_2(t) >  S_1(t)$ at the one-sided 2.5\% level.
The upper right panel is the 95\% central confidence interval width for the transformed limits (transformation is $\beta/(2-\beta)$ and allows fair comparisons with finite limits).
The lower panels represent the one-sided error rates of the one-sided 97.5\% confidence intervals.
Scenario~1 and the last 4 times of Scenario~4 represent the null case when $S_2(t)=S_1(t)$.  The power for those null case points may be read off of the lower left panel, since in the null case the power is error in the one-sided test.
\label{sfig-RE16}
}

\end{figure}

\begin{figure}
\includegraphics[width=6.0in]{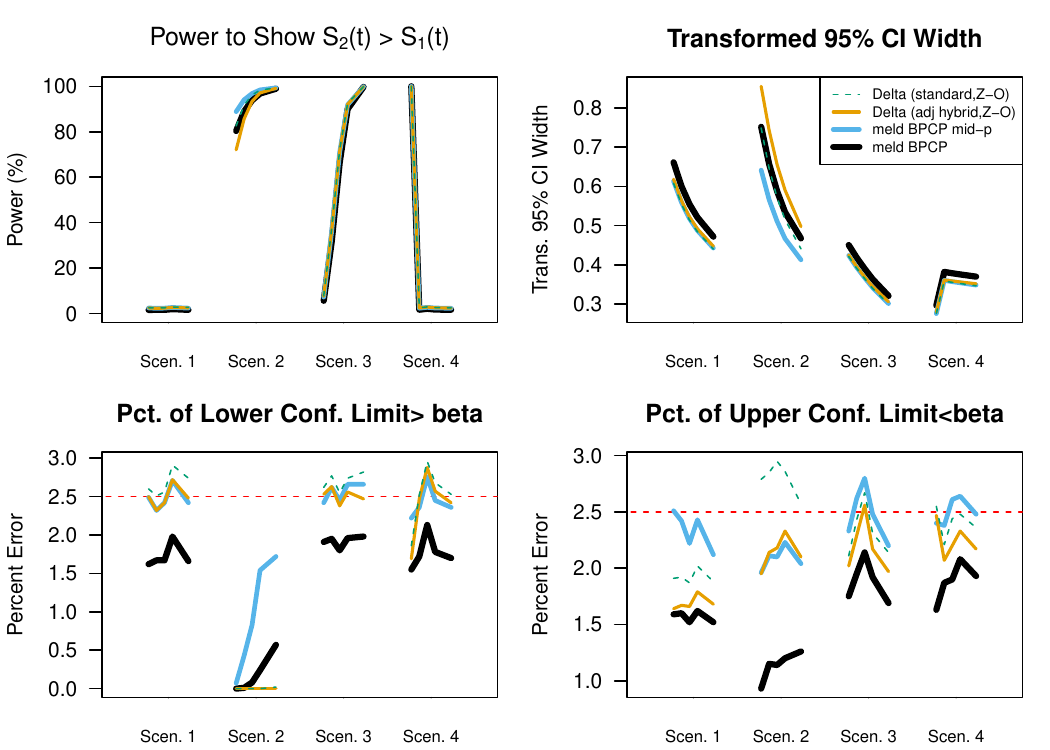}

\caption{
{\bf Simulation Results for $\beta=1-\log(S_2(t))/\log(S_1(t))$ with $n_1=100, n_2=200$ with light censoring.}
In each panel the four lines connect the 5 milestone times for each of the 4 methods.
 Upper left panel is the power to show $S_2(t) >  S_1(t)$ at the one-sided 2.5\% level.
The upper right panel is the 95\% central confidence interval width for the transformed limits (transformation is $\beta/(2-\beta)$ and allows fair comparisons with finite limits).
The lower panels represent the one-sided error rates of the one-sided 97.5\% confidence intervals.
Scenario~1 and the last 4 times of Scenario~4 represent the null case when $S_2(t)=S_1(t)$.  The power for those null case points may be read off of the lower left panel, since in the null case the power is error in the one-sided test.
\label{sfig-RE17}
}

\end{figure}

\begin{figure}
\includegraphics[width=6.0in]{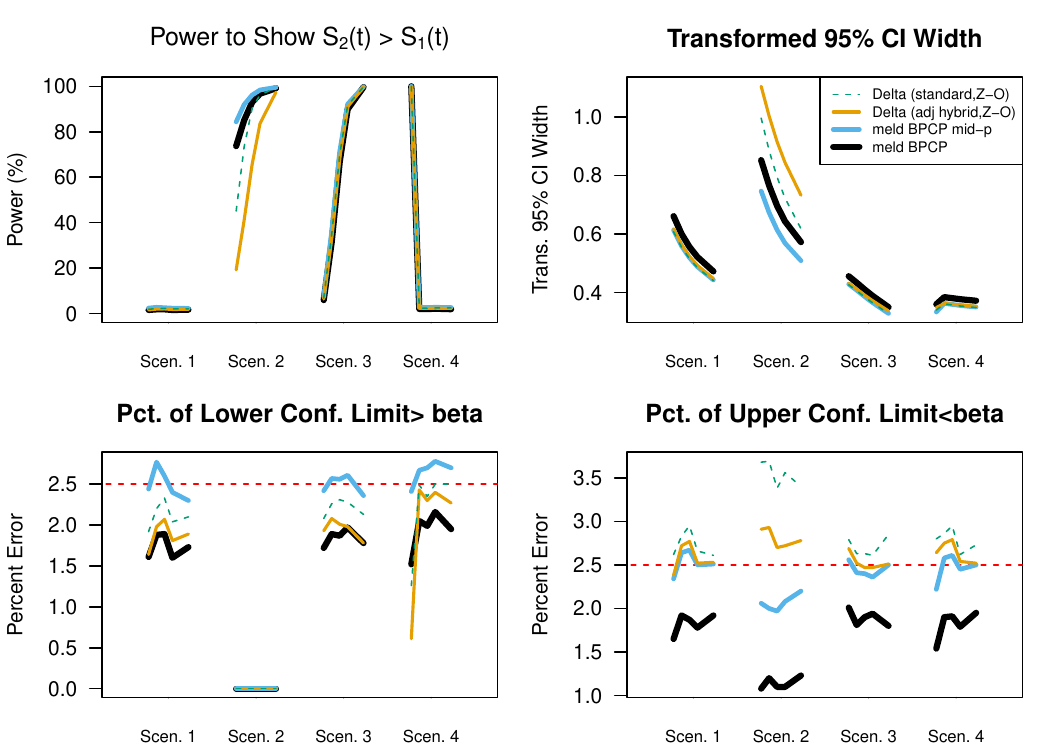}

\caption{
{\bf Simulation Results for $\beta=1-\log(S_2(t))/\log(S_1(t))$ with $n_1=200, n_2=100$ with light censoring.}
In each panel the four lines connect the 5 milestone times for each of the 4 methods.
 Upper left panel is the power to show $S_2(t) >  S_1(t)$ at the one-sided 2.5\% level.
The upper right panel is the 95\% central confidence interval width for the transformed limits (transformation is $\beta/(2-\beta)$ and allows fair comparisons with finite limits).
The lower panels represent the one-sided error rates of the one-sided 97.5\% confidence intervals.
Scenario~1 and the last 4 times of Scenario~4 represent the null case when $S_2(t)=S_1(t)$.  The power for those null case points may be read off of the lower left panel, since in the null case the power is error in the one-sided test.
\label{sfig-RE18}
}

\end{figure}


\clearpage

\section{Simulating a Discrete Problem}
\label{sec-simDiscrete}

In this section, for studying the two-sample problem, we focus on
a case where there may  substantial differences in each of the one-sample confidence intervals. Specifically,  we consider a two-sample problem with $n=300$ in each arm, but by the milestone time $t=5$,
$90\%$ of the study population is censored before the milestone time, leaving small numbers at risk.
Suppose there are three classes of individuals, (1) those that always fail at time 1,
(2) those that are effected by treatment and fail at time 4 if randomized to the treated arm
 but fail at time 2 if randomized to the control arm, and (3) those that have the events after time 6.
 Let the proportions in the population of the classes be $p_A$ (always fail), $p_E$ (effected),
 or $p_N$ (never fail before time 6). Suppose there is independent censoring, where
 $\pi_c=0.90$ of the study population are censored at time 3, and the rest are censored at time 6.
 In this study, the treatment can only delay the event in some people.
 Suppose we test at milestone time $t=5$ to see if there is any long-term effect of the treatment.
 In this case $S_1(t)=S_2(t)$ and there is no effect, but because there is a lot of censoring between the
 times when the treatment is having an effect, there may be differences between the methods.

 For the simulation, we set $p_A=0.50$ and try 41 different values for $p_E$ ($p_E=0,0.01,0.02,$ $ \ldots,$ $0.40$). The expected number at risk in each arm just before
 $t=5$ will be
  $n(1-p_A-p_E)(1-\pi_c) = 3,3.3,\ldots, 15$. We simulated 1,000 data sets for each case.
 We use the efficacy with log(S) estimand which corresponds to the complementary log-log transformation
 (see Table~\ref{tab:betas}) and is expected to have the best power.

For each of the four methods, we calculate the $95\%$ central confidence interval and get the proportion
where the lower limit is greater than $\beta_{els} = \beta_e=0$ and the proportion where the upper limit is less than $\beta_{els} = \beta_e=0$. This corresponds to nominal one-sided alpha levels of 2.5\%.

\begin{figure}[H]
\includegraphics[width=6.0in]{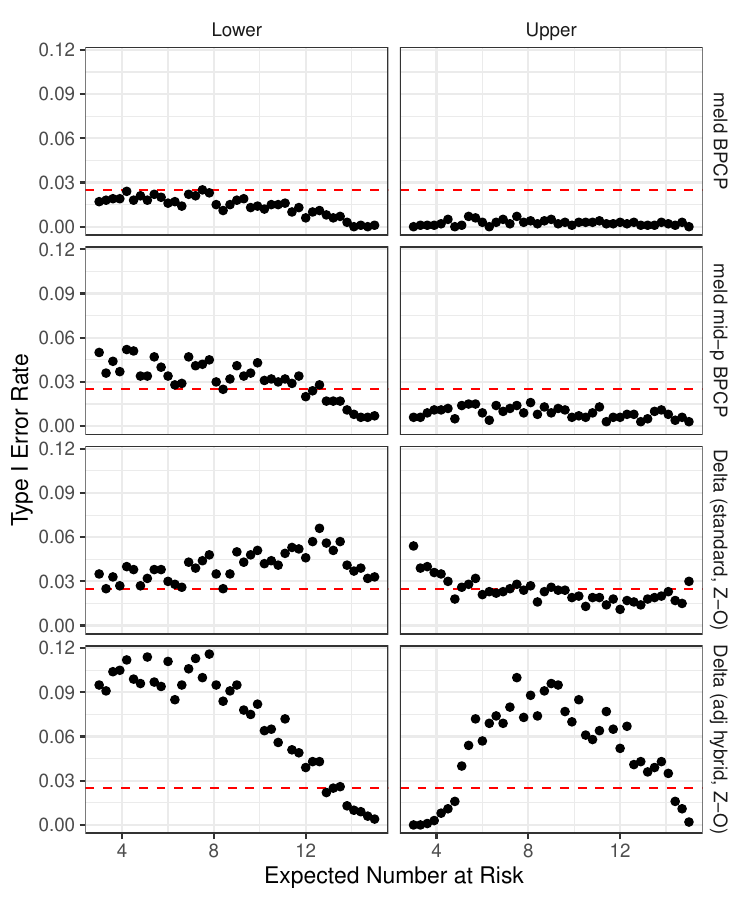}

\caption{
Simulated type I error rates for discrete simulation with censoring for $\beta=1-\log(S_2(t))/\log(S_1(t))$.
Number of replicates per case is $1,000$.
The lower column is the proportion that reject $H_0: \beta \leq 0$
when the lower limit of the 95\% central interval for $\beta$ is larger than $0$.
The  upper column is the proportion that reject $H_0: \beta \geq 0$
when the upper limit of the 95\% central interval for $\beta$ is smaller than $0$.
The four methods are meld BPCP, meld mid-p BPCP,
standard delta method with zero-one adjustments, and 
adjusted hybrid method with zero-one adjustments.
Red dotted lines represent the nominal 2.5\% one-sided $\alpha-$level.
\label{sfig-simDiscreteResults}
}

\end{figure}

  Figure~\ref{sfig-simDiscreteResults} gives the simulation results. Although $S_1(t)=S_2(t)$ at $t=5$,
  before that they are different, so the two one-sided tests are not symmetric.
  Once again, the melded BPCP is the only method that is appears valid (simulated type I error rate is less than $\alpha$ in every scenario). Of the other methods, melded mid-p BPCP appears to be valid for the upper (reject when the upper limit is less than $0$), but not for the lower (reject when the lower limit is greater than $0$).
 The standard delta method using zero-one adjustments appears to bound the type I error rate for the upper limit except for very small expected number at risk, but at the lower limit has errors at rates 3-6\% (similar to the melded mid-p BPCP). The delta method using the adjusted hybrid with zero-one adjustments performs the worst, with simulated type I error rates up to about 9\% or 11\% for the upper and lower limits respectively.




\end{document}